\documentclass[
a4paper,
twocolumn,
amsmath,
aps,pra,reprint,showpacs,
]{revtex4-1}
\usepackage[utf8]{inputenc}
\usepackage{amsmath}
\usepackage{amsfonts}
\usepackage{amssymb}
\usepackage{graphicx}
\usepackage{epstopdf}

\newcommand{\HH}{{\cal H}}
\newcommand{\bE}{{\bf E}}
\newcommand{\br}{{\bf r}}
\newcommand{\bv}{{\bf v}}
\newcommand{\re}{{\rm Re}}
\newcommand{\mm}{\,\,{\rm mm}}
\newcommand{\au}{\,\,{\rm a.u.}}
\newcommand{\Wcm}{\,\,{\rm W}{\rm cm}^{-2}}
\newcommand{\cc}{\,\,{\rm cm}^{-3}}

\begin{document}
\title{Classical versus quantum views of intense laser pulse propagation in gases}
\author{S.A. Berman$^{1,2}$,  C. Chandre$^2$, J. Dubois$^2$, M. Perin$^1$, T. Uzer$^1$}
\affiliation{$^1$School of Physics, Georgia Institute of Technology, Atlanta, Georgia 30332-0430, USA}
\affiliation{$^2$Aix Marseille Univ, CNRS, Centrale Marseille, I2M, Marseille, France}

\begin{abstract}
We study the behavior of reduced models for the propagation of intense laser pulses in atomic gases.
The models we consider incorporate ionization, blueshifting, and other nonlinear propagation effects in an {\it ab initio} manner, by explicitly taking into account the microscopic electron dynamics.
Numerical simulations of the propagation of ultrashort linearly-polarized and elliptically-polarized laser pulses over experimentally-relevant propagation distances are presented.
We compare the behavior of models where the electrons are treated classically with those where they are treated quantum-mechanically.
A classical equivalent to the ground state is found, which maximizes the agreement between the quantum and classical predictions of the single-atom ionization probability as a function of laser intensity. 
We show that this translates into quantitative agreement between the quantum and classical models for the laser field evolution during propagation through gases of ground-state atoms. 
This agreement is exploited to provide a classical perspective on low- and high-order harmonic generation in linearly-polarized fields.
In addition, we demonstrate the stability of the polarization of a nearly-linearly-polarized pulse using a two-dimensional model.
\end{abstract}

\maketitle

\section{Introduction}
The propagation of intense, low-frequency laser pulses through gases triggers a variety of highly nonlinear, nonperturbative phenomena, such as high-harmonic generation (HHG) \cite{Brab00,Gaar08}, terahertz (THz) generation \cite{Amic08,Mart15}, and filamentation \cite{Berg07,Schu17}.
These phenomena intrinsically tie together two disparate length scales: the microscopic scale, defined by the coupling of individual atoms (or molecules) to the electromagnetic field, and the macroscopic scale, defined by the coupling of the electromagnetic field to the mean polarization induced across the entire gas.
Further, the self-consistent interaction between the gas particles and the field plays a paramount role in these processes.
In the case of HHG and THz generation, the observed spectra depend sensitively on the frequency dependence of long-distance phase matching, which hinges on the reshaping of the laser field during propagation by the radiation emitted by ionizing atoms; see \cite{Gaar08,Popm10,John18} for examples in HHG and \cite{Rodr10,Karp09,Babu10} for examples in THz generation.
Similarly, filamentation is born out of the interplay between the radiation of bounded electrons and that of the tunnel-ionized electrons during propagation \cite{Berg07}.
Therefore, theories of these phenomena must bridge the gap between the microscopic electron dynamics and the macroscopic evolution of the laser field during propagation.
The coupling between the microscopic and macroscopic dynamics is the main subject of this paper.

The most accurate description of the laser-gas system is the Maxwell-Schr\"odinger model \cite{Lori07,Lori11}.
This is a first-principles model consisting of Maxwell's equations in three-dimensions for the macroscopic electromagnetic field, with source terms obtained from the microscopic electronic wave functions of the gas atoms.
Due to the high dimensionality and vast separation of scales, simulations of this model under realistic conditions are only feasible using super-computers at present \cite{Lori07,Lori11,Farr11}.
Therefore, reduced models are desirable.
The most popular consist of dimensionally-reduced unidirectional propagation equations for the electomagnetic field \cite{Kole04}, which are typically coupled to models of tunneling ionization \cite{Geis99} and nonlinear susceptibility \cite{Brab97,Kole13} for the low-frequency part of the atomic response, and a semi-classical trajectory model for the high-frequency part \cite{Lewe94,Gaar08}.
The approximations on the atomic response in these models can miss important features of ionization and the generation of THz and low-order harmonic radiation \cite{Xion14,Bree17}, which become particularly important when propagation effects are taken into account \cite{Chri00,Briz13}.

As an intermediate alternative, dimensionally-reduced propagation models which retain a first-principles description of the atomic response may be employed.
For example, a reduced Maxwell-Schr\"odinger model was derived by assuming plane wave solutions of Maxwell's equations, while retaining a first-principles, three-dimensional  Schr\"odinger description of the atomic wave function \cite{Chri98,Shon00}.
This model can be further reduced by considering reduced spatial dimensions for the wave function \cite{Lori08,Lori12,Lyto16}.
These kinds of models can be computationally tractable, and they provide an {\it ab initio} description of numerous ubiquitious propagation effects, including nonlinear susceptibility, ionization losses, dynamical blueshifting \cite{Lee01,Gaar06}, and high-pressure phase-matching \cite{Shon00}.
Surprisingly, despite the advantages of these reduced models, their behavior with experimentally relevant combinations of gas density and propagation length has not been widely explored.
One reason for this is undoubtedly the opaque physical picture provided by the quantum description of the electrons.
This fundamental difficulty persists even on the microscopic level, prompting countless research efforts to date to focus primarily on single-atom quantum dynamics in intense laser fields.
On the contrary, in certain situations, macroscopic phase-matching effects can render the single-atom radiation spectrum unrecognizable \cite{Farr11}, underscoring the importance of taking macroscopic effects into account.

Recently, we introduced a purely classical model that complements reduced Maxwell-Schr\"odinger models and allows the interpretation of high-harmonic spectra in terms of the dynamics of the electron in phase space \cite{Berm18,Berm18_2}.
Our model goes beyond the typical classical-trajectory description of atomic electrons in external fields \cite{Kula93,Cork93,Band90,Both09} by incorporating the self-consistent coupling to the macroscopic electromagnetic field.
Hence, it provides an alternative first-principles description of the coupled electron-laser dynamics that is more physically transparent than the quantum description, while remaining computationally tractable.
We are thus able to perform numerical simulations of our classical model and the corresponding quantum model for atomic gases of realistic density and length.
We showed that the field spectra predicted by our classical model and the quantum model are in quantitative agreement for low frequencies, when the atomic electron is initialized in an ionized state and the field is linearly polarized (LP).
Though the quantum model is necessary for calculating the intensity and phase of the high-harmonic radiation, we showed that the classical model allows the explanation of fine and unexpected features of the quantum spectrum, such as the extension of the high-harmonic cutoff due to propagation effects \cite{Berm18}.

In this paper, we extend our classical model to account for both ground-state atoms and elliptically-polarized (EP) laser fields.
We show that care must be taken in selecting the ground-state initial condition of the classical model, in order to avoid an instability during propagation which is not present in the quantum model.
Then, we investigate the behavior of the classical model with an optimized initial condition {\it vis-a-vis} the quantum model for self-consistently calculated laser pulses propagating through ground-state atomic gases of experimentally-relevant density and length.
Good agreement between the two models is demonstrated for the case of ultrashort LP pulses over a range of laser intensities.
Specifically, the classical model exhibits agreement with the quantum model for low-order harmonics  originating from bound electrons and provides insight into the transition to tunneling currents \cite{Sere14,Babu17} and other nonperturbative processes \cite{Beau16,Yun18} as the primary mechanism of low-order harmonic generation with increasing laser intensity.
Lastly, the classical and quantum models are used to investigate the stability of the polarization of an initially nearly-LP pulse throughout propagation.
Both models predict that the polarization remains steady, justifying the use of maximally-reduced one-dimensional models for the propagation of purely LP pulses.

The paper is organized as follows.
In Sec.~\ref{sec:frame}, we describe the reduced models that we simulate throughout the paper and the observables we use to quantify the results.
In Sec.~\ref{sec:gs}, we show how to build the classical model in such a way that it provides good quantitative agreement with the corresponding quantum model for the propagation of LP pulses through gases of ground-state atoms with experimentally relevant densities and millimeter-scale propagation distances.
In Sec.~\ref{sec:hhspec}, we use the reduced models to examine harmonic generation in LP fields, both from ground-state atoms and pre-ionized atoms.
In Sec.~\ref{sec:EP}, we report the results of simulations of the reduced models with two-dimensional atoms in a nearly-LP field.
We conclude in Sec.~\ref{sec:concl}.
In the appendix, we provide details related to the numerical implementation of the reduced models.
Atomic units are used throughout, unless stated otherwise.

\section{Method}\label{sec:frame}
\subsection{Reduced models}
The reduced models we consider describe the evolution of the electric field ${\bf E}(z,\tau) = E_x(z,\tau)\hat{\bf x} + E_y(z,\tau)\hat{\bf y}$ of a laser pulse propagating in the $z$-direction.
We employ a coordinate frame moving at the speed of light $c$ with the incident laser pulse, i.e.\ $\tau = t-z/c$.
In both the classical and quantum models, the evolution equation for the field is given by
\begin{equation}\label{eq:EOMfield}
\partial_z {\bf E} = \frac{2\pi\rho}{c}\overline{\bf v}(z,\tau),
\end{equation}
where $\rho$ is the number density of the gas and $\overline{\bf v} = \overline{v}_x \hat{\bf x} + \overline{v}_y \hat{\bf y}$ is the mean dipole velocity of the atoms.
This equation is derived in the dipole approximation, under the additional assumptions that the electromagnetic fields are plane waves whose only spatial dependence is on the propagation coordinate $z$, and that backward propagating waves are negligible, i.e.\ the unidirectional approximation \cite{Shon00,Berm18_2}.
The plane wave assumption is an idealization, but it greatly reduces the computational complexity compared to approaches where the fields depend on two \cite{Farr11} or three \cite{Lori07} spatial coordinates.
In all examples considered in this paper, we also assume that $\rho$ is constant, independent of $z$.
In Eq.~\eqref{eq:EOMfield}, the evolution parameter is $z$, and it may be solved as an initial-value problem with initial condition ${\bf E}(0,\tau) = {\bf E}_0(\tau)$.
One simply needs to specify how to compute $\overline{\bf v}$ at a given $z$ for the electric field at that position, ${\bf E}(z,\tau)$.
We integrate Eq.~\eqref{eq:EOMfield} using a finite-difference scheme with a fixed spatial step $\Delta z$, as described in detail in App.~\ref{sec:numericsUni}.

\begin{figure}
\includegraphics[width=0.5\textwidth]{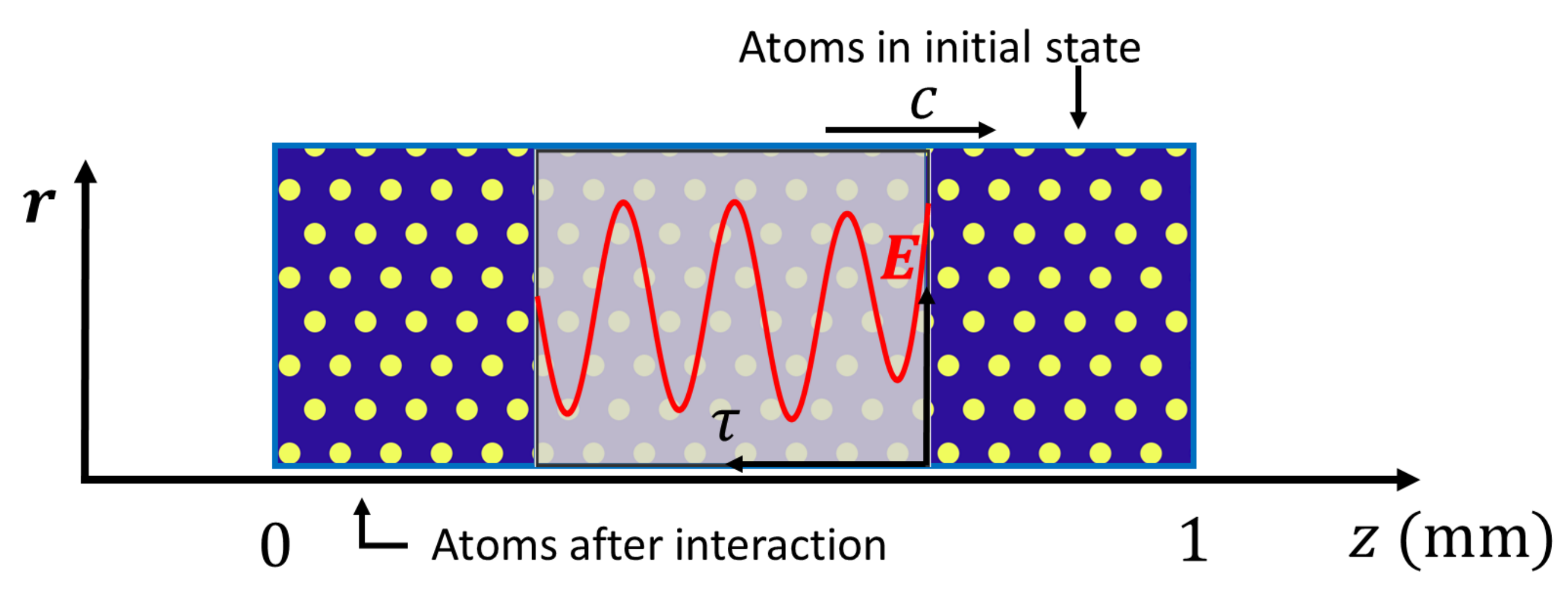}
\caption{Schematic of the reduced model. The time-dependence of the laser electric field $\bE(z,\tau)$ evolves as the pulse position advances in $z$ through the gas.}\label{fig:schematic}
\end{figure}

Here, we restrict our attention to the case where the atoms can be treated in the single-active-electron approximation. 
That is, we assume they consist of a singly-charged ion and an electron.
Further, we assume the electrons only move in the $x$-$y$ plane, i.e.\ the laser polarization plane.
This too is an idealization, but again reduces the computational complexity compared to models describing the three-dimensional (3D) electron motion \cite{Chri98,Shon00}.
Then, in the quantum model, the electron is described by the wave function $\psi(\br,z,\tau)$, where $\br = (x,y)$ is the electron position relative to the ion.
The Schr\"odinger equation governing the evolution of $\psi$ in the dipole approximation and the mean dipole velocity are then given by
\begin{subequations}
\begin{align}\label{eq:EOMq}
i \partial_\tau \psi & = \left[-\frac{1}{2}\nabla^2 + V(r) + \bE(z,\tau)\cdot \br\right]\psi, \\ 
\overline{\bv}(z,\tau) & = \overline{\bv}_0(z) - \int_{0}^\tau \bigg[ \bE(z,\tau') \nonumber \\ \label{eq:DipQ}
& + \int \nabla V(r) |\psi(\br,z,\tau')|^2 {\rm d}^2 \br \bigg] {\rm d} \tau',
\end{align}
\end{subequations}
where $\nabla = (\partial_x,\partial_y)$, $r = |\br|$, and $V$ is the electron-ion interaction potential, which we take to be the soft-Coulomb potential $V(r) = -(r^2+a^2)^{-1/2}$ \cite{Java88,Beck12}.
The quantity $a$ is the softening parameter, which may be adjusted according to the atom one is modeling.
The initial dipole velocity $\overline{\bv}_0(z)$ is given by
\begin{equation*}\label{eq:DipQ0}
\overline{\bv}_0(z) = -i \int \psi^*(\br,z,0) \nabla \psi(\br,z,0) {\rm d}^2 \br.
\end{equation*}
In Eq.~\eqref{eq:DipQ}, we have used Ehrenfest's theorem to express the dipole velocity as the time integral of the dipole acceleration, and we have assumed that the wave function is normalized at all times, i.e.\ $\int |\psi(\br,z,\tau')|^2 {\rm d}^2 \br = 1$.
In numerical simulations, parts of the wave packet eventually leave the computational domain, due to ionization.
Hence, using this form of the dipole velocity ensures that the ``free electron" parts of the wave function continue to contribute through the $\bE$ term in Eq.~\eqref{eq:DipQ}.
This way, we avoid the need of a separate equation to account for free electron effects \cite{Lori11}.
We obtain an approximate numerical solution of Eq.~\eqref{eq:EOMq} using a split-operator method, described in more detail in App.~\ref{A:Schr}.

In the classical model, the dipole velocity is computed by averaging over an ensemble of electron trajectories with a probability distribution on the phase space $f(\br,\bv,z,\tau)$.
The Liouville equation governing the evolution of $f$ in the dipole approximation and the mean dipole velocity are then given by \cite{Berm18,Berm18_2}
\begin{subequations}
\begin{align}\label{eq:EOMc}
& \partial_\tau f  = -\bv \cdot \nabla f + [\nabla V + \bE(z,\tau)]\cdot \partial_{\bv} f, \\ \label{eq:DipC}
& \overline{\bv}(z,\tau) = \int \bv f(\br,\bv,z,\tau) {\rm d}^2 \br {\rm d}^2 \bv.
\end{align}
\end{subequations}
Equation \eqref{eq:EOMc} can be viewed as a classical-trajectory Monte Carlo (CTMC) model \cite{Band90,Both09} in the limit of an infinite number of trajectories.
Importantly, this limit makes our classical model deterministic, not stochastic.
Furthermore, it goes beyond traditional single-atom CTMC calculations with an external field by including macroscopic effects via the coupling to Eq.~\eqref{eq:EOMfield}.
Note that, to obtain approximate numerical solutions of Eq.~\eqref{eq:EOMc}, we integrate an ensemble of deterministically selected trajectories, which are characteristics of Eq.~\eqref{eq:EOMc}, as described in App.~\ref{sec:Liouville}.

Thus, to compute $\overline{\bv}$ at a given $z$ in the quantum model, Eq.~\eqref{eq:EOMq} must be integrated in time, from $\tau=0$ to $\tau=\tau_f$, with the electric field $\bE(z,\tau)$ at that $z$ and an initial condition $\psi(\br,z,0) = \psi_0(\br,z)$.
Here, $\tau_f$ is the final integration time, a fixed parameter.
With the solution $\psi(\br,z,\tau)$ in hand, $\overline{\bv}$  may be evaluated with Eq.~\eqref{eq:DipQ}, and the field equation Eq.~\eqref{eq:EOMfield} may be advanced in $z$.
In the classical model, on the other hand, $\overline{\bv}$ is obtained by integrating Eq.~\eqref{eq:EOMc} from $\tau=0$ to $\tau=\tau_f$ with initial condition $f(\br,\bv,z,\tau) = f_0(\br,\bv,z)$ and applying Eq.~\eqref{eq:DipC}.
The final time $\tau_f$ is selected according to the the latest time of interest in the moving frame.
For example, suppose the initial laser pulse starts at $\tau=0$ and has a duration $T_m$.
If the pulse travels with a group velocity significantly less than $c$ and one is interested in the electron dynamics at the end of the pulse, then one should choose a $\tau_f > T_m$, because in the moving frame, the pulse will end at later and later times $\tau>T_m$ as it propagates.
In general, selecting larger values of $\tau_f$ does not influence the results for times $\tau < \tau_f$; for more details, see App.~\ref{sec:numericsUni}.

The previously described models contain two spatial dimensions for the electric field and electron motion, which is the minimum model dimension for studying EP laser pulses.
For LP pulses, an even simpler model may be considered: a 1D electric field with 1D electron motion along the laser polarization direction.
For most of the paper, we focus on this case, taking $x$ to be the laser polarization direction.
Hence, the model equations are obtained from those given above by the substitutions $\bE \rightarrow E$, $\br \rightarrow x$, and $\bv \rightarrow v$, where we omit $x$ subscripts.
Also, the wave function and distribution function are assumed to be on reduced configuration and phase spaces, respectively, i.e.\ $\psi = \psi(x,z,\tau)$ and $f = f(x,v,z,\tau)$.
This constitutes a further dimensional reduction and hence a further reduction in computational complexity.
We are able to integrate both the quantum and classical 1D models on ordinary desktop computers for realistic sets of parameters in a reasonable amount of time (on the order of hours).

\subsection{Observables}
We assess the behavior of the 1D models by looking at the electric field energy density, electron energy, instantaneous carrier frequency, and high-harmonic spectrum throughout propagation.
Because the field spectrum is typically dominated by a narrow range of frequencies around the laser fundamental $\omega_L$, even after propagation, the field energy and instantaneous carrier frequency mainly reflect this part of the spectrum.
For the 2D models, we assess their behavior by computing the spatiotemporal dependence of the ellipticity of the field $\bE(z,\tau)$.
The definitions of these quantities are given in the following sections.
Note that all the observables pertaining to the 1D models can be readily generalized to the 2D case.

\subsubsection{Electric field energy density}
We define the time-averaged field energy density for the 1D model in the moving frame as 
\begin{equation}\label{eq:Uem}
U_{\rm EM}(z) = \frac{1}{4\pi \tau_f}\int_0^{\tau_f} E(z,\tau)^2 {\rm d}\tau.
\end{equation}
In the lab frame, the conservation of energy allows one to relate the instantaneous field energy to the instantaneous electron energy \cite{Berm18_2}.
This is not possible in the moving frame, but nevertheless the change in $U_{\rm EM}(z)$ may be related to the change in particle energy using Eq.~\eqref{eq:EOMfield}.
In particular, multiplying both sides by $E$ yields
\begin{equation}\label{eq:power}
\partial_z\left(\frac{E^2}{4\pi}\right) = \frac{\rho}{c}\overline{v}(z,\tau)E(z,\tau).
\end{equation}
This equation provides a local energy conservation law in the moving frame, analogous to Poynting's theorem, stating that the change in the field energy density is equal and opposite to the power $-\rho \overline{v} E$ supplied by the field to the electrons.
Integrating Eq.~\eqref{eq:power} over $\tau$ yields
\begin{equation}\label{eq:energy}
\partial_z U_{\rm EM} = -\frac{\rho}{c \tau_f} \Delta \overline{H}(z),
\end{equation}
where $\Delta \overline{H}(z) \equiv \overline{H}(z,\tau_f) - \overline{H}(z,0)$ is the change in mean electron energy $\overline{H}(z,\tau)$, defined precisely below, between times $\tau=0$ and $\tau=\tau_f$ for the electrons located at $z$.
Equations \eqref{eq:power} and \eqref{eq:energy} present one of the advantages of using an {\it ab initio} reduced model: we have exact energy conservation laws coming from first principles \cite{Berm18_2} rather than {\it a posteriori} considerations \cite{Brab00}, relating the energy of the electrons to the energy of the field.
These are useful in practice as a measure of the accuracy of the numerical simulations, as shown in App.~\ref{sec:numericsUni}.

\subsubsection{Electron energy}
For the 1D quantum model, the mean electron energy $\overline{H}$ is defined as the expectation value of the electron Hamiltonian operator in the absence of the electric field,
\begin{equation}\label{eq:energyQ}
\overline{H}(z,\tau) = \int \psi^*(x,z,\tau) \left[-\frac{1}{2}\partial_x^2 + V(x)\right] \psi(x,z,\tau) {\rm d} x.
\end{equation}
Meanwhile, in the classical model it is defined as the ensemble-average of the corresponding classical electron energy $H(x,v) = \frac{v^2}{2} + V(x)$, i.e.\
\begin{equation}\label{eq:energyC}
\overline{H}(z,\tau) = \int H(x,v)f(x,v,z,\tau){\rm d}x{\rm d}v.
\end{equation}
In practice, Eq.~\eqref{eq:energyQ} may be inconvenient to implement, because ionized parts of the electronic wave function may escape outside of the finite computational domain.
Thus, Eq.~\eqref{eq:energyQ} does not account for the energy of this part of the wave function and therefore underestimates the true electron energy.
This effect can be mitigated by choosing large enough computational domains.
On the other hand, this drawback is not present for the implementation of Eq.~\eqref{eq:energyC} because the numerical scheme we choose for solving the Liouville equation consists of integrating the electron trajectories.
 
\subsubsection{Instantaneous carrier frequency}
For computing the instantaneous carrier frequency in the 1D model, we use the Wigner-Ville transform \cite{Boas15} of the electric field
\begin{equation}\label{eq:WVtransform}
W(\tau,\omega;z) = \frac{1}{\pi} \int_{-\infty}^\infty \hat{E}^*(z,\tau - \tau')\hat{E}(z,\tau + \tau') e^{-2 i \omega \tau'} {\rm d}\tau',
\end{equation}
where the asterisk denotes the complex conjugate and $\hat{E}(z,\tau)$ is the analytic representation of the field $E(z,\tau)$.
The analytic representation is, roughly speaking, the inverse Fourier transform of the positive-frequency part of a function's Fourier transform, i.e.\
\begin{align*}
\hat{E}(z,\tau) & = \frac{1}{\pi}\int_0^\infty \tilde{E}(z,\omega) e^{i\omega \tau} {\rm d}\omega,\,\,{\rm where} \\
\tilde{E}(z,\omega) & = \int_{-\infty}^\infty E_p(z,\tau) e^{-i\omega\tau} {\rm d}\tau.
\end{align*}
Here, $E_p$ refers to the field $E$ after post-processing, which may be necessary to perform a meaningful Fourier analysis.
For instance, post-processing may consist of windowing the field with the function $w(\tau)$, in which case $E_p(z,\tau) = E(z,\tau)w(\tau-\tau_c)$.
We specify the post-processing applied for each example we consider.
 
The analytic representation, itself complex, is a useful representation of the real field because it satisfies $E_p(z,\tau) = \re[\hat{E}(z,\tau)]$. Thus, it naturally decomposes the  field into its amplitude, $|\hat{E}(z,\tau)|$, and phase, $\arg[\hat{E}(z,\tau)]$ \cite{Boas15}.
The Wigner transform (which uses $E$ instead of $\hat{E}$ in Eq.\eqref{eq:WVtransform}) has proven effective at analyzing frequency-related propagation effects \cite{Lee01,Hong02}, and we have found that the Wigner-Ville transform is even better suited to this task, especially in the case where $E$ contains multiple frequency components.
At a given $z$, $W(\tau,\omega;z)$ provides information on the frequency content of $E$ at time $\tau$.
When $E$ consists of multiple frequency components,  $W(\tau,\omega;z)$ typically contains several peaks at a given $\tau$, one for each component.
The instantaneous carrier frequency $\omega_c(z,\tau)$ is defined as the frequency such that $W(\tau,\omega;z)$ is the maximum for a given $z$ and $\tau$ \cite{Hong02}.
Further, we define the maximum instantaneous carrier frequency as $\omega_{\rm max}(z) = \max_\tau \omega_c(z,\tau)$.

\subsubsection{High-harmonic spectrum}
We assess high harmonic generation using several methods.
On the field side, we evaluate the field spectrum $|\tilde{E}(z,\omega)|^2$ throughout propagation.
Also, to understand the coherent buildup of radiation in a particular frequency band $[\omega_a,\omega_b]$, we track the evolution of the frequency-filtered analytic field \cite{Gaar08}, i.e.\
\begin{equation}\label{eq:analytic_signal}
\hat{E}_{ab}(z,\tau) = \frac{1}{\pi}\int_{\omega_a}^{\omega_b} \tilde{E}(z,\omega) e^{i\omega \tau} {\rm d}\omega.
\end{equation}

On the particle side, we study the evolution of the spectrogram of the Coulombic part of the dipole acceleration $d_a(\tau) = -\overline{\partial_x V} (\tau)$ in the quantum model, which provides information on the time-frequency properties of the emission \cite{Pukh03,Yako03}.
These spectrograms are related to the statistics of recollisions in the classical model \cite{Berm18}.
We monitor recollisions by computing a quantity $R(\kappa,\tau;z)$ that we call the recollision flux.
This quantity is a measure of the probability of a recollision with kinetic energy $\kappa$ occurring at time $\tau$ for the atoms at $z$, and is defined
\begin{equation}
R(\kappa,\tau;z) =  \int f(x,v,z,\tau) \Theta(x_c-|x|) \Theta(\kappa_c - |v^2/2-\kappa|)\,\mathrm{d}x\mathrm{d}v,
\end{equation}
where $\Theta$ is the Heaviside step function.
The quantities $x_c$ and $\kappa_c$ are adjustable parameters, with $x_c$ being the threshold for recollision and $\kappa_c$ controlling the kinetic energy bin size.
We take $x_c = 5\,\,\mathrm{a.u.}$ and adjust $\kappa_c$ based on the kinetic energy scale of a given simulation.
To gain deeper insight into the classical dynamics, we also visualize $f(x,v,z,\tau)$ itself and examine the electron trajectories which underlie it.

In the quantum model, we also look at the phase of the high-harmonic radiation as a function of $z$.
The source of the radiation is the dipole velocity of the atoms $\overline{v}$.
We compute its phase by first computing the analytic dipole velocity $\hat{\overline{v}}_{ab}(z,\tau)$ in the frequency range of interest, by applying Eq.~\eqref{eq:analytic_signal} to $\overline{v}(z,\tau)$.
Now, the phase of the complex $\hat{\overline{v}}_{ab}(z,\tau)$ tells us the phase of the emitted radiation.
The phase has a natural evolution at the carrier frequency of the radiation in this frequency range, so we must subtract this off to observe variations in the phase about this reference phase.
We define the reference phase $\phi_{\rm ref}(\tau)$ as
\begin{equation*}
\phi_{\rm ref}(\tau) = \arg[\hat{\overline{v}}_{ab}(0,0)] + \left(\frac{1}{\tau_f}\int_0^{\tau_f} \frac{\rm d}{{\rm d} \tau'}\arg[\hat{\overline{v}}_{ab}(0,\tau')] {\rm d}\tau' \right)\tau.
\end{equation*}
Hence, we define the phase of the high-harmonic emission $\phi_{ab}(z,\tau)$ as 
\begin{equation*}
\phi_{ab}(z,\tau) = \arg[\hat{\overline{v}}_{ab}(z,\tau)]-\phi_{\rm ref}(\tau).
\end{equation*}

\subsubsection{Ellipticity}
We evaluate the ellipticity of the 2D electric field $\bE(z,\tau)$ using the Stokes parameters \cite{Anto96}.
The Stokes parameters are defined for a complex field, $\hat{\bE} = (|\hat{E}_x|e^{-i \phi_x}, |\hat{E}_y|e^{-i \phi_y})$, where $\phi_x = -\arg \hat{E}_x$ and $\phi_y = -\arg \hat{E}_y$, as
\begin{subequations}\label{eq:stokes}
\begin{align}
& s_0 = |\hat{E}_x|^2 + |\hat{E}_y|^2, \\
& s_1 = |\hat{E}_x|^2 - |\hat{E}_y|^2, \\
& s_2 = 2 |\hat{E}_x| |\hat{E}_y| \cos(\phi_y - \phi_x), \\
& s_3 = 2 |\hat{E}_x| |\hat{E}_y| \sin(\phi_y - \phi_x).
\end{align}
\end{subequations}
With these definitions, the relation $s_0^2 = s_1^2 + s_2^2 + s_3^2$ is satisfied, and physically it corresponds to completely polarized light.
From the Stokes parameters, one can calculate the parameters characterizing the polarization ellipse, namely, the angle $\theta$ of the major axis of the ellipse with respect to the $x$-axis, and the ellipticity $\xi$, the ratio of the minor to major axes of the ellipse (with a positive sign signifying counter-clockwise rotation).
In experiments, the Stokes parameters can be measured.
However, the measurements correspond to space and time averages of the field, which is generally partially polarized.
Since we compute $\bE(z,\tau)$, we know the exact spatiotemporal dependence of the field and do not need to average.
This allows us to define instantaneous polarization ellipse angles and ellipticities $\theta(z,\tau)$ and $\xi(z,\tau)$.
We do this by using the analytic representation of the calculated field $\hat{\bE}(z,\tau)$ (see Eq.~\eqref{eq:analytic_signal}) in Eqs.~\eqref{eq:stokes}, leading to space- and time-dependent Stokes parameters.
Then, using the relations
\begin{align*}
\tan 2\theta = \frac{s_2}{s_1}, \\
\sin 2 \chi = \frac{s_3}{s_0}, \\
\xi = \tan \chi,
\end{align*}
we obtain $\theta(z,\tau)$ and $\xi(z,\tau)$.

\section{Propagation of a pulse through a gas of one-dimensional ground-state atoms}\label{sec:gs}
\subsection{Selection of initial conditions}\label{sec:ics}
\begin{figure*}
\includegraphics[width=0.75\textwidth]{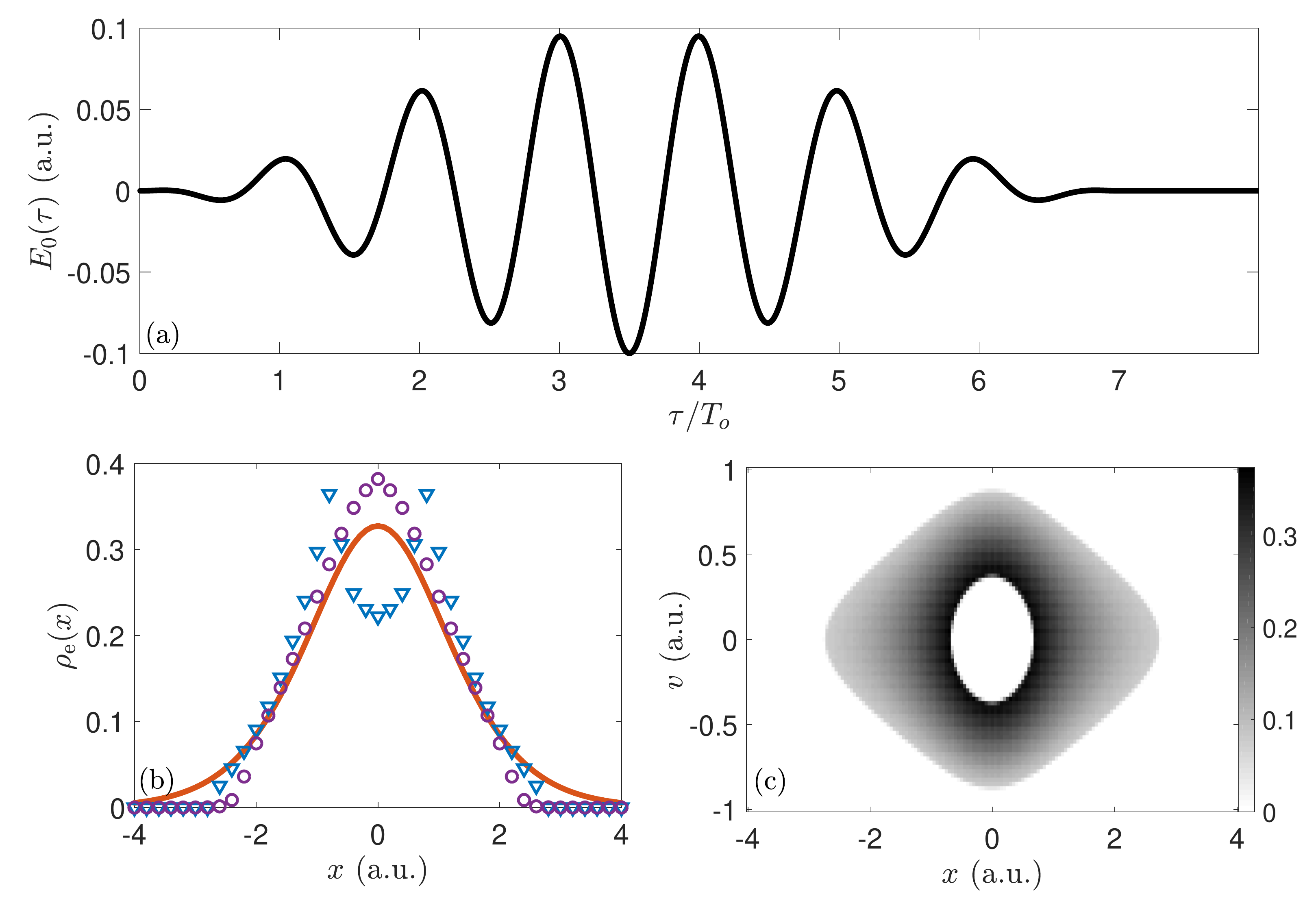}
\caption{Initial conditions of the ground-state simulation. (a) Initial electric field $E_0(\tau)$ (see Eq.~\eqref{eq:incident}), with a peak intensity $I=3.5\times10^{14} \Wcm$. (b) Initial microscopic electron density $\rho_{\rm e}(x)$. The solid orange curve is for the quantum model, the blue triangles are for the classical model with initial energy distribution $g_1$, and the purple circles are for the classical model with $g_\sigma$.  (c) Initial electron phase space distribution $f_0(x,v)$ for the classical model with $g_1$, with the probability density indicated by the linear color scale.}\label{fig:gs_ICs}
\end{figure*}
We begin by applying the 1D reduced models to the propagation of a laser pulse through a gas of ground-state atoms.
The initial conditions are plotted in Fig.~\ref{fig:gs_ICs}.
We take an incident laser pulse given by
\begin{equation}\label{eq:incident}
E_0(\tau) = \begin{cases}
E_0 \sin^2(\frac{\pi \tau}{T_m})\cos(\omega_L \tau) & {\rm for}\,\,\ 0 < \tau < T_m, \\
0 & {\rm for}\,\, T_m < \tau < \tau_f,
\end{cases}
\end{equation}
where $E_0$ is the maximum field amplitude, $\omega_L$ is the laser frequency, and $T_m$ is the duration of the laser pulse.
We relate $E_0$ in atomic units to the peak intensity of the pulse $I$ in $\Wcm$ using $E_0 = 5.338\times 10^{-9} \sqrt{I}$.
For the rest of the parameters, we fix $\omega_L = 0.0378\,\,{\rm a.u.}$, $T_m=7T_o$, and $\tau_f=8T_o$, where $T_o = 2\pi/\omega_L$ is one optical cycle.
These values correspond to a laser wavelength $\lambda_L=1.2\,\,\mu{\rm m}$, and a FWHM pulse duration of $T_m/2 = 14\,\,{\rm fs}$.
We consider positions of the laser pulse between $z=0$ and $z=1\,\,{\rm mm}$, where the gas is assumed to have a constant density $\rho$.
Everywhere else is assumed to be vacuum, as illustrated in Fig.~\ref{fig:schematic}, and the field does not evolve in those regions.

For the quantum model, the initial state of the electron is taken as the ground state of Eq.~\eqref{eq:EOMq} in the absence of the electric field.
We take the softening parameter $a = \sqrt{2}$, which means the ground state has energy $I_p=-0.5\,\,{\rm a.u.}$ and the form \cite{Majo18}
\begin{equation*}
\psi_0(x) = N_\psi \left(1 + \sqrt{x^2 +2}\right)\exp\left[-\sqrt{x^2+2}\right],
\end{equation*}
where $N_\psi$ is a normalization constant.
Note that we take $\psi_0$ to be independent of $z$, because the initial state of the atoms is assumed to be uniform.

For the classical model, a variety of options have been considered for designing a suitable initial phase space distribution $f_0(x,v)$ corresponding to the quantum ground state, given that it is not possible to obtain one in a strictly self-contained manner.
We aim to maximize the quantitative agreement between the classical and quantum models, where the main difficulty is getting the classical model to exhibit a similar intensity-dependent ionization probability as the quantum model.
In this respect, the most na\"ive option for the classical ground state---a microcanonical ensemble at the quantum ground-state energy $h=I_p$---does not perform well for the one-dimensional (1D) SAE model, in part because the onset of ionization occurs too suddenly \cite{Rich96,Both09}.
As an alternative, we take a distribution of initial energies, $h \in [h_{\rm min},h_{\rm max}]$ with a probability density $g(h)$, leading to a distribution function of the form \cite{Both09}
\begin{equation}\label{eq:classicalGs}
f_0(x,v) = \int_{h_{\rm min}}^{h_{\rm max}} g(h) N_h \delta(h - H(x,v)) {\rm d}h.
\end{equation}
Meanwhile, $N_h$ is a normalization constant such that 
\begin{equation}\label{eq:energyNorm}
N_h \int \delta(h-H(x,v)) {\rm d}x{\rm d}v = 1,
\end{equation}
and its calculation is discussed in App.~\ref{sec:Liouville}.
Note that, because $f_0$ can be written as a function of $H$, which is conserved along a trajectory in the absence of the electric field, it is a stationary state of the field-free Liouville equation \eqref{eq:EOMc}.

\begin{figure*}
\includegraphics[width=0.85\textwidth]{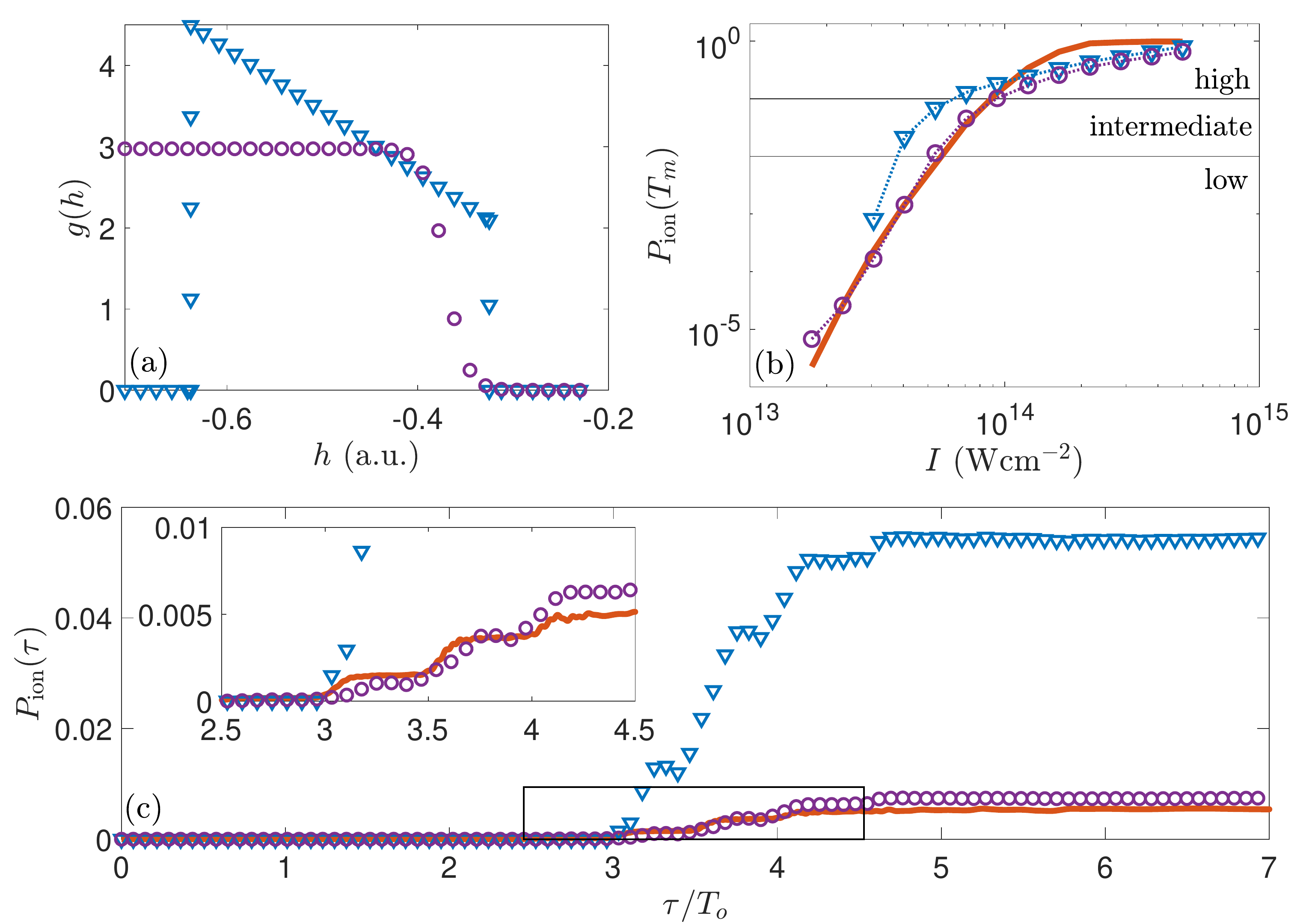}
\caption{Classical energy distributions and ionization probability. The blue triangles correspond to the classical model with $g_1$, the purple circles correspond to the classical model with $g_\sigma$, and the orange solid curves correspond to the quantum model. (a) Distribution of initial energies $g(h)$  for the classical model. (b) Probability of ionization at the end of the laser pulse $P_{\rm ion}(T_m)$ as a function of the peak intensity $I$ of the incident pulse. Regions of low ionization probability (less than $1\%$), intermediate ioniation probability ($1$--$10\%$), and high ionization probability (greater than $10\%$) are indicated. (c) Time-dependent ionization probability for an incident pulse with peak intensity $I=5\times 10^{13}\Wcm$. The inset shows a magnification of the box.}\label{fig:energy_ionization}
\end{figure*}
In the first example, we choose the energy distribution to emulate the one employed in Ref.~\cite{Both09}.
There, the energy distribution is obtained by truncating the energy distribution of the Wigner distribution function of the corresponding quantum ground state.
The resulting classical calculations of the intensity-dependent ionization probability are shown to have reasonable agreement with the quantum calculations. 
We select the same energy range as in Ref.~\cite{Both09}, so that $h_{\rm min,1} = -0.638\au$ and $h_{\rm max,1} = -0.325\au$
Their truncated-Wigner energy distribution appears nearly linear, so we estimate it by the distribution $g_1(h) = a_0 + a_1 h$, with $a_1 = -7.656\au$ and $a_0=-0.4915\au$ chosen such that $g_1(h)$ is normalized to one.
Our distribution satisfies $\overline{h} = \int h g_1(h){\rm d}h = -0.5011 \au \approx I_p$.
This suggests it is a good approximation to the distribution of Ref.~\cite{Both09}, for which the mean energy of the classical ensemble $\overline{h}$ is exactly equal to the quantum ground state energy $I_p$.

The distribution $g_1$ is plotted in Fig.~\ref{fig:energy_ionization}a, along with the microscopic electron density $\rho_e(x)$ in Fig.~\ref{fig:gs_ICs}b and the phase space distribution in Fig.~\ref{fig:gs_ICs}c.
For the classical model, $\rho_e(x) = \int f_0(x,v) {\rm d}v$, while for the quantum model, $\rho_e(x) = |\psi_0(x)|^2$, and this is also plotted in Fig.~\ref{fig:gs_ICs}b for comparison.
The quantum electron density exhibits a single peak at the origin, typical of the ground-state electron density.
Meanwhile, the classical electron density with $g_1$ as the energy distribution exhibits two sharp peaks, symmetric about $x=0$.
This resembles the shape of the electron density of a highly excited yet bound quantum eigenstate, which is typically flanked by two sharp peaks around the classical turning points of the binding potential.
However, the location of the peaks here are closer to $x=0$ than the turning points of the classical state with the quantum ground state energy $I_p$, which would be at $|x|=1.41\au$
Hence, the shape of the classical electron density with $g_1$ is not simply identified with that of the quantum ground state nor any of the excited states.

Besides this qualitative difference between the classical and quantum representations of the electron ground state, the two models also differ quantitatively on their predictions of the intensity-dependent ionization probability, plotted in Fig.~\ref{fig:energy_ionization}c.
We define the time-dependent ionization probability $P_{\rm ion}(\tau)$ as the probability of finding the electron with $|x| > 10\,\,{\rm a.u.}$ at time $\tau$.
Then, the probability of the electron being ionized at the end of the pulse is $P_{\rm ion}(T_m)$.
In Fig.~\ref{fig:energy_ionization}b, we compare the variation of this quantity as a function of the peak-intensity $I$ of the incident laser pulse given by Eq.~\eqref{eq:incident} for the quantum model and the classical model with initial energy distribution $g_1$.
Evidently, for this laser frequency and pulse duration, there is much room for improvement in terms of the quantitative agreement between intensity-dependent ionization probabilities.
For example, for intensities below $I=3\times10^{13}\Wcm$, the classical model with $g_1$ predicts no ionization, whereas the probability is nonzero in the quantum case.
Then, between $I=3\times10^{13}\Wcm$ and $I=10^{14}\Wcm$, the classical model with $g_1$ overestimates the ionization probability compared to the quantum case, and for even higher intensities it reverts to underestimating it.
We shall show below that these discrepancies impact the agreement between the quantum and classical models for the evolution of the laser pulse during propagation.

\begin{figure*}
\includegraphics[width=0.85\textwidth]{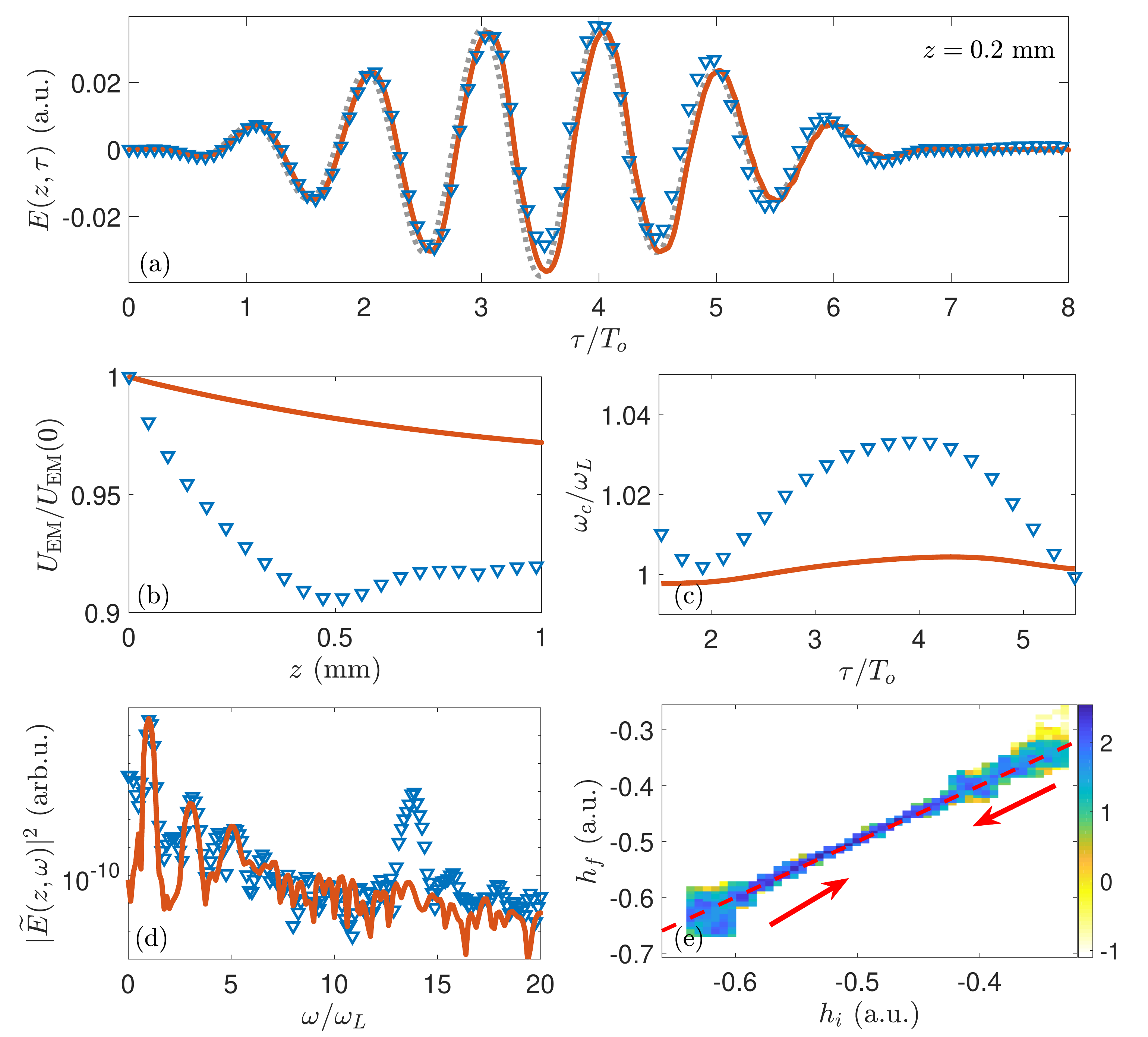}
\caption{Results of pulse propagation through $1\,\,{\rm mm}$ of ground-state atoms with density $\rho = 2\times10^{19}\,\,{\rm cm}^{-3}$, with a peak incident pulse intensity of $I=5\times 10^{13}\Wcm$ and carrier frequency $\omega_L=0.0378\au$ The orange solid curves correspond to the quantum model, while the blue triangles correspond to the classical model with $g_1$. (a) Time-dependent electric field $E(z,\tau)$ at $z = 0.2\,\,{\rm mm}$. The grey dotted curve is the initial field $E_0(\tau)$. (b) Normalized time-averaged pulse energy density $U_{\rm EM}$ as a function of $z$. (c) Instantaneous carrier frequency $\omega_c(z,\tau)$ at $z=0.2\mm$. (d) Power spectrum of the filtered electric field $|\widetilde{E}(z,\omega)|^2$ at $z = 0.2\,\,{\rm mm}$. (e) Probability density for transitioning from a state with initial energy $h_i$ to a state with final energy (at $\tau=\tau_f$) $h_f$ for the classical model at $z=0.2\,\,{\rm mm}$. The density is indicated by a logarithmic color scale. The red dashed line is $h_f=h_i$. The spatial step used is $\Delta z = 1.3\lambda_L = 1.57\,\,\mu{\rm m}$.}\label{fig:BPfail}
\end{figure*}
Given these initial conditions, we simulate the propagation of the pulse from $z=0$ to $z=1\,\,{\rm mm}$ for a gas with density $\rho = 2\times 10^{19}\,\,{\rm cm}^{-3}$, corresponding to a room-temperature gas at atmospheric pressure, using the quantum model and the classical model with initial energy distribution $g_1$.
We choose the peak intensity of the incident pulse as $I = 5\times10^{13}\Wcm$, which is a low ionization probability regime.
Here, the quantum model gives  $P_{\rm ion}(T_m) \approx 0.5 \%$ and the classical model with $g_1$ gives $P_{\rm ion}(T_m) \approx 5\%$ for the incident pulse, as shown in Fig.~\ref{fig:energy_ionization}c.
In Fig.~\ref{fig:BPfail}a, we compare the electric fields $E(z,\tau)$ at $z=0.2\,\,{\rm mm}$ computed from each model. 
Up to this distance, the dominant propagation effects are captured by both models.
For example, the group velocity of the pulse is noticeably less than $c$ in both calculations.
In the moving frame, this is evidenced by the temporal shift of the pulse to the right of the initial pulse $E_0(\tau)$, as seen in Fig.~\ref{fig:BPfail}a.
This effect is captured equally well by the quantum and classical calculations for $\tau/T_o < 3$, and the agreement between the fields for larger $\tau$ is fair.
For larger times, both calculations also predict a time-dependent blueshift, as seen in the instantaneous carrier frequency $\omega_c(z,\tau)$ at $z=0.2\mm$, plotted in Fig.~\ref{fig:BPfail}c.
The classical calculation yields a much larger blueshift than the quantum calculation, because blueshifting is caused by ionized electrons \cite{Kim02,Gaar06}, and the classical model has a higher ionization probability than the quantum model for this $I$.
The higher ionization probability is consistent with higher ionization losses observed in the classical calculation compared to the quantum calculation, as evidenced by the more rapidly decreasing pulse energy density $U_{\rm EM}(z)$, plotted in Fig.~\ref{fig:BPfail}b. 
In summary, the classical and quantum calculations for these low-frequency observables are qualitatively similar up to $z=0.2\mm$, owing their quantitative disagreement to the discrepancy in ionization probability.

\begin{figure*}
\includegraphics[width=0.85\textwidth]{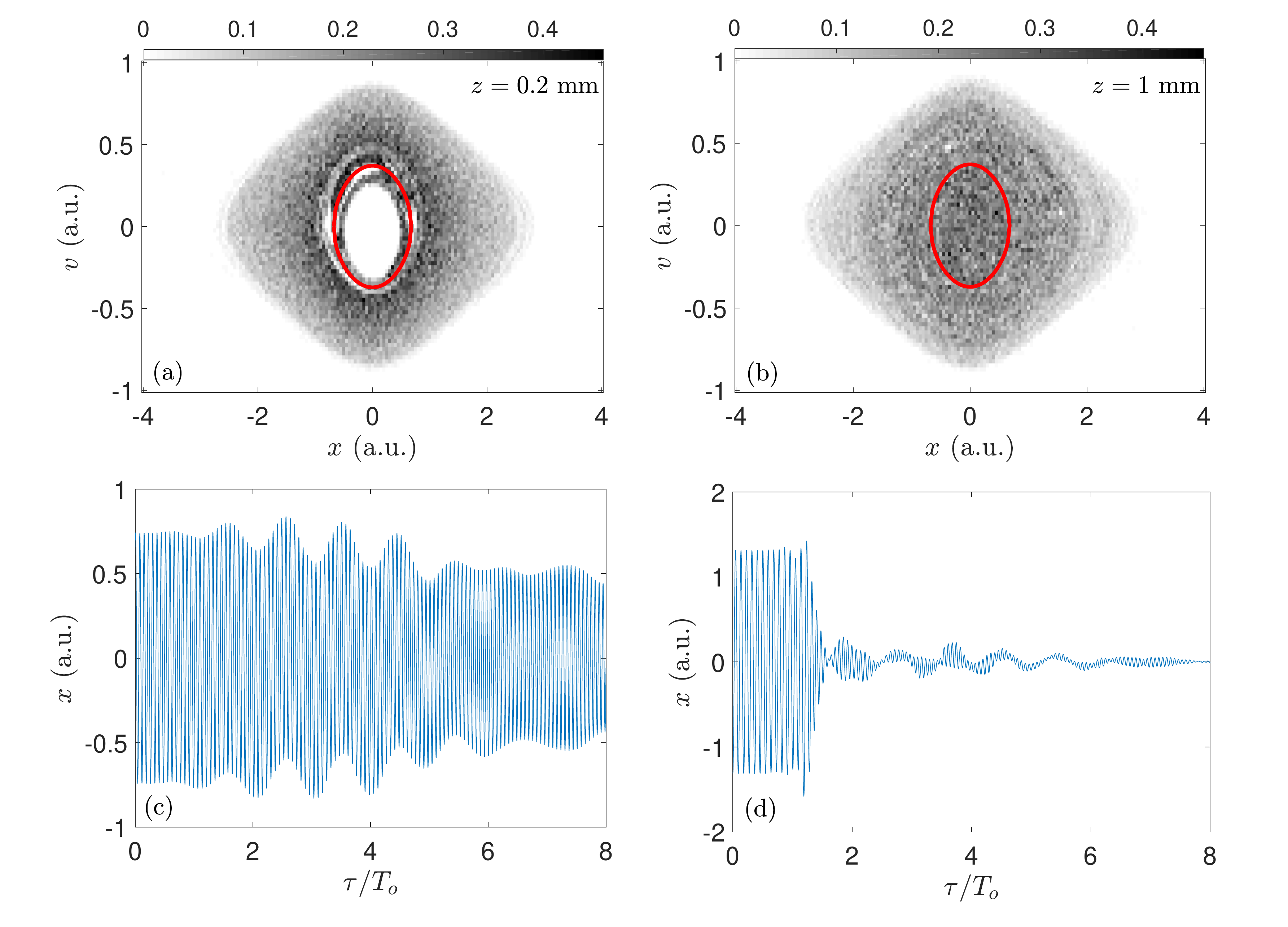}
\caption{Electron energy loss in the classical model with the parameters of Fig.~\ref{fig:BPfail}. (a),(b) Electron distribution function at the end of the laser pulse $f(x,v,z,\tau_f)$. The red curve indicates the initial minimum energy $H(x,v) = h_{\min,1}$. (c),(d) Electron trajectory $x(\tau)$ which ends with the smallest energy $h_f$. Panels (a) and (c) are computed at $z=0.2\,\,{\rm mm}$, while (b) and (d) are computed at $z=1\,\,{\rm mm}$.}\label{fig:BPfail_phase}
\end{figure*}
However, for larger $z$, the classical calculation significantly departs from the quantum calculation.
One symptom of the problem is seen in Fig.~\ref{fig:BPfail}b, where $U_{\rm EM}$ actually starts increasing at around $z=0.5\mm$.
By energy conservation, i.e.\ Eq.~\eqref{eq:energy}, this implies the mean electron energy must experience a net decrease.
While electron energy loss is virtually nonexistent for the atoms at $z=0$, such behavior manifests itself in the course of pulse propagation.
We see the precursors to this behavior by $z=0.2\mm$, as shown in Fig.~\ref{fig:BPfail}e and Fig.~\ref{fig:BPfail_phase}a.
In Fig.~\ref{fig:BPfail}e, we have plotted the joint distribution of the initial electron energy $h_i$ and final electron energy $h_f$, i.e.\ at the end of the pulse, computed from the classical model.
We see that classically, it is possible for an electron to lose energy, evidenced by the nonzero probability for $h_f < h_i$, i.e. below the dashed line.
In particular, states with energies lower than $h_{\min,1}$ become populated by the end of the pulse.
These are seen clearly in the distribution function $f(x,v,z,\tau_f)$ at $z=0.2\,\,{\rm mm}$, shown in Fig.~\ref{fig:BPfail_phase}a, where the states inside of the red ring have an energy $h_f < h_{\min,1}$.

Further, the electron energy loss becomes more severe as $z$ increases, as seen in $f(x,v,z,\tau_f)$ and the joint $h_i$-$h_f$ distribution at $z = 1\,\,{\rm mm}$ in Fig.~\ref{fig:BPfail_phase}b and Fig.~\ref{fig:BPfail1}c, respectively.
Eventually, the energy lost by these electrons outweighs the energy gained by the ionized electrons, leading to the increase in field energy seen in Fig.~\ref{fig:BPfail}b around $z= 0.5\,\,{\rm mm}$.
In the quantum model, this cannot possibly happen when all the electrons are initialized in the lowest possible energy state, i.e.\ the ground state.
There, the pulse is always losing energy throughout propagation because the gas is always strictly gaining energy by excitation and ionization of the atoms.
Hence, a net energy increase at any point throughout the pulse propagation is an unphysical effect that we would like to avoid when using the classical model for a gas of ground-state atoms.

\subsection{Improving the classical model}
\begin{figure*}
\includegraphics[width=0.85\textwidth]{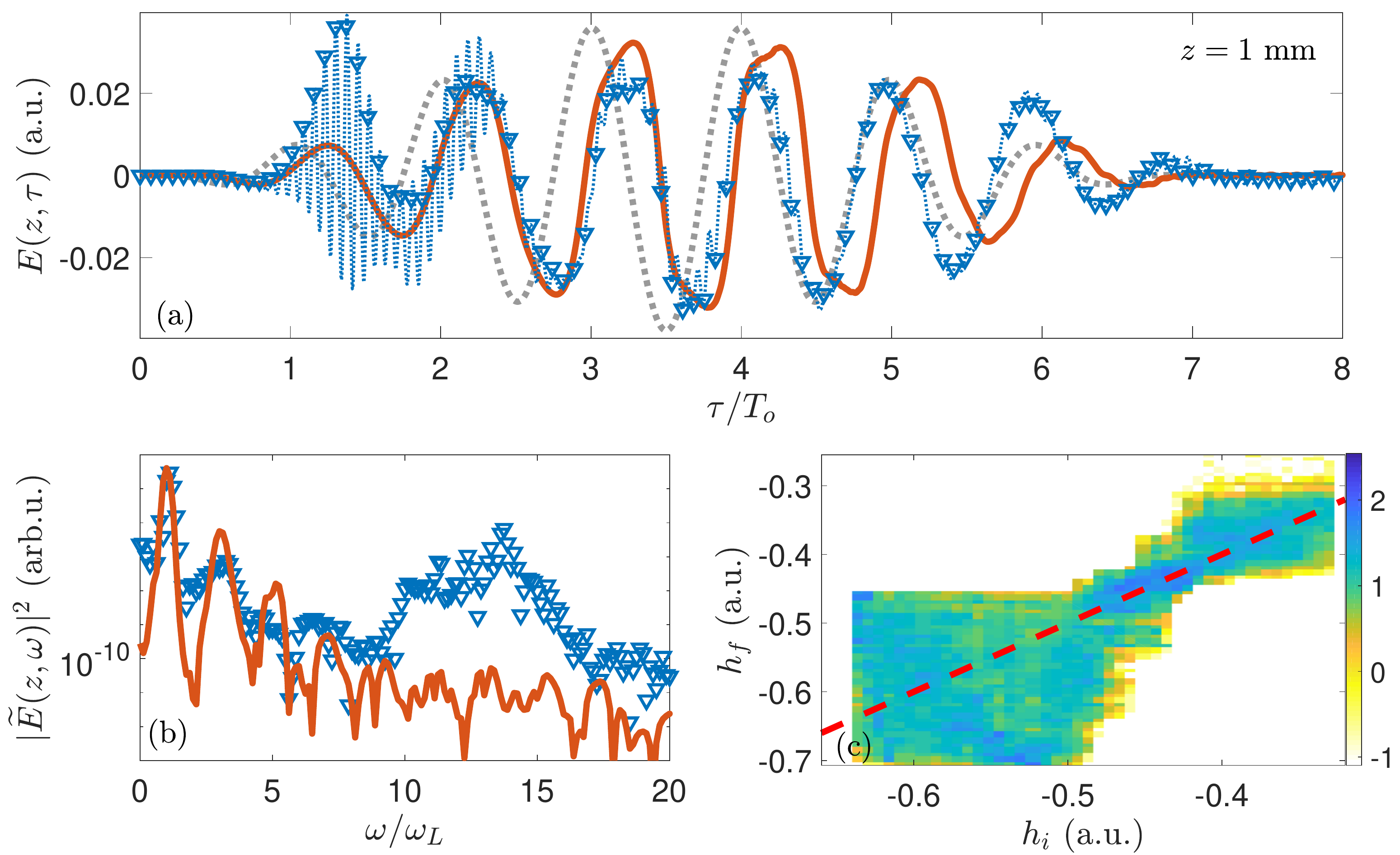}
\caption{Same calculation as Fig.~\ref{fig:BPfail}, with the results shown at $z=1\,\,{\rm mm}$. The orange solid curves correspond to the quantum model, while the blue triangles and thin dotted lines correspond to the classical model with $g_1$. (a) Time-dependent electric field $E(z,\tau)$. The grey dotted curve is the initial field $E_0(\tau)$. (b) Power spectrum of the filtered electric field $|\widetilde{E}(z,\omega)|^2$. (c) Probability density for transitioning from a state with initial energy $h_i$ to a state with final energy (at $\tau=\tau_f$) $h_f$ for the classical model. The density is indicated by a logarithmic color scale. The red dashed line is $h_f=h_i$.}\label{fig:BPfail1}
\end{figure*}
The agreement between the quantum and classical models may be improved by altering the initial energy distribution $g$ used in the classical model.
Looking more closely at the mechanism for the anomalous electron energy loss in the classical model provides some intuition on what refinements need to be made to $g$ to effect this improvement.
In Fig.~\ref{fig:BPfail}e and \ref{fig:BPfail1}c, we have plotted the joint distribution of $h_f$ and $h_i$ for electrons at $z=0.2\,\,{\rm mm}$ and $z=1\,\,{\rm mm}$, respectively.
Here, we see that the electrons most likely to lose energy are those with initial energies near the two possible extremes, $h_{{\rm min},1}$ and $h_{{\rm max},1}$.
In between these extremes, the energies of the electrons remain more or less at their initial values, a signature of bounded electron motion \cite{Rich96} on invariant tori \cite{Maug09_1,Maug10_2}.
As propagation proceeds, electron energy loss becomes more and more probable.
The range of energies at which this happens gradually creeps inward from both extremes, as indicated by the arrows on Fig.~\ref{fig:BPfail}e and confirmed by Fig.~\ref{fig:BPfail1}c.

At the same time, we observe a resonant-like growth of the electric field modes with frequencies near $\omega = 14\omega_L$, visible in the spectra of the field $|\widetilde{E}(z,\omega)|^2$ at $z = 0.2\,\,{\rm mm}$ in Fig.~\ref{fig:BPfail}d and $z=1\,\,{\rm mm}$ in Fig.~\ref{fig:BPfail1}b.
The classical spectrum displays a broad peak at these modes, whose intensity grows rapidly in $z$ and leads to a highly distorted electric field for $z > 0.2\,\,{\rm mm}$.
For example, the classically-calculated field at $z=1\,\,{\rm mm}$, plotted in Fig.~\ref{fig:BPfail1}a, exhibits large, rapid oscillations for $\tau < 2.5T_o$ due to the radiation near $\omega=14\omega_L$.
This is in stark contrast to the field of the quantum calculation. 
Indeed, the quantum spectrum lacks a conspicuous peak around $\omega = 14\omega_L$ at $z=0.2\,\,{\rm mm}$, shown in Fig.~\ref{fig:BPfail}d, as well as at larger values of $z$, as shown in Fig.~\ref{fig:BPfail1}b for $z=1\,\,{\rm mm}$.
In response to an incident quasi-monochromatic field, the classical single-atom model is known to exhibit radiation at frequencies near that of the field-free bounded electron motion \cite{Band92,Leop93}.
Hence, it is plausible that the classical atoms radiate strongly at the frequency $\nu$ corresponding to the electron orbit with energy $h_{\min,1}$, because the initial electron energy distribution $g_1(h)$ is effectively peaked at $h_{\min,1}$.
The energy-dependent frequency of the field-free orbits $\nu(h)$ can be approximated  by expanding the exact expression for $\nu$ for small $h-h^*$ \cite{Maug10_2}, where $h^* = -1/a$ is the energy of the equilibrium at $x=0$ and the lower bound on the electron energy in the classical model.
In this case, the expansion to leading order (see Ref.~\cite{Maug10_2} for more details) is
\begin{equation}\label{eq:omegaA}
\nu(h) \approx a^{-3/2} - \frac{9}{8}a^{-1/2}(h - h^*).
\end{equation}
Using Eq.~\eqref{eq:omegaA}, we obtain $\nu(h_{\min,1})\approx 14\omega_L$, in striking agreement with the location of the broad peak in Fig.~\ref{fig:BPfail}c.

Thus, it appears that the radiation generated near this frequency builds up in the early part of the gas (for small $z$), until it is strong enough to interact resonantly with the electrons naturally oscillating at those same frequencies in the later parts of the gas.
This likely causes the electron energy loss observed at increasing $z$, beginning with the electrons with energies $h_i \approx h_{\min,1}$, while continuing to feed the growth of radiation at frequencies near $14\omega_L$.
Evidence of this is shown in Figs.~\ref{fig:BPfail_phase}c and \ref{fig:BPfail_phase}d, where we have plotted the trajectories $x(\tau)$ of the electron with the smallest energy at $\tau_f$, at $z=0.2\,\,{\rm mm}$ and $z=1\,\,{\rm mm}$, respectively.
At $z=0.2\,\,{\rm mm}$, the electron experiences a gradual energy loss, inferred from the gradually decreasing amplitude of the oscillations.
Eventually, this becomes a sudden drop in energy, as seen for the electron trajectory at $z=1\,\,{\rm mm}$ in between $\tau=T_o$ and $\tau=2T_o$.
It is also during this time period that the high-frequency oscillations in the classically-calculated electric field are particularly prominent, as seen in Fig.~\ref{fig:BPfail1}a.
This suggests that these oscillations are at a similar frequency but out of phase with the electron motion in Fig.~\ref{fig:BPfail_phase}d, leading to the electron's rapid energy loss.
 
Given these observations, we propose to improve the classical model by judiciously selecting another initial energy distribution $g$ to mitigate the pitfalls of $g_1$.
On the one hand, we want a distribution that improves the agreement between the classical and quantum  intensity-dependent ionization probabilities $P_{\rm ion}(T_m)$.
This ensures the classical model can mimic the quantum model with respect to the blueshift, ionization losses, and subluminal group velocity of the pulse.
On the other, we want a distribution which is not sharply peaked at an energy greater than $h^*$.
Avoiding this may prevent the amplification of the radiation of bounded electrons that subsequently also appears to trigger their energy loss.

We have found that the sigmoid distribution meets these criteria.
The distribution is given by
\begin{equation}\label{eq:gSigma}
g_\sigma(h) = \frac{N_\sigma}{1 + \exp{[k(h - h_m)}]},
\end{equation}
defined on the energy range $[h^*, h_{\max,\sigma}]$, with $N_\sigma$ a normalization constant.
We choose $h_{\max,\sigma}=-0.23\,\,{\rm a.u.}$
The free parameters $k$ and $h_m$ are optimized to maximize the agreement between the classical and quantum predictions for $P_{\rm ion}(T_m)$ for the values of intensity plotted in Fig.~\ref{fig:energy_ionization}, yielding $k=93.22\,\,{\rm a.u.}$ and $h_m=-0.3709\,\,{\rm a.u.}$
The distribution $g_\sigma$ with the optimized parameters is plotted in Fig.~\ref{fig:energy_ionization}a, and the microscopic electron density $\rho_e(x)$ obtained from $g_\sigma$ is plotted in Fig.~\ref{fig:gs_ICs}b.
Compared to the electron density of $g_1$, the electron density of $g_\sigma$ is more similar to its quantum counterpart, exhibiting a single peak at the origin.
Hence, $g_\sigma$ provides a more physically reasonable representation of $\rho_e(x)$ than $g_1$.
Furthermore, in Fig.~\ref{fig:energy_ionization}b, we see that the ionization probabilities of $g_\sigma$ agree very well with the quantum ones for ionization probabilities below about $10\%$, in stark contrast to those of $g_1$.
For higher ionization probabilities, the performance of $g_\sigma$ is similar to $g_1$, with the ionization probabilities actually being slightly lower in this range.
To illustrate the improved performance of $g_\sigma$ compared to $g_1$ in the low ionization probability regime, we compare the quantum and classical predictions of $P_{\rm ion}(\tau)$ in Fig.~\ref{fig:energy_ionization}c for an external pulse with $I=5\times10^{13}\Wcm$.
Clearly, the $g_\sigma$ ionization probability is much closer to the quantum calculation than that of $g_1$ for all times $\tau>3T_o$, when significant ionization begins to take place.

\begin{figure*}
\includegraphics[width=0.85\textwidth]{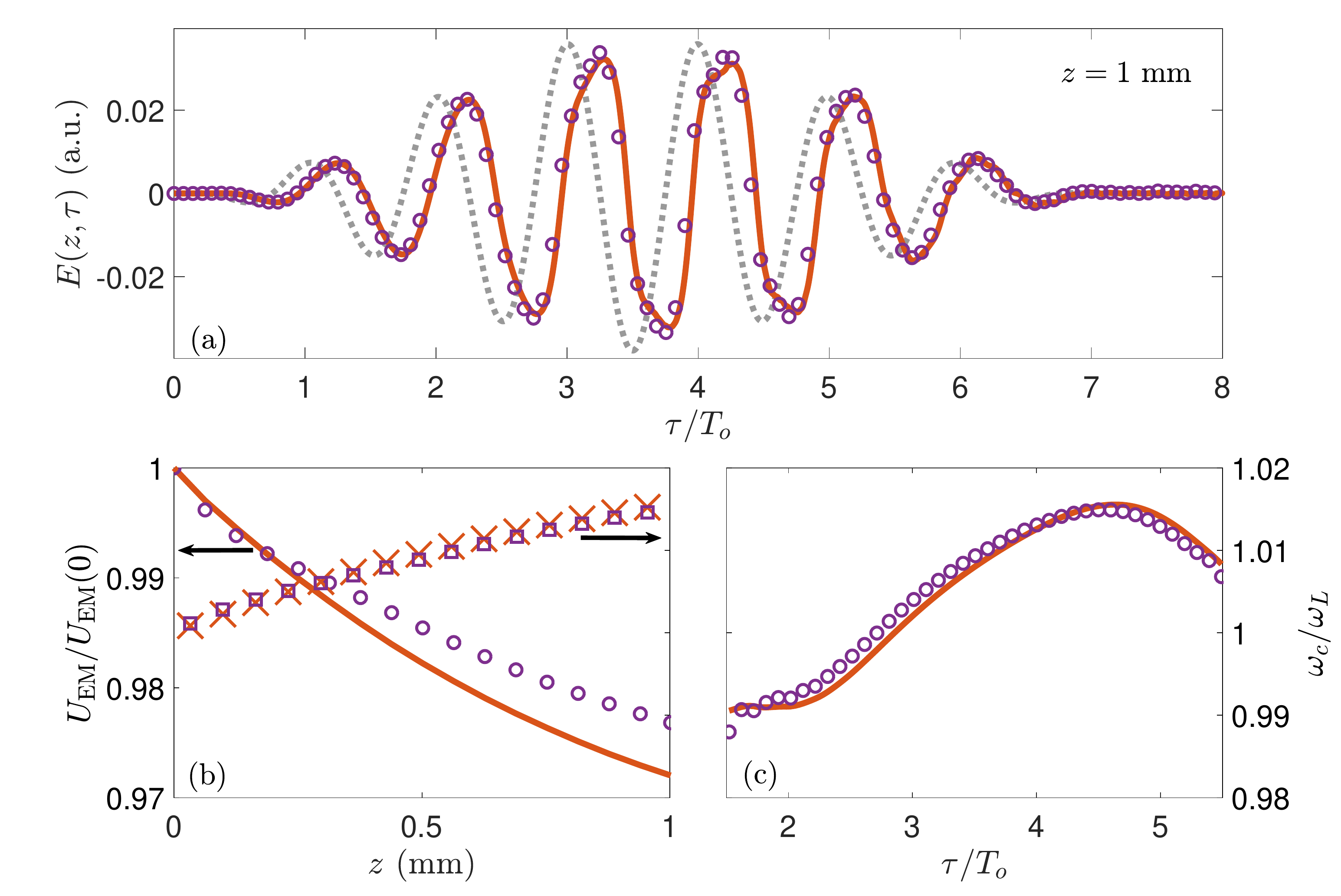}
\caption{Results of pulse propagation through $1\,\,{\rm mm}$ of ground-state atoms with density $\rho = 2\times10^{19}\,\,{\rm cm}^{-3}$, peak incident pulse intensity $I=5\times10^{13}\Wcm$ and carrier frequency $\omega_L=0.0378\au$ The solid orange curves and crosses correspond to the quantum model, while the purple circles and squares correspond to the classical model with $g_\sigma$. (a) Time-dependent electric field $E(z,\tau)$ at $z = 1\,\,{\rm mm}$. The grey dotted curve is the initial field $E_0(\tau)$. (b) Normalized pulse energy $U_{\rm EM}$ (curve and circles, left axis) and maximum instantaneous carrier frequency $\omega_{\rm max}$ (crosses and squares, right axis) as a function of $z$. The scale of the right axis is the same as in panel (c). (c) Time-dependent carrier frequency $\omega_c(z,\tau)$ at $z = 1\,\,{\rm mm}$. The spatial step used is $\Delta z = 1.3\lambda_L = 1.57\,\,\mu{\rm m}$.}\label{fig:fundSigI5E13}
\end{figure*}
\begin{figure*}
\includegraphics[width=0.8\textwidth]{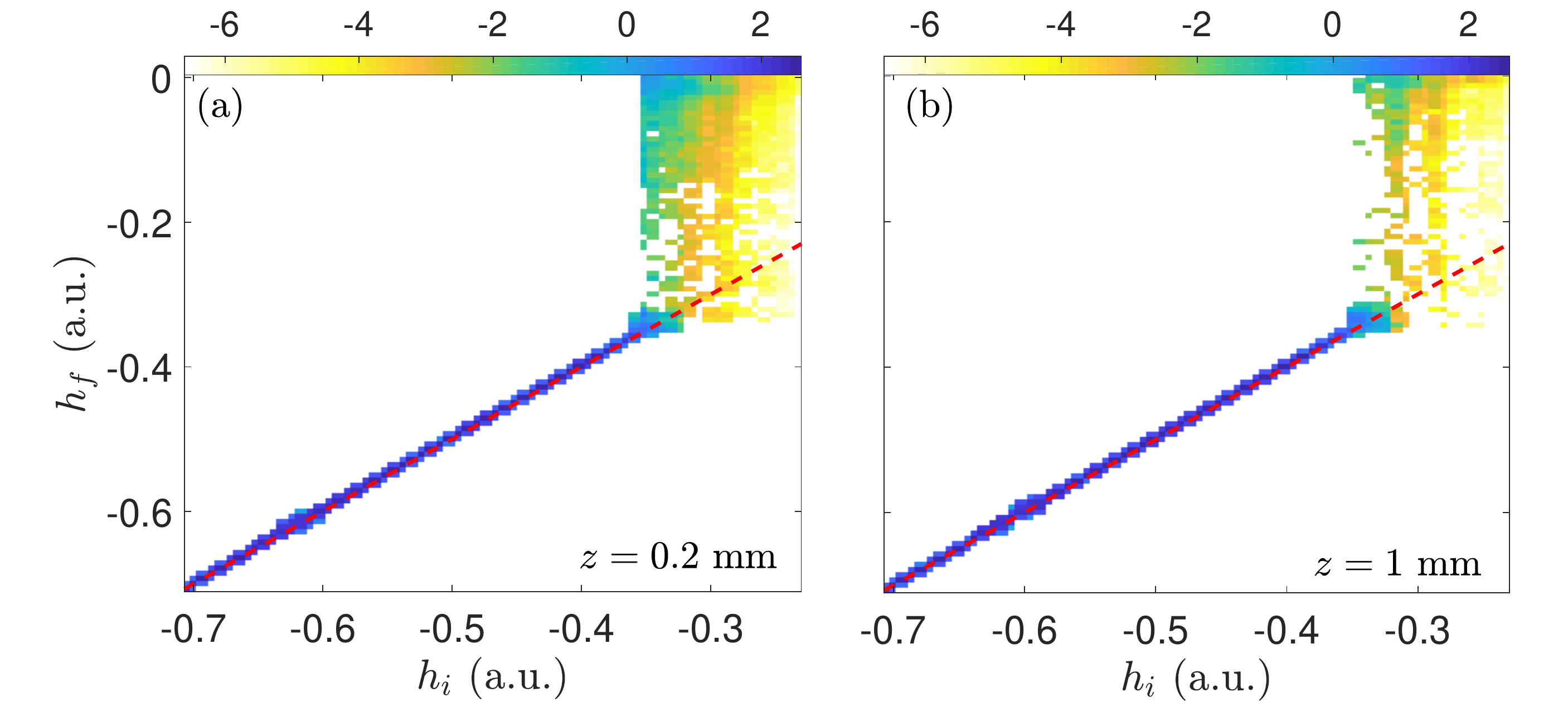}
\caption{Final versus initial energy distributions at $z=0.2\mm$ (a) and $z=1\mm$ (b) for the classical model using $g_\sigma$ as the initial energy distribution. The pulse propagation parameters are the same as for Fig.~\ref{fig:fundSigI5E13}. The dashed red line indicates $h_f=h_i$.}\label{fig:Ef_dist}
\end{figure*}

Now, going beyond the single-atom perspective, we show that using $g_\sigma$ instead of $g_1$ also improves the classical propagation simulations, allowing millimeter-scale propagation without the pulse energy ever increasing or a resonance at a bound-electron frequency developing.
We show the results of the classical propagation calculation with $g_\sigma$ under the same conditions as Fig.~\ref{fig:BPfail} in Figs.~\ref{fig:fundSigI5E13} and \ref{fig:Ef_dist}.
The quantum and classical calculations agree well for the time-dependent electric field at $z=1\mm$ (Fig.~\ref{fig:fundSigI5E13}a), as well as the ionization losses and instantaneous frequency (Figs.~\ref{fig:fundSigI5E13}b and \ref{fig:fundSigI5E13}c, respectively).
At the same time, we no longer observe a significant probability of energy loss among the low-energy bounded electrons at any point during propagation.
This is seen by comparing the joint $h_i$-$h_f$ distributions with $g_\sigma$, plotted in Fig.~\ref{fig:Ef_dist}, to those with $g_1$, plotted in Fig.~\ref{fig:BPfail}e and Fig.~\ref{fig:BPfail1}c.
We thus confirm that our strategy of matching ionization probabilities and appropriately shaping the classical initial energy distribution succeeds in eliminating unphysical effects from the classical model while simultaneously improving the quantitative agreement with the quantum model.
Henceforth, when we refer to the classical model, we mean the classical model with $g_\sigma$ as the initial energy distribution.

\subsection{Increasing the ionization probability}
\begin{figure*}
\includegraphics[width=0.85\textwidth]{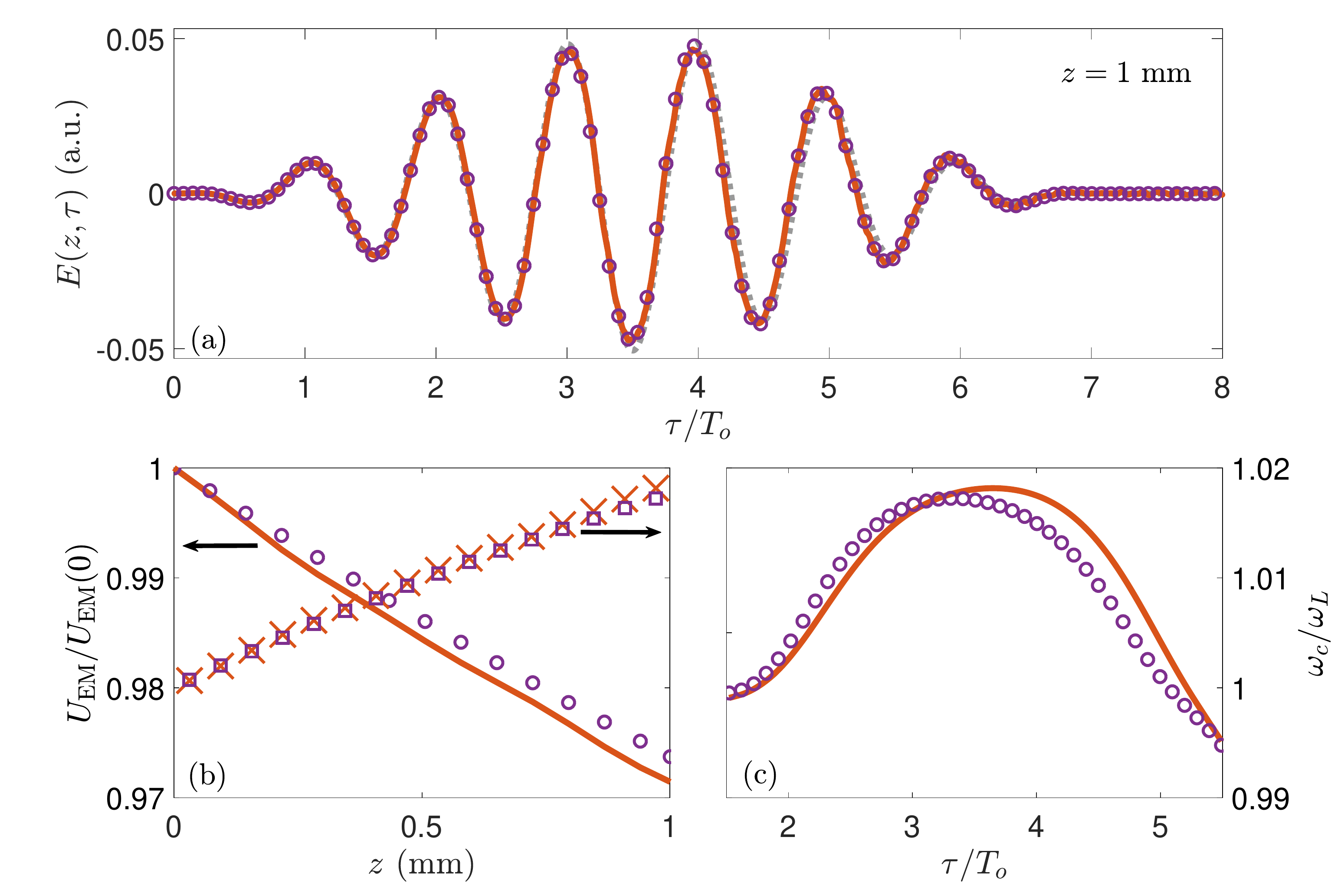}
\caption{Results of pulse propagation through $1\,\,{\rm mm}$ of ground-state atoms with density $\rho = 10^{18}\,\,{\rm cm}^{-3}$, peak incident pulse intensity $I=9\times10^{13} \Wcm$ and carrier frequency $\omega_L=0.0378\au$ The solid orange curves and crosses correspond to the quantum model, while the purple circles and squares correspond to the classical model with $g_\sigma$. (a) Time-dependent electric field $E(z,\tau)$ at $z = 1\,\,{\rm mm}$. The grey dotted curve is the initial field $E_0(\tau)$. (b) Normalized pulse energy $U_{\rm EM}$ (curve and circles, left axis) and maximum instantaneous carrier frequency $\omega_{\rm max}$ (crosses and squares, right axis) as a function of $z$. The scale of the right axis is the same as in panel (c). (c) Time-dependent carrier frequency $\omega_c(z,\tau)$ at $z = 1\,\,{\rm mm}$. The spatial step used is $\Delta z = 2\lambda_L = 2.4\,\,\mu{\rm m}$.}\label{fig:fundSigI9E13}
\end{figure*}
To conclude this section, we report on the correspondence between quantum and classical pulse propagation simulations in intermediate and high ionization probability regimes.
Figure \ref{fig:fundSigI9E13} shows the results of the propagation calculations in an intermediate ionization probability regime.
Here, a pulse with initial peak intensity $I=9\times10^{13} \Wcm$ propagates through $1\,\,{\rm mm}$ of a gas with density $\rho = 10^{18}\,\,{\rm cm}^{-3}$.
The density has been reduced by a factor $20$ compared to the previous simulations because the ionization probability is initially $P_{\rm ion}(T_m) \approx 10\%$, approximately $20$ times higher than at $I = 5\times10^{13} \Wcm$.
Hence, scaling down the density by this factor maintains the ionized electron density at the same level as the previous calculations, suggesting that the ionization-driven propagation effects here are comparable to the previous case.
For this set of parameters, we again observe a high level of quantitative agreement between the quantum and classical calculations for the time-dependent electric fields (Fig.~\ref{fig:fundSigI9E13}a), ionization losses (Fig.~\ref{fig:fundSigI9E13}b), and time-dependent blueshift (Fig.~\ref{fig:fundSigI9E13}b and \ref{fig:fundSigI9E13}c).
This provides evidence of the robustness of the classical model with $g_\sigma$ with respect to a range of incident laser pulse intensities.
In the intermediate ionization probability regime, both the quantum and classical calculations show that the laser field is not substantially reshaped during propagation, despite a maximum ionized electron density of about $10^{17}\,\,{\rm cm}^{-3}$.
This is due to a balance between neutral atom dispersion and free electron dispersion.
These conditions are favorable for the phase-matching of high harmonic radiation \cite{Popm10}, and the coherent buildup of this radiation is investigated in Sec.~\ref{sec:hhspec_gs}.

\begin{figure*}
\includegraphics[width=0.85\textwidth]{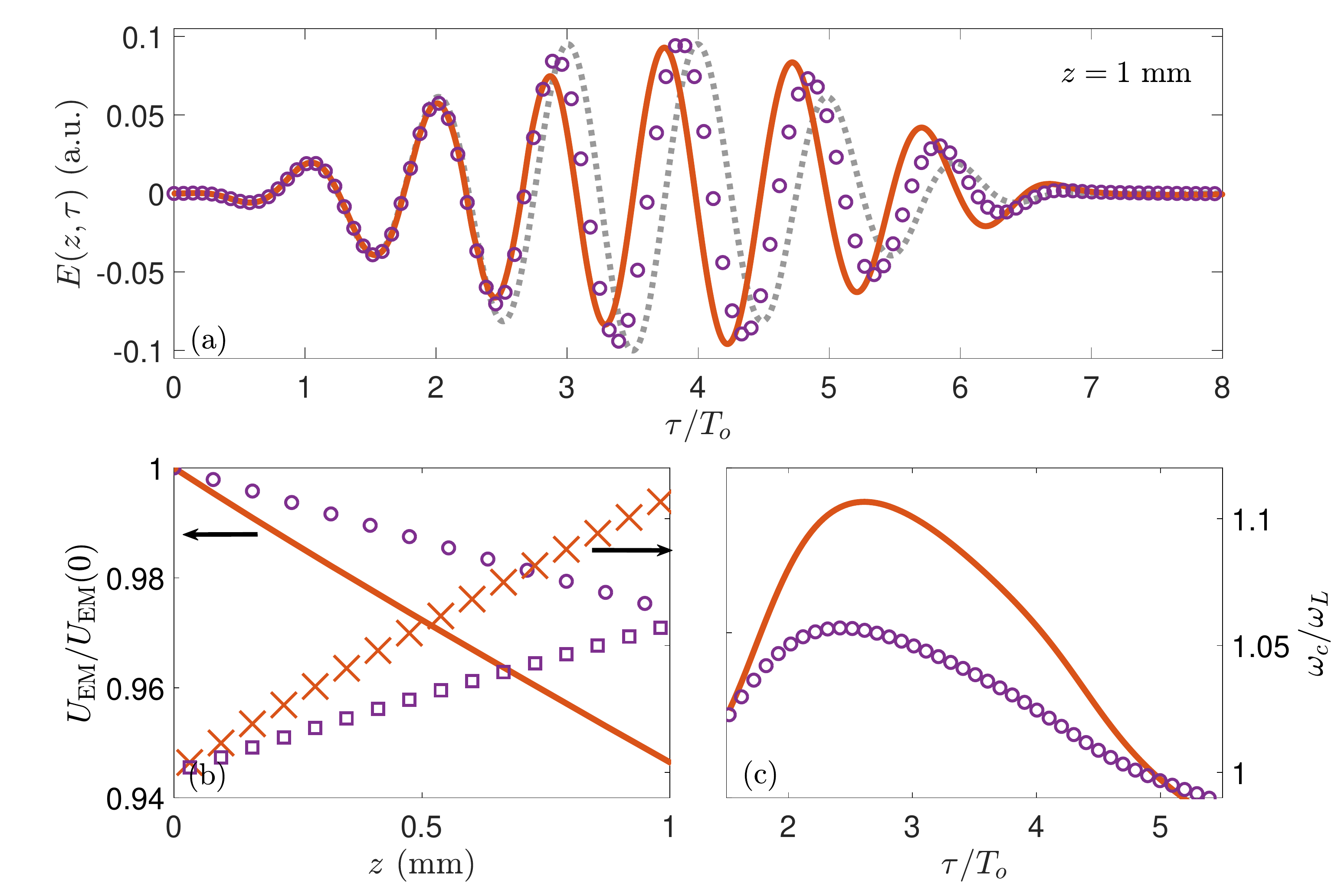}
\caption{Results of pulse propagation through $1\,\,{\rm mm}$ of ground-state atoms with density $\rho = 5\times10^{17}\,\,{\rm cm}^{-3}$, peak incident pulse intensity $I=3.5\times10^{14} \Wcm$ and carrier frequency $\omega_L=0.0378\au$ The solid orange curves and crosses correspond to the quantum model, while the purple cirlces and squares correspond to the classical model with $g_\sigma$. (a) Time-dependent electric field $E(z,\tau)$ at $z = 1\,\,{\rm mm}$. The gray dotted curve is the initial field $E_0(\tau)$. (b) Normalized pulse energy $U_{\rm EM}$ (curve and circles, left axis) and maximum instantaneous carrier frequency $\omega_{\rm max}$ (crosses and squares, right axis) as a function of $z$. The scale of the right axis is the same as in panel(c). (c) Time-dependent carrier frequency $\omega_c(z,\tau)$ at $z = 1\,\,{\rm mm}$. The spatial step used is $\Delta z = 2.6 \lambda_L = 3.2\,\,\mu{\rm m}$.}\label{fig:fundSig}
\end{figure*}

Figure \ref{fig:fundSig} shows the results of a propagation calculation in a high ionization probability regime.
Here, a pulse with initial peak intensity $I=3.5\times10^{14} \Wcm$ propagates through $1\,\,{\rm mm}$ of a gas with density $\rho = 5\times10^{17}\,\,{\rm cm}^{-3}$.
This leads to an initial ionization probability of $P_{\rm ion}(T_m) \approx 100\%$ for the quantum model and $P_{\rm ion}(T_m) \approx 50 \%$ for the classical model.
Given this discrepancy in ionization probability, we see in Fig.~\ref{fig:fundSig} that the quantitative agreement between the quantum and classical calculations is reduced compared to the low and intermediate ionization probability regimes.
Qualitatively, however, both calculations give the same results.
For the field $E(z,\tau)$ at $z=1\mm$ plotted in Fig.~\ref{fig:fundSig}, the quantum and classical calculations overlap up to $\tau\approx2.5T_o$, and both predict a phase advance during the latter-half of the pulse compared to the initial pulse $E_0(\tau)$.
This phase advance, a signature of the negative dispersion of free electrons, also manifests itself in the dramatic time-dependent blueshift, seen in both calculations of $\omega_c(z,\tau)$ at $z=1\mm$ (Fig.~\ref{fig:fundSig}c).
The maximum carrier frequency $\omega_{\max}$ and the ionization losses are larger in the quantum model, as shown in Fig.~\ref{fig:fundSig}b, because the ionization probability is higher than in the classical model.

For short pulses, a high ionization probability is typically attained in the barrier-suppression regime, where the potential barrier in the combined Coulomb and maximum laser fields is depressed below the quantum ground-state energy \cite{Brab00}.
At intensities above the barrier-suppression intensity $I_{\rm BS}$, ionization takes place by over-the-barrier ionization, rather than tunneling or multiphoton ionization.
For this system with $I_p=-0.5\,\,{\rm a.u.}$, we have $I_{\rm BS} = 1.4 \times 10^{14} \Wcm$, and it turns out that this is approximately where the intensity-dependent ionization probabilities of the classical model with $g_\sigma$ depart from those of the quantum model (Fig.~\ref{fig:energy_ionization}b).
Even though over-the-barrier ionization is essentially classical, our initial energy distribution is not effective in this regime because of the apparent drawback of having a distribution which is peaked at an energy greater than $h^*$.
In order to avoid the bound-electron resonant interaction that plagues the classical model at low intensities, we specifically designed $g_\sigma$ to populate classical states with energies $h < I_p$, as seen in Fig.~\ref{fig:energy_ionization}a.
The lower the energy of the state, the higher peak intensity required to ionize it \cite{Rich96}.
Thus, many of these states remain bounded even for $I > I_{\rm BS}$, while in the quantum case the atom becomes fully ionized.
Consequently, when tuning the classical model to accurately capture propagation effects, there is a trade-off between accuracy in the low-to-intermediate ionization probability regime and accuracy in the high ionization probability regime.
For an example of an alternative initial condition distribution which is tailored specifically to the high ionization probability regime and the corresponding propagation calculations, we refer the reader to Ref.~\cite{BermThesis}.
However, the classical model with this distribution exhibits the unphysical bound-electron resonance behavior described in Sec.~\ref{sec:ics} at lower laser intensities.

\section{Dynamics of the harmonic spectrum}\label{sec:hhspec}
We have seen that matching the quantum and classical single-atom intensity-dependent ionization probabilities leads to a good agreement between the two models for the field evolution in the macroscopic gas.
It turns out that this agreement can be improved even further in the context of a numerical experiment known as the scattering experiment \cite{Prot96,Sand99,Kamo14}, in which the electron is initialized in a pre-ionized scattering state.
With the ionization step artificially removed, the quantum and classical propagation calculations become nearly indistinguishable for the dominant frequency component of the field \cite{Berm18}.
This ensures maximal correspondence between the quantum and classical electron dynamics throughout propagation.
That makes this scenario, hereafter referred to as the scattering-propagation experiment, a natural starting point for exploring the mechanisms of harmonic radiation phenomena, which is the subject of this section.
After a close examination of HHG in the scattering-propagation experiment, we consider low- and high-order harmonic generation from ground-state atoms.

\subsection{The scattering-propagation experiment}\label{sec:scattering}
\begin{figure*}
\includegraphics[width=0.85\textwidth]{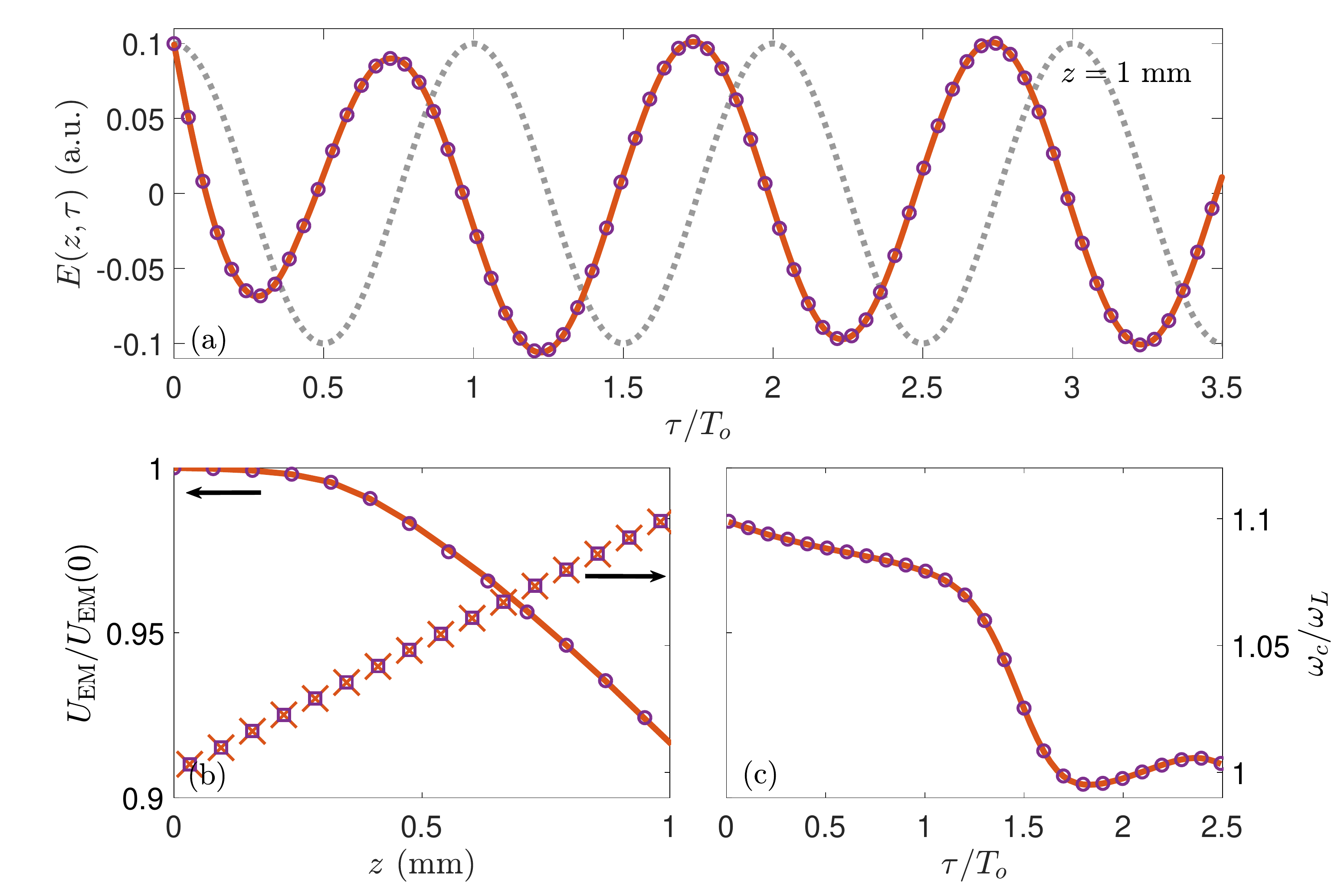}
\caption{Results of pulse propagation through $1\,\,{\rm mm}$ of atoms prepared in a scattering state (see text) with density $\rho = 5\times10^{17}\,\,{\rm cm}^{-3}$, peak incident pulse intensity $I=3.5\times10^{14} \Wcm$ and carrier frequency $\omega_L=0.0378\au$ The orange curves and crosses correspond to the quantum model, while the purple circles and squares correspond to the classical model. (a) Time-dependent electric field $E(z,\tau)$ at $z = 1\,\,{\rm mm}$. The dashed curve is the initial field ${\cal E}_0(\tau)$. (b) Normalized pulse energy $U_{\rm EM}$ (curve and circles, left axis) and maximum instantaneous carrier frequency $\omega_{\rm max}$ (crosses and squares, right axis) as a function of $z$. The scale of the right axis is the same as in panel (c). (c) Time-dependent carrier frequency $\omega_c(z,\tau)$ at $z = 1\,\,{\rm mm}$. The spatial step used is $\Delta z = 2.6 \lambda_L = 3.2\,\,\mu{\rm m}$.}\label{fig:fundScat}
\end{figure*}

\subsubsection{Initial conditions}
We again consider the propagation of the laser field from $z=0$ to $z=1\mm$, through a gas of density $\rho=5\times 10^{17}\,\,{\rm cm}^{-3}$.
Because we are not concerned with the gradual ionization of the atom here, we are not obliged to use a realistic pulse shape for the initial electric field.
Thus, we initialize the field as a simple monochromatic wave, $E_0(\tau) = E_0 \cos(\omega_L \tau)$, and we reduce the final time to $\tau_f = 3.5T_o$.
We take the same field parameters as in the high ionization probability regime, i.e.\ $E_0=0.1\au$ for an intensity of $I=3.5\times 10^{14}\Wcm$.
The electron is initialized as a Gaussian wave packet at rest, centered at the quiver radius $E_0/\omega_L^2$, as in Refs.~\cite{Prot96,Sand99,Kamo14}.
Thus, for the quantum case, the initial wave function is 
\begin{equation*}
\psi_0(x) = \left(\frac{\gamma^2}{\pi}\right)^{1/4} \exp\left[-\frac{\gamma^2}{2}\left(x-\frac{E_0}{\omega_L^2}\right)^2\right],
\end{equation*}
where the parameter controlling the wave packet width is chosen as $\gamma = 0.2236\au$ \cite{Sand99}.
Meanwhile, in the classical case, the corresponding initial distribution function is also a Gaussian wave packet, with identical position and velocity spreads to the quantum wave packet, i.e.
\begin{equation*}
f_0(x,v) = \frac{1}{\pi} \exp\left[-\gamma^2\left(x-\frac{E_0}{\omega_L^2}\right)^2 -\frac{v^2}{\gamma^2}\right]
\end{equation*}
Unlike in the ground-state case, we have not performed any adjustments to the classical distribution to optimize the agreement between the quantum and classical propagation calculations---indeed, this $f_0$ is precisely the Wigner transform of $\psi_0$ \cite{Zago12}.
As we shall see below, excellent agreement may already be obtained with this distribution.

\subsubsection{Evolution of the dominant component of the field}
Figure \ref{fig:fundScat} shows the results of the pulse propagation calculations for the total field, energy loss, and blueshift.
For the calculations of $\omega_c(z,\tau)$, the post-processed field $E_p$ was defined on the interval $\tau \in [-3T_o,3.5T_o]$, with a $z$-independent, smoothly ramped-up oscillation for $\tau < 0$, followed by $E(z,\tau)$ multiplied by a window which sends the field smoothly to zero over the last computed laser cycle.
Precisely, $E_p$ is given by
\begin{equation}
E_p(z,\tau) = 
\begin{cases}
E_0\cos^2\left(\frac{\pi\tau}{6T_o}\right) \cos(\omega_L\tau) & \text{for } \tau < 0 \\
E(z,\tau) & \text{for } 0 \leq \tau < 2.5 T_o \\
E(z,\tau) \cos^2\left( \frac{\pi(\tau - 2.5 T_o)}{2T_o} \right) & \text{for } \tau \geq 2.5 T_o.
\end{cases}
\end{equation}
This post-processing prescription allowed the computation of a clean Wigner-Ville transform that leads to an instantaneous carrier frequency which clearly captures the blueshift concentrated between $0 < \tau < T_o$, as seen by looking at Figs.~\ref{fig:fundScat}a and \ref{fig:fundScat}c.
We observe that in the scattering-propagation experiment, the classical and quantum calculations are in excellent agreement for the observables reflecting the dominant frequency component of the field, even better than in the ground-state case (Figs.~\ref{fig:fundSigI5E13} and \ref{fig:fundSigI9E13}).
This corroborates our assertion that the main source of the discrepancy the classical and quantum ground-state calculations is the description of ionization.
Indeed, by beginning in a fully ionized state instead of the ground state, we observe a massive improvement in the agreement between the two calculations for an incident pulse with the same peak laser intensity.

\subsubsection{Evolution of the high harmonic spectrum}
\begin{figure*}
\includegraphics[width=0.7\textwidth]{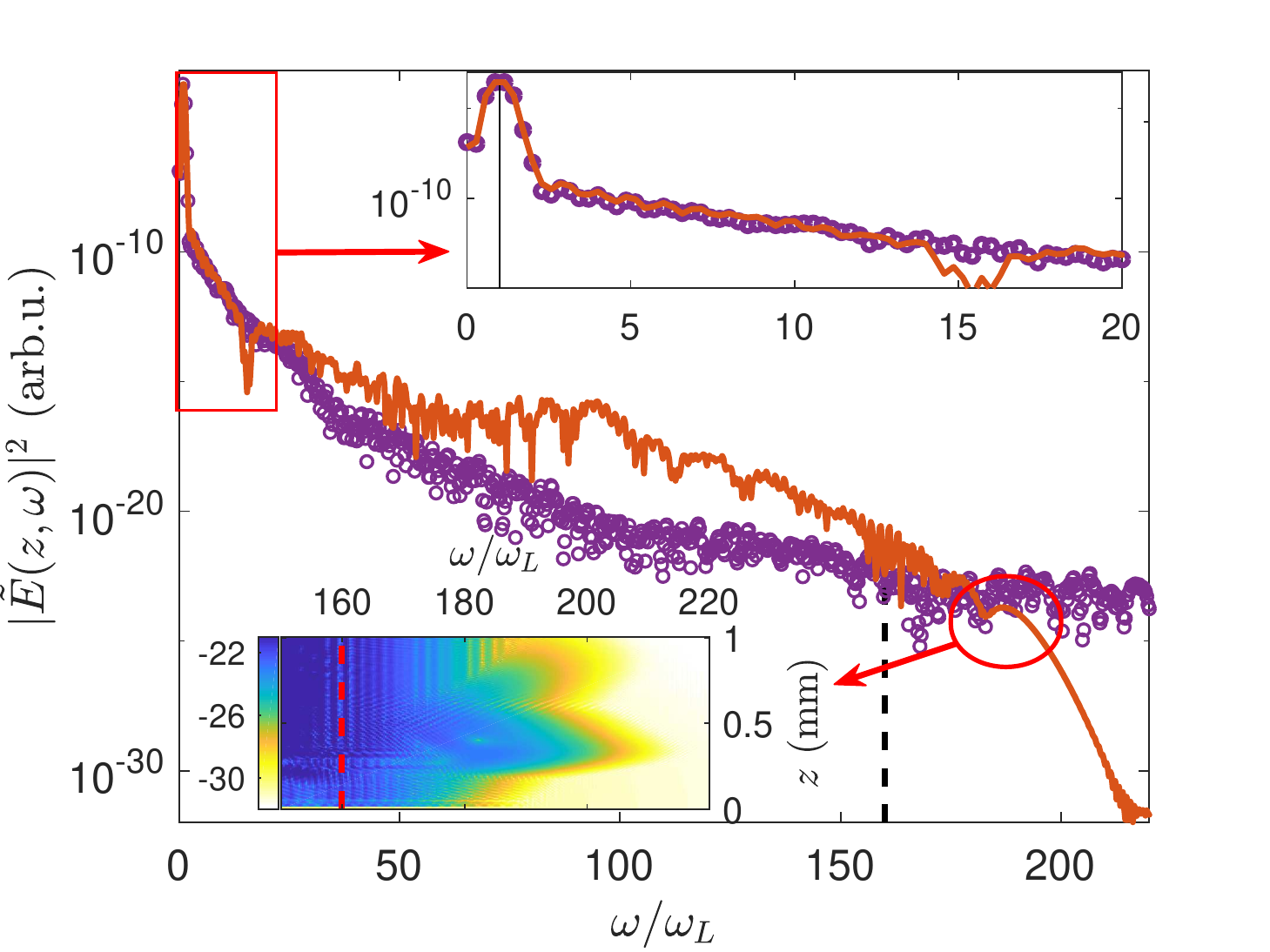}
\caption{High harmonic spectra for the scattering-propagation experiment at $z=0.37\mm$, with the parameters of Fig.~\ref{fig:fundScat}. The orange solid curves correspond to the quantum model and the purple circles correspond to the classical model. The dashed lines indicate $3.17U_p + |I_p|$. Upper inset: magnification of the spectrum of the low-order harmonics. The vertical line is $\omega=\omega_L$. Lower inset: Spectrum of the harmonics in the cutoff region for the quantum model as a function of $z$. The harmonic intensities are indicated by the logarithmic color scale.}\label{fig:specScat}
\end{figure*}
Figure \ref{fig:specScat} shows the power spectra of the electric fields of the classical and quantum models at $z = 0.37\mm$.
For these spectra, the post-processing consisted of applying a $\sin^4$ window to the calculated electric fields.
The classical and quantum spectra agree well for the low-order harmonics, as shown in the upper inset of Fig.~\ref{fig:specScat}.
However, a high harmonic plateau and cutoff are only observed in the quantum model, 
as observed in single atom calculations \cite{Sand99}.
This confirms the fundamental role played by quantum interference effects for high harmonic emission, even when propagation effects are taken into account.
In the quantum spectrum, we see that the cutoff is extended well past the usual $3.17U_p + |I_p|$ cutoff law, where $U_p=E_0^2/4\omega_L^2$ is the ponderomotive energy, which is valid for SAE atoms in monochromatic fields.
The lower inset of Fig.~\ref{fig:specScat} shows the evolution of the cutoff region throughout propagation.
While $3.17U_p + |I_p| \approx 160\omega_L$ is a reasonable approximation of the cutoff for small $z$, the cutoff increases significantly during propagation, reaching about $180\omega_L$ before receding again.
To understand this anomalous cutoff extension driven by the pulse propagation, we examine the electron dynamics.

\begin{figure*}
\includegraphics[width=0.85\textwidth]{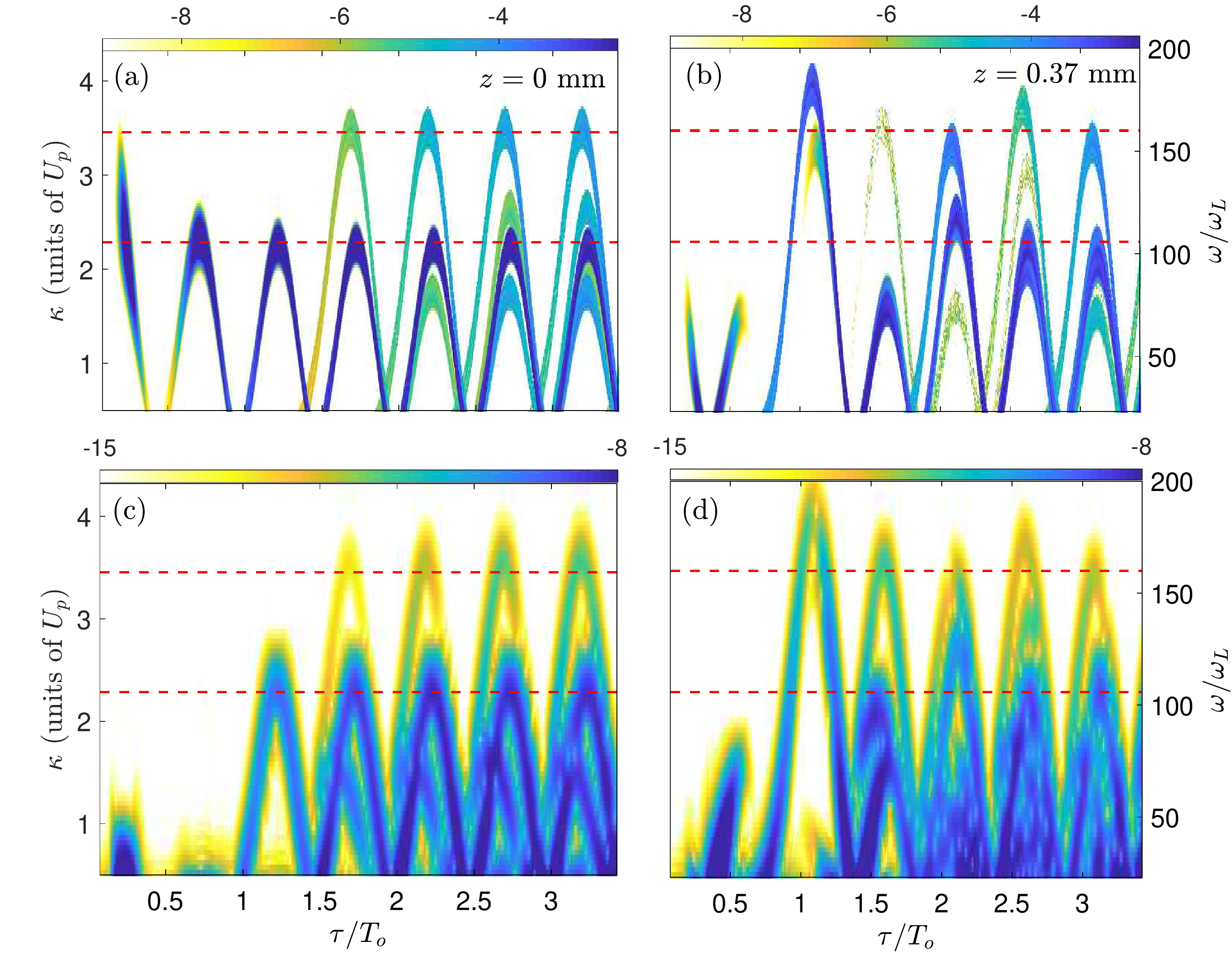}
\caption{Electron dynamics at $z=0$ (a),(c) and $z=0.37\mm$ (b),(d) for the scattering-propagation experiment, with the parameters of Fig.~\ref{fig:fundScat}. (a),(b) Recollision flux $R(\kappa,\tau;z)$ from the classical model. (c),(d) Spectrogram of the dipole acceleration $d_a(\tau)$ from the quantum model. The spectrograms were computed using a $\cos^4$ window of duration $0.15T_o$. The dotted lines indicate $2U_p+|I_p|$ and $3.17U_p+|I_p|$. The left axes, indicating the recollision kinetic energy $\kappa$, are related to the right axes, indicating the radiated frequency $\omega$, by $\kappa = \omega$.}\label{fig:recollisionsScat}
\end{figure*}
The correspondence between recollisions and radiation is clearly seen when comparing the recollision flux of the classical model with the spectrogram of the dipole acceleration from the quantum model.
High-harmonic emission occurs when multiple electron energy states are simultaneously occupied near the core, leading to interference in the quantum model at frequencies equal to the difference in energy between all possible pairs of states \cite{Pukh03,Kohl10}.
Traditionally, one conceives of high harmonic radiation as occurring from the interference between a recolliding electron with kinetic energy $\kappa$ and the ground state of total energy $I_p$ \cite{Cork93,Lewe94}, which, estimating the potential energy of the recolliding electron as $I_p$, leads to radiation at the frequency $\omega \approx (\kappa + I_p) - I_p = \kappa$.
One may also obtain high harmonic emission from the interference of recollisions of two different energies, $\kappa_1$ and $\kappa_2$, at a frequency $\omega = |\kappa_1-\kappa_2|$ \cite{Kohl10}.

In Fig.~\ref{fig:recollisionsScat}, we observe both kinds of emission.
First, we focus on Figs.~\ref{fig:recollisionsScat}a and \ref{fig:recollisionsScat}c at $z=0$, where the classical and quantum atoms are driven by an identical electric field, $E_0(\tau)$.
For $\tau < T_o$, even though there are recollisions, as seen in Fig.~\ref{fig:recollisionsScat}a, they mainly occur at a single energy at each time, and the ground state is initially completely empty in the scattering experiment setup. 
Thus, no high harmonic emission is observed in the quantum model (Fig.~\ref{fig:recollisionsScat}c).
Beginning at $\tau \gtrsim T_o$, part of the electron wave packet becomes trapped near the core \cite{Sand99}, leading to the population of the ground state \cite{Zago12}.
Subsequently, high-harmonic emission from the interference of recolliding electrons with the trapped electrons is evident from the direct correspondence between the spectrogram of Fig.~\ref{fig:recollisionsScat}c and the recollision flux of Fig.~\ref{fig:recollisionsScat}a, particularly for the families of recollisions with a maximum kinetic energy near $2U_p+|I_p|$ and those with a maximum kinetic energy near $3.17U_p+|I_p|$.
This second family of recollisions does not emerge until $\tau \gtrsim 1.5T_o$, and once it does, we also observe high-harmonic emission arising from the interference between these two families of recollisions.
These are the dark blue stripes of radiation with a frequency decreasing in time from about $75\omega_L$ to $25\omega_L$ that appear every half laser cycle for $\tau > 1.5T_o$.

By $z=0.37\mm$, the electric fields in the quantum and classical models are now different, as they have been driven by different dipole velocities in Eq.~\eqref{eq:EOMfield}.
Nevertheless, the dominant component of the fields agree so closely throughout propagation, as shown in Fig.~\ref{fig:fundScat}, that we continue to observe a close correspondence between the quantum and classical electron dynamics.
This is reflected by the comparisons of the classical recollision flux and the quantum dipole acceleration spectrogram at $z=0.37\mm$ in Figs.~\ref{fig:recollisionsScat}b and \ref{fig:recollisionsScat}d.
In particular, we still observe high-harmonic emission in the quantum case with a timing and frequency matching the timing and energy of the classical recollisions.
We also observe radiation from the interference between different families of recollisions, which is most clearly seen at $\tau \approx 1.5T_o$.
Comparing the electron dynamics at $z=0.37\mm$ and $z=0$, we notice two striking changes.
The first is that, at $z=0.37\mm$, recollision-driven radiation is observed for $\tau < T_o$, whereas it was not at $z=0$.
This implies that electron trapping near the core occurs earlier in the laser pulse as propagation proceeds.
The second is that, at $z=0.37\mm$ around $\tau = T_o$, we observe recollisions and their corresponding radiation at energies that significantly exceed the usual $3.17U_p+|I_p|$ harmonic cutoff, whereas this does not occur at $z=0$.
These recollisions drive the extension of the high-harmonic cutoff that we observed in the electric field spectrum of the quantum calculation, as seen in Fig.~\ref{fig:specScat}.
Next, by studying the electron dynamics in phase space using the classical model, we identify the mechanism of the cutoff extension.
\begin{figure*}
\includegraphics[width=0.85\textwidth]{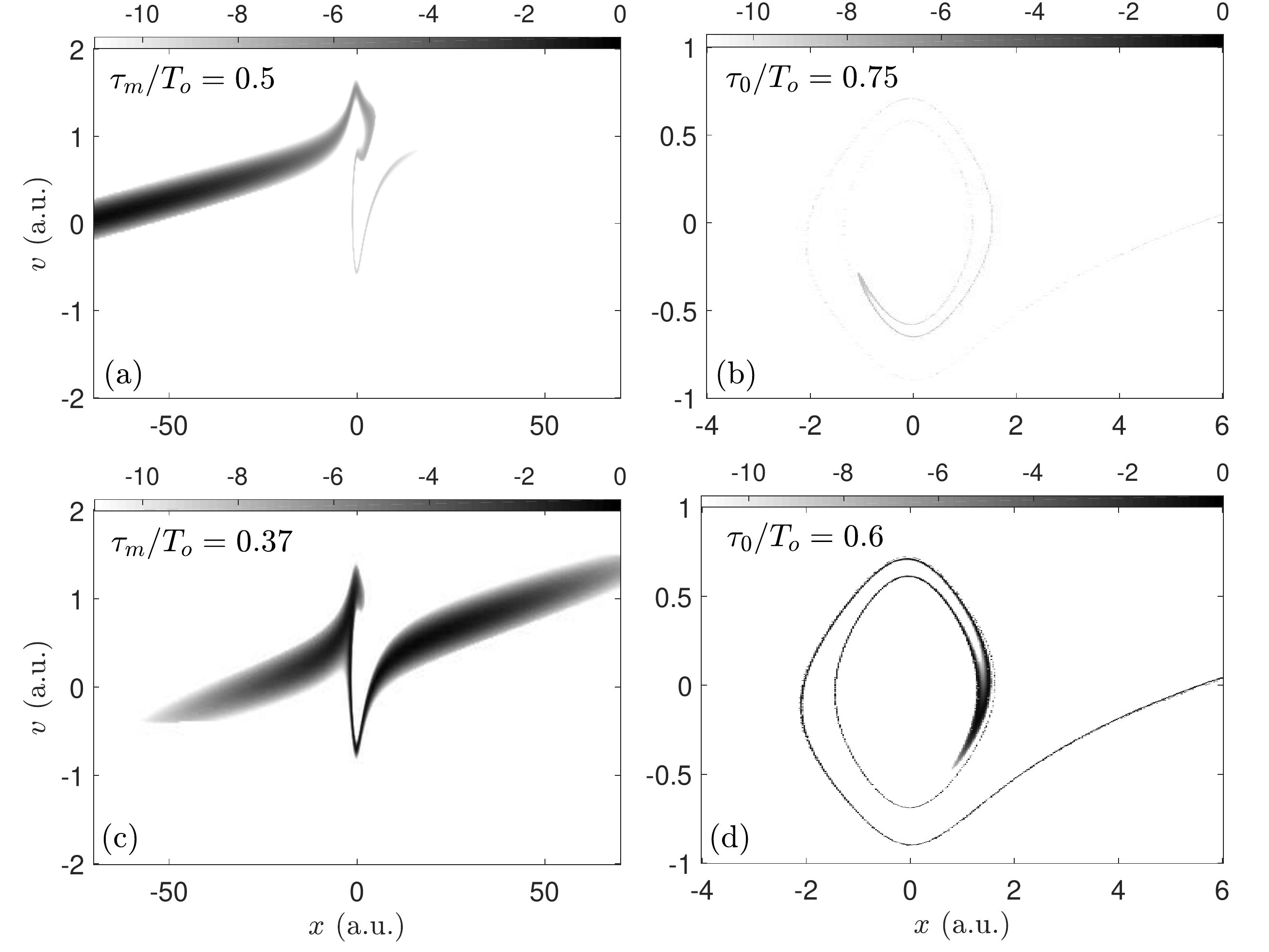}
\caption{Snapshots of the distribution function $f(x,v,z,\tau)$ at $z=0\mm$ (a),(b) and $z=0.5\mm$ (c),(d), with the parameters of Fig.~\ref{fig:fundScat}. (a),(c) The distribution function $f(x,v,z,\tau_m)$ at time $\tau_m$, the time of the first field intensity maximum after the start of the pulse. (b),(d) The distribution function $f(x,v,z,\tau_0)$ at time $\tau_0$, the first zero of the field following $\tau_m$.}\label{fig:snaps}
\end{figure*}

Figure \ref{fig:snaps} shows snapshots of the electron distribution function $f(x,v,z,\tau)$ at particular times $\tau$ and propagation positions $z$.
In the scattering-propagation experiment, the electron wave packet always begins on the right side of the core and is initially accelerated towards it by the laser field.
We define $\tau_m$ as the first maximum of the instantaneous field intensity $E(z,\tau)^2$ after $\tau=0$; hence, the laser force is maximal and pointing opposite to its initial direction.
As propagation proceeds, the blueshift causes the laser field to reverse direction earlier in the pulse, i.e.\ $\partial_z \tau_m < 0$.
This causes the center of the wave packet at time $\tau_m$ to be displaced to the right, as seen by comparing Fig.~\ref{fig:snaps}a and \ref{fig:snaps}c.
For a range of intermediate values of $z$ the wave packet is thus nearly centered over the ion, with the electron velocities distributed about zero, as illustrated in Fig.~\ref{fig:snaps}c at $z=0.5\mm$.
Electrons near the core with low kinetic energy have a high probability of becoming trapped \cite{Zago12}, and indeed a trapped part of the wave packet is clearly observed in the subsequent snapshot of the distribution function at $\tau_0$ in Fig.~\ref{fig:snaps}d, where $\tau_0$ is the time of the first zero of the field after $\tau_m$.
Comparing this with the distribution function $f(x,v,0,\tau_0)$ in Fig.~\ref{fig:snaps}b, we see that indeed, the probability of trapping by this time has greatly increased during propagation.
Hence, bound-states become populated for $\tau < T_o$, and thus recollision-driven high-harmonic radiation for $\tau < T_o$ becomes possible after propagation, whereas it is not at $z=0$.

\begin{figure*}
\includegraphics[width=0.9\textwidth]{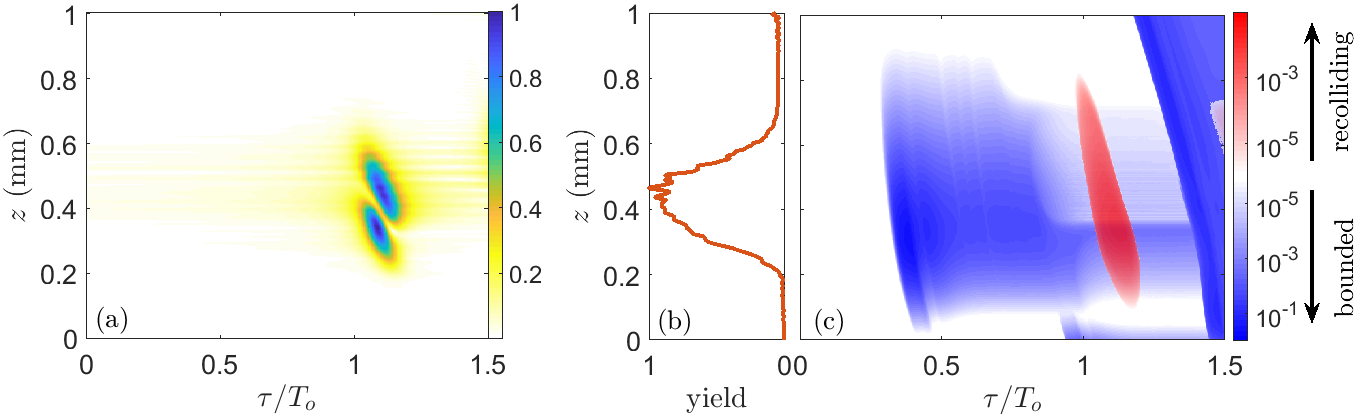}
\caption{Evolution of the anomalously high-harmonic radiation, with the parameters of Fig.~\ref{fig:fundScat}. (a) Normalized spatiotemporal amplitude profile $|\hat{E}_{ab}(z,\tau)|$ of the radiation in the $\omega > 175\omega_L$ frequency band, computed from the electric field of the quantum model. (b) Normalized yield of radiation with $\omega > 175\omega_L$, computed from the electric field of the quantum model. (c) Population of bound states $P_{b}$ and population of recolliding states $P_r$ with energies $\kappa > 3.78 U_p$ as a function of $z$ and $\tau$, calculated from the classical model. The population of bound states is indicated by the logarithmic blue color scale, and the population of recolliding states is indicated by the logarithmic red color scale.}\label{fig:cutoffExtension}
\end{figure*}
Besides enabling the recollision-driven radiation for $\tau < T_o$, these trapped states are also important for the emergence of the anomalously high-frequency radiation.
In Fig.~\ref{fig:cutoffExtension}c, we show calculations of the bound state populations $P_b(z,\tau)$ and the anomalously high energy recollisions $P_r(z,\tau)$ from the classical model.
We estimate the bound-state population from the recollision flux by integrating $R(\kappa,\tau,z)$ over low kinetic energies, while we obtain the probability of anomalously high recollision by integrating the same over the highest kinetic energies.
Specifically, we define $P_b$ and $P_r$ as 
\begin{align*}
& P_b(z,\tau) = \int_0^{\kappa_b} R(\kappa,\tau,z){\rm d}\kappa \,\,\,\text{for } \kappa_b = 0.11 U_p \\
& P_r(z,\tau) = \int_{\kappa_r}^\infty R(\kappa,\tau,z){\rm d}\kappa \,\,\,\text{for } \kappa_r = 3.78 U_p.
\end{align*}
This choice of $\kappa_b$ ensures that the instantaneous energy of the counted electrons is negative, while this choice of $\kappa_r$ corresponds to radiation at $\omega_r = 175\omega_L$.
Here, it is clear that the propagation induces the population of bound electron states earlier in the pulse.
Indeed, for $z=0$, these states are not occupied until $\tau \approx T_o$, while by $z=0.1\mm$, they become occupied by $\tau = 0.5T_o$.
Likewise, it is evident that the high-energy recollisions emerge only around $z=0.2\mm$.

Figure ~\ref{fig:cutoffExtension}b shows the yield of the harmonics greater than $\omega_r$ from the quantum calculation, i.e.\ the integral of the power spectrum $\int_{\omega_r}^\infty |\tilde{E}(z,\omega)|^2{\rm d}\omega$.
We see that these modes are either amplified or absorbed for $0.2\mm <  z < 0.7\mm$; at other positions, they are comparatively quiescent.
This range corresponds exactly to the range of $z$ in which we observe an overlap in the bound state population and the recolliding population.
Indeed, it is only when both of these states are occupied that interference between them can occur in the quantum model, leading to the emission (or absorbtion) of these modes of the field.
Furthermore, we show the time-profile of these modes as a function of $z$ in Fig.~\ref{fig:cutoffExtension}a, by plotting the amplitude of the analytic representation $|\hat{E}_{ab}(z,\tau)|$ computed for the modes $\omega > \omega_r$.
We see that these high-harmonics are generated at the same times $\tau$ as the recollisions.
However, they have a nontrivial spatiotemporal evolution, indicating rapidly evolving phase-matching conditions.

\begin{figure*}
\includegraphics[width=0.9\textwidth]{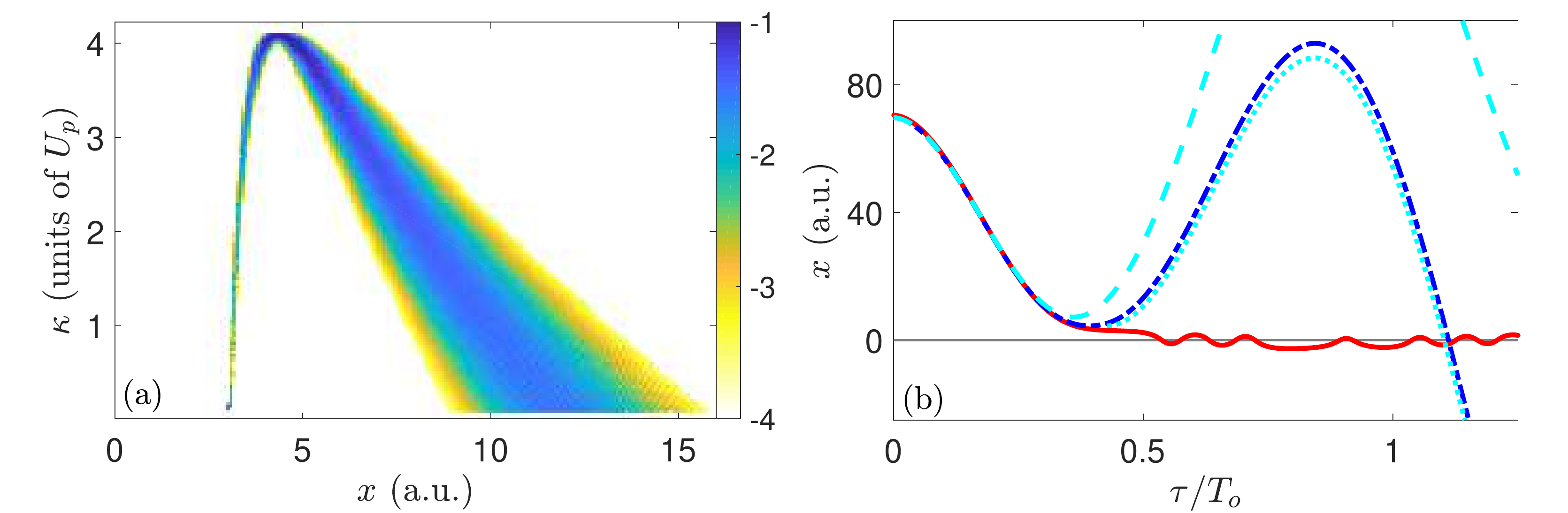}
\caption{Trajectory analysis of $4U_p$ recollisions at $z = 0.37\mm$ using the classical model, with the parameters of Fig.~\ref{fig:fundScat}. (a) Joint probability distribution of $(x,\kappa)$ in a logarithmic scale, for electrons which come to rest at $x$ near the first extremum of the laser field, i.e.\ near times $\tau = \tau_m$, and then recollide with kinetic energy $\kappa$. (b) A typical trajectory from the classical model with $\kappa > 4U_p$ (dash-dotted blue line), a typical trapped trajectory (solid red line), the SFA trajectory with the same initial conditions as the $4U_p$ trajectory (dashed cyan line), and the SFA trajectory initiated at a time $\tau_0$ near time $\tau = \tau_m$ with the maximum recollision kinetic energy (dotted cyan line). The thin black line is $x=0$.} \label{fig:trajAnalysis}
\end{figure*}
By looking at the electron trajectories, we determine the mechanism of the increase of the recollision energy beyond the usual $3.17U_p+|I_p|$ cutoff.
We have focused on the trajectories belonging to the first family of recollisions containing the anomalously high-energy recollisions, with $\kappa > \kappa_r$ exceeding $4U_p$, at $z=0.37\mm$.
We have observed that most of these electrons come to rest at some position $x > 0$ near to the core, before achieving their first recollision.
In Fig.~\ref{fig:trajAnalysis}a, we have plotted the joint probability distribution of $x$ and $\kappa$, the kinetic energy of each electron's subsequent recollision.
We see that the electron's $x$ is highly correlated with its $\kappa$, and in particular the highest energy recollisions come to rest at about $x = 4.5\au$, very close to the core.
A typical example of such a trajectory, with $\kappa > 4U_p$, is plotted in Fig.~\ref{fig:trajAnalysis}b.
Because these trajectories approach the core with low kinetic energy, they very nearly become trapped there.
This is evidenced by its similarity of this recolliding trajectory to the typical trapped trajectory, also plotted in Fig.~\ref{fig:trajAnalysis}b.
Since these anomalously high energy recollisions come so close to the core that they barely escape trapping, one may expect the Coulomb field to play a central role in the increase in energy of these trajectories.

We assess the role of the Coulomb field through two calculations based on the strong-field approximation (SFA), one in which the Coulomb field is neglected entirely \cite{Cork93}, and one in which it is treated as a perturbation \cite{Kamo14}.
In the first calculation, we compute the trajectory of the electron with the same initial conditions $(x_0,v_0)$ as the $4U_p$ trajectory plotted in Fig.~\ref{fig:trajAnalysis}b, but neglecting the Coulomb field.
Hence, $x_{\rm SFA}(\tau) = x_0 + v_0\tau - \int_0^\tau \int_0^{\tau'} E(z,\tau'') {\rm d}\tau''{\rm d}\tau'$, where $z = 0.37\mm$, and this trajectory is plotted in Fig.~\ref{fig:trajAnalysis}b. 
We see that $x_{\rm SFA}$ agrees well with the true trajectory until about $\tau = \tau_m = 0.39T_o$, the extremum of the electric field, at which point the close encounter with the core takes place.
In the second calculation, we fix $x_0=4.5\au$ and $v_0=0$, and find the initial time $\tau_0$ near $\tau_m$ such that the subsequent recollision kinetic energy of the SFA trajectory with these initial conditions, $\kappa_{\rm SFA}$, is at a maximum.
This results in $\kappa_{\rm SFA} = 3.69 U_p$, and the corresponding trajectory is also plotted in Fig.~\ref{fig:trajAnalysis}b.
It is seen to be quite close to the true high-energy recolliding trajectory.
Furthermore, the effect of the Coulomb field on the return kinetic energy may be included perturbatively \cite{Kamo14}.
This yields a maximum return kinetic energy of simply $\kappa_{\rm SFA} - V(0) = 4.1U_p$.
This value of the maximum kinetic energy is in excellent agreement with the maximum energy recollision we observed for this $z$ (see Figs.~\ref{fig:recollisionsScat}b and \ref{fig:recollisionsScat}d).

Therefore, while the Coulomb field has a decisive effect on the electron dynamics, it is not responsible for the increase in the high-harmonic cutoff energy {\it per se}.
The maximum cutoff energy is well-predicted by a Coulomb-perturbed SFA with $E(z,\tau)$, the propagated electric field.
Because the Coulomb perturbation is always present and independent of the field, the increase in energy must be solely due to the change of shape of the field.
In other words, the SFA cutoff for an initially monochromatic field of $\kappa_{\rm SFA} = 3.17U_p$ becomes $\kappa_{\rm SFA} = 3.69U_p$ after propagation to $z=0.37\mm$ due to the accumulated radiation at other frequencies, and this causes the increase in the cutoff energy.
Nevertheless, the Coulomb interaction near $\tau_m$, though brief, is critical for making these higher energy recollisions accessible to the electrons \cite{Berm15}.
Indeed, the only way that the trajectory shown in Fig.~\ref{fig:trajAnalysis}b can bring back $4U_p$ to the core is by becoming momentarily trapped there; the SFA trajectory with the same initial condition, also shown in Fig.~\ref{fig:trajAnalysis}b, does not come back to the core at all.

\subsection{Harmonic generation from ground-state atoms}\label{sec:hhspec_gs}
Now, we return to the more realistic case of pulse propagation through ground-state atoms, focusing  on the low- and intermediate-ionization probability regimes.
We study both low-order and high-order harmonic generation.
In each case, we investigate the extent to which the classical model allows us to understand the results of the quantum calculations.

\subsubsection{Low-order harmonic generation}
\begin{figure*}
\includegraphics[width=0.9\textwidth]{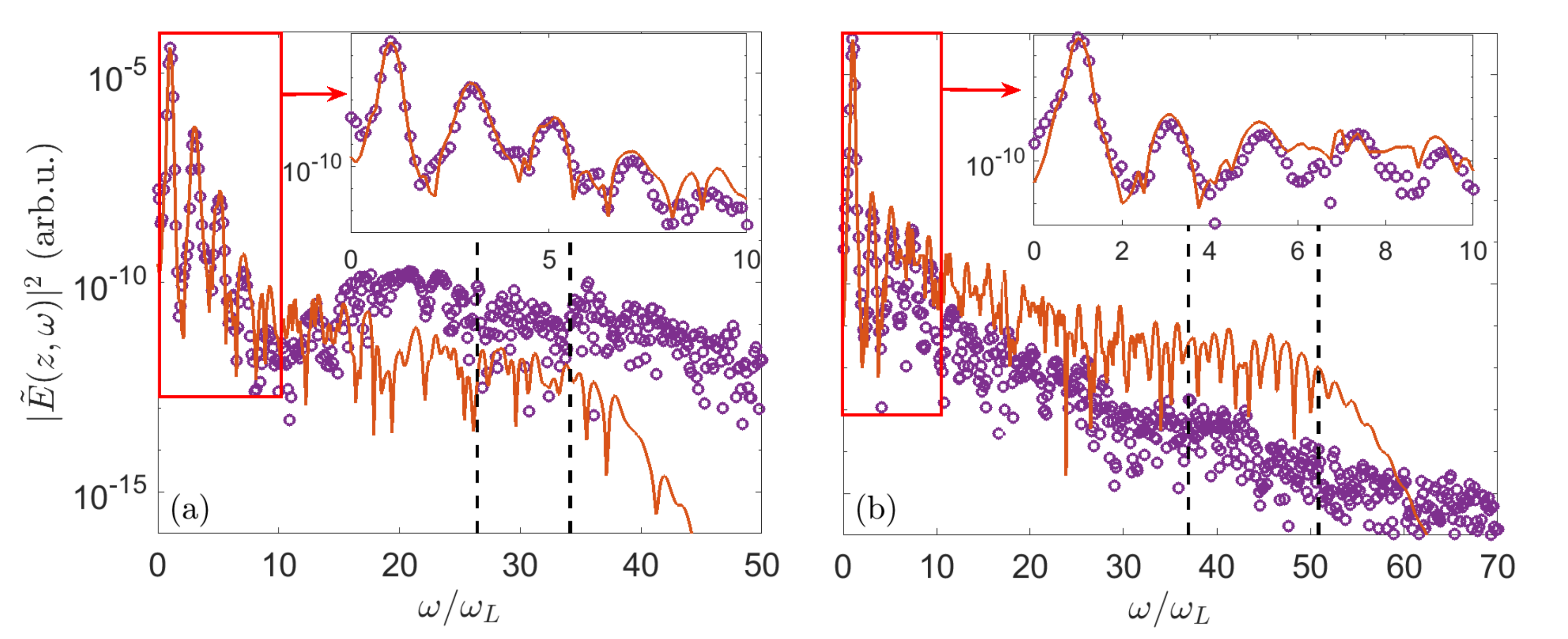}
\caption{High harmonic spectra of the quantum and classical models at $z=1\mm$ in the low-ionization probability regime (a) and the intermediate-ionization probability regime (b). The orange curves correspond to the quantum model while the purple circles correspond to the classical model with $g_\sigma$. The dashed lines indicate the frequencies $2U_p+|I_p|$ and $3.17U_p+|I_p|$. The gas densities are chosen to give a similar peak free-electron density in each case. (a) Incident pulse peak intensity $I = 5\times10^{13} \Wcm$, carrier frequency $\omega_L=0.0378\au$, and gas density $\rho = 2\times 10^{19} \cc$, as in Fig.~\ref{fig:fundSigI5E13}. (b) Incident pulse peak intensity $I = 9\times 10^{13} \Wcm$, carrier frequency $\omega_L=0.0378\au$, and gas density $\rho = 10^{18} \cc$, as in Fig.~\ref{fig:fundSigI9E13}. The vertical axes are directly comparable.}\label{fig:specGs}
\end{figure*}
\begin{figure*}
\includegraphics[width=0.85\textwidth]{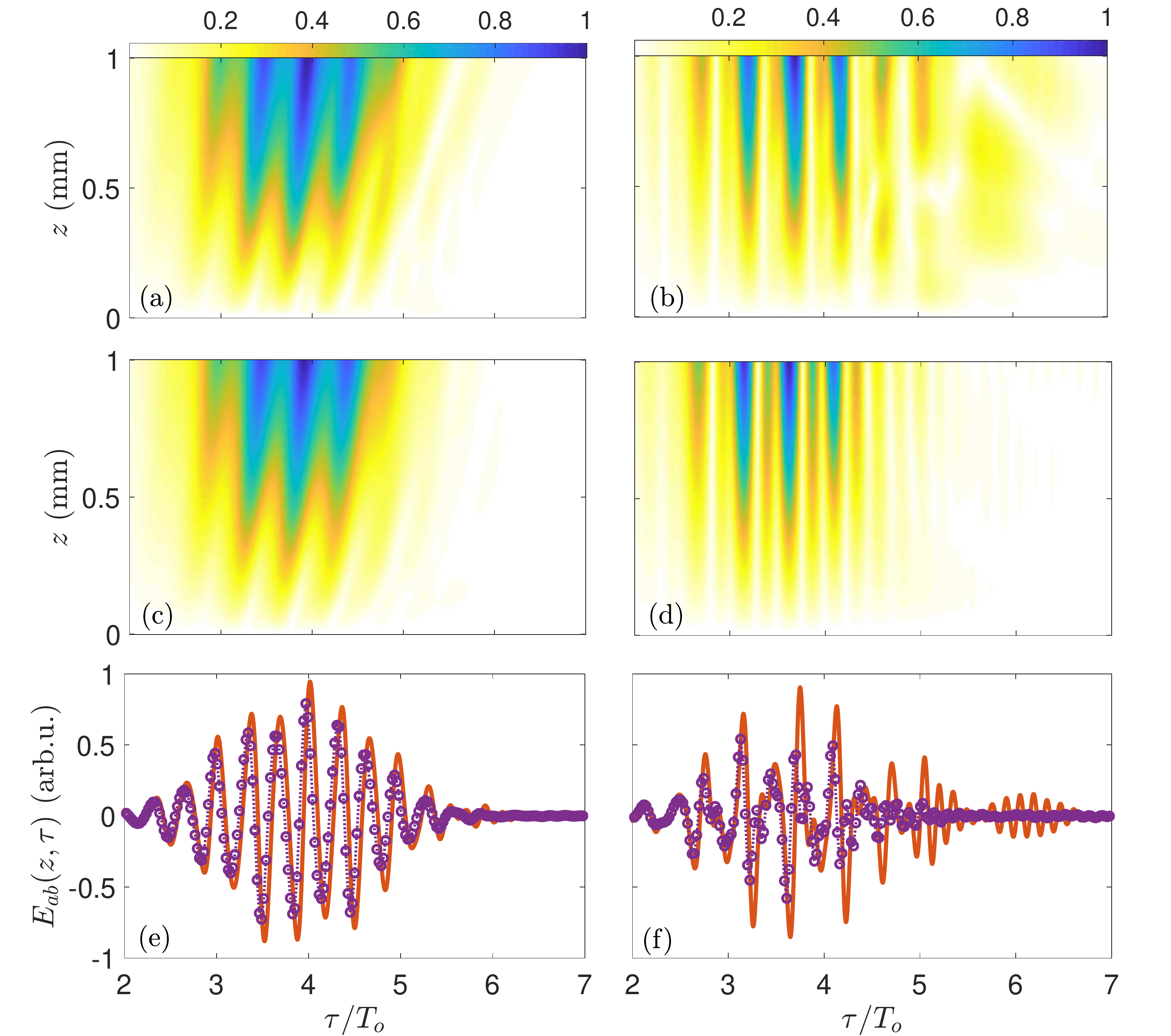}
\caption{Spatiotemporal evolution of low-order harmonics. Frequencies in the range $[2\omega_L,8\omega_L]$ were considered in the calculation of the filtered analytic field $\hat{E}_{ab}(z,\tau)$, where the post-processing consists of multiplying $E$ by a $\sin^2$ window. The left panels are for the low ionization probability case, while the right panels are for the intermediate ionization probability case, with the parameters of Fig.~\ref{fig:specGs}. (a),(b) Harmonic amplitude $|\hat{E}_{ab}(z,\tau)|$ from the quantum model. (c),(d) Harmonic amplitude $|\hat{E}_{ab}(z,\tau)|$ from the classical model with $g_\sigma$. (e),(f) Time-dependent harmonic field $\re[\hat{E}_{ab}(z,\tau)]$ at $z=1\mm$. The solid orange curves are the quantum model, while the purple circles and dotted lines are the classical model with $g_\sigma$. In both sets of panels, the fields are normalized by the maximum harmonic amplitude obtained in the quantum model.}\label{fig:low_harmonics}
\end{figure*}

Figure \ref{fig:specGs} shows the spectra of the field $|\tilde{E}(z,\omega)|^2$ at $z=1\mm$ for the quantum and classical models after propagation through ground-state gases, in low and intermediate ionization probability regimes.
Similarly to the scattering-propagation experiment (Fig.~\ref{fig:specScat}), we see a good agreement for the low frequencies, and no agreement for the high frequencies.
We attribute the discrepancy in the structure of the high-harmonic spectrum to quantum interference effects, as before.
However, the agreement for the low-order harmonics in this case is actually quite remarkable, because here, most of the electrons are bound instead of ionized.
To probe the extent of this agreement deeper, we study the spatiotemporal evolution of the harmonic radiation between $2\omega_L$ and $8\omega_L$, as shown in Fig.~\ref{fig:low_harmonics}.
In the low ionization probability regime, the classical and quantum models agree for the amplitude of the radiation in this frequency band in both $z$ and $\tau$, as seen by comparing Fig.~\ref{fig:low_harmonics}a to Fig.~\ref{fig:low_harmonics}c, particularly for $\tau < 6T_o$.
They also match in phase, as seen in the filtered time-dependent field at $z=1\mm$ in Fig.~\ref{fig:low_harmonics}e.
This suggests that the generation mechanism for these low-order harmonics is the same in both cases.
In the intermediate ionization probability regime, the spatiotemporal evolution of the filtered field's amplitude and phase is close in the classical and quantum cases for $\tau < 3T_o$, but for larger $\tau$ there are significant discrepancies.
This suggests that another mechanism of low-order harmonic generation takes over in the quantum case in this regime.

Because the prominent low-order harmonics of Fig.~\ref{fig:specGs} are absent in the scattering spectrum in Fig.~\ref{fig:specScat}, we conclude that these harmonics are due to the presence of bound electrons.
Aside from interfering with the recolliding electrons, the large population of bound states contributes to harmonic radiation in two ways: through the nonlinear response of bound electrons to the field \cite{Band90,Band92,Uzdi10,Bahl17,Wahl11}, and through the tunneling current \cite{Sere14}.
The radiation from the latter contribution is also known as Brunel radiation \cite{Brun90,Babu17}.
These two mechanisms of bound state radiation make large contributions to the response of the atoms at the fundamental frequency and the low harmonic orders \cite{Sere14}, i.e.\ those magnified in the insets of Fig.~\ref{fig:specGs}.
Because we tuned the classical model to match the ionization probabilities of the quantum model, we should expect the Brunel contribution of both models to be similar.
On the other hand, if the classical and quantum models also agreed for the nonlinear response of bound electrons, this would be an added bonus.
The advantage of the classical model over the quantum model is that by looking in phase space, we can distinguish the contributions of bound electrons versus ionizing electrons.
This allows us to confirm that in fact, the classical model does capture the bound electron radiation, at least in the low ionization probability regime.

As can be inferred from Fig.~\ref{fig:Ef_dist}, the electrons in the classical model follow very different kinds of trajectories depending on their initial energy $h_i$.
Most bound electron trajectories are certain to end in a state with energy $h_f$ very close to $h_i$.
In Fig.~\ref{fig:Ef_dist}, there is clearly a critical initial energy $h_c$ which separates trajectories that are certain to remain bound from those which have a significant probability of ionizing, i.e.\ ultimately ending with an energy $h_f > 0$.
Note that, in Fig.~\ref{fig:Ef_dist}, we only represent $h_f<0$, but the region of $h_i$ where we observe a wide-ranging distribution of $h_f$ far from $h_i$ is also the one from which ionization takes place.
We define $h_c$ as the smallest energy $h_i$ such that there is a nonzero probability of ending in a final state with energy $h_f > V(10) = -0.099 \au$, i.e.\ the energy of an electron at rest at $x=10\au$, and we obtain $h_c(z)$ numerically.
Subsequently, we split the distribution function into two parts: the part consisting of electrons with $h_i < h_c$ and the part consisting of electrons with $h_i \geq h_c$.
By averaging over the latter distribution, we obtained the classical tunneling current $\overline{v}_t(z,\tau)$, which is the mean dipole velocity of the electrons likely to ionize, emulating quantum tunneling.
Averaging over the former distribution gives the classical bound current $\overline{v}_b(z,\tau)$.
Because the total distribution function is the sum of these two distribution functions, the total classical current (or mean dipole velocity) is $\overline{v}=\overline{v}_b + \overline{v}_t$.
Notably, the classical model excludes a third term in the current due to quantum interference between bounded and ionizing states \cite{Sere14}.

\begin{figure*}
\includegraphics[width=0.85\textwidth]{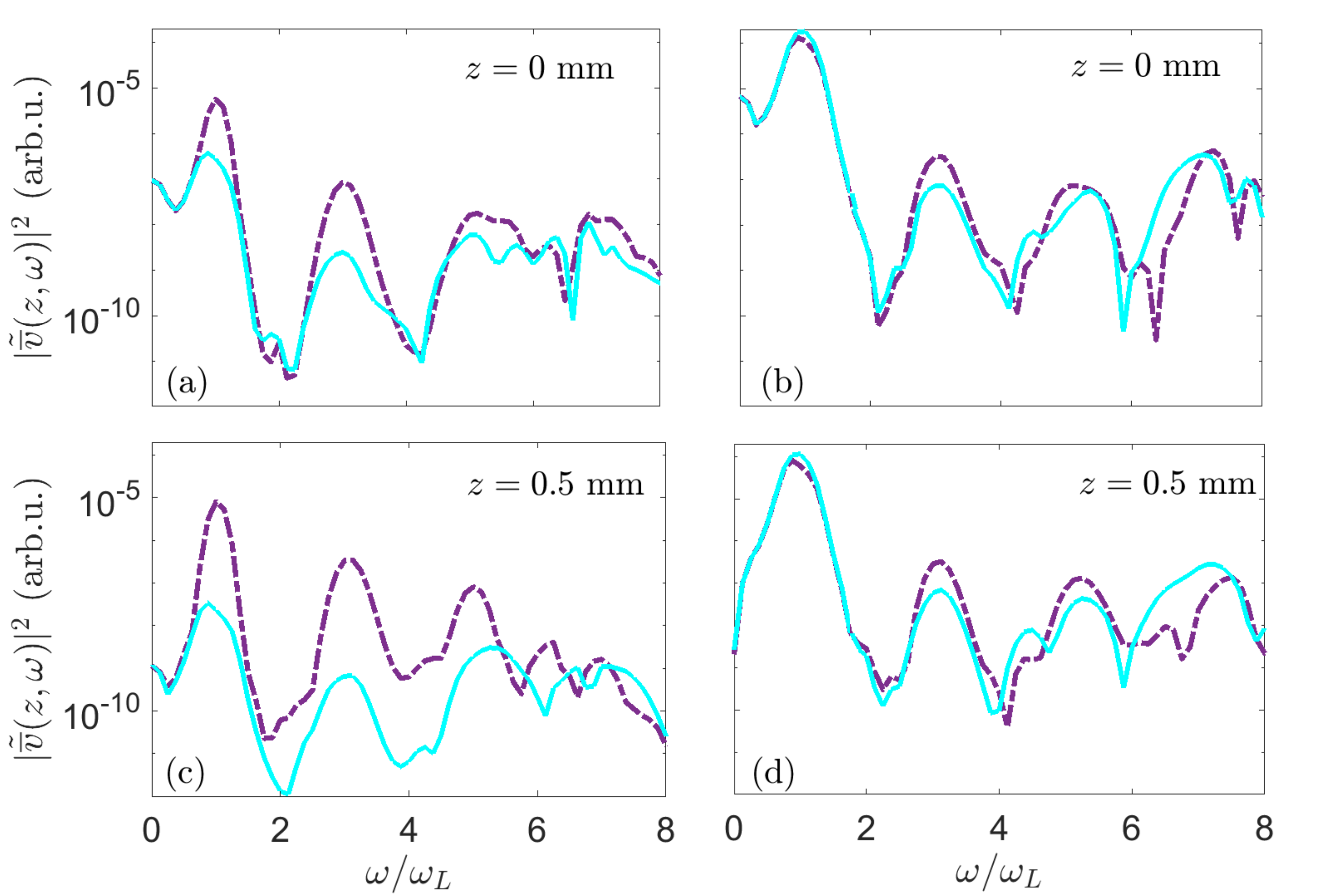}
\caption{Evolution of the classical tunneling current spectrum in the low ionization probability case (a),(c) and the intermediate ionization probability case (b),(d), with the parameters of Fig.~\ref{fig:specGs}. The dash-dotted purple curves are the full dipole velocity spectrum $|\tilde{\overline{v}}(z,\omega)|^2$ from the classical model, while the solid cyan curves are the spectrum of the classical tunneling current $|\tilde{\overline{v}}_t(z,\omega)|^2$. A $\sin^2$ window was applied to the velocities for the computation of the spectrum.}\label{fig:specTun}
\end{figure*}
Figure \ref{fig:specTun} compares the spectra of the tunneling current and the total current for $z=0$ and $z=0.5\mm$ in both the low- and intermediate-ionization regimes.
In the low ionization probability regime, where the initial ionization probability is about $0.5\%$, the tunneling current makes a small contribution to the first and third harmonics at $z=0$, as show in Fig.~\ref{fig:specTun}a, and this extends to the fifth harmonic by $z=0.5\mm$, as shown in Fig.~\ref{fig:specTun}c.
Therefore, most of the low-order harmonic radiation throughout propagation in the low ionization probability regime is due to the bound-electron radiation, as opposed to Brunel radiation.
The good agreement between the classical and quantum ionization probabilities implies that the classical tunneling current is in agreement with the tunneling current, and correspondingly the bound currents are in agreement as well.
Thus, we conclude that in the low ionization probability regime, the primary mechanism of low-order harmonic generation, at least up to fifth order, is the bound-electron nonlinearity.
Furthermore, the classical model and quantum model are in agreement for the generation and propagation of these harmonics, as shown in Fig.~\ref{fig:low_harmonics}.

For the intermediate-ionization regime, on the other hand, Figs.~\ref{fig:specTun}b and \ref{fig:specTun}d indicate that the tunneling current is comparable to the total current.
Indeed, it seems to dominate the total current at frequencies near $\omega_L$, and for the other harmonics the two currents are comparable, indicating the bound radiation and Brunel radiation are also comparable.
However, in this regime the agreement between the classical and quantum low-order harmonics for $\tau < 3T_o$ (Fig.~\ref{fig:low_harmonics}b, \ref{fig:low_harmonics}d, and \ref{fig:low_harmonics}f) gives way to gradually worse agreement for larger $\tau$.
The agreement for smaller $\tau$, when the probability of ionization is still relatively small, suggests that the radiation due to the bound electron motion in the classical and quantum models are still in agreement, as in the low ionization probability regime.
At the same time, we also expect the tunneling current and thus the Brunel radiation to be in agreement.
Hence, the discrepancy must be due to quantum interference effects, which at these low-harmonic orders may come from low-energy recollisions \cite{Xion14} and electron trapping in excited states \cite{Beau16,Yun18}.

\subsubsection{High-order harmonic generation}
\begin{figure*}
\includegraphics[width=0.85\textwidth]{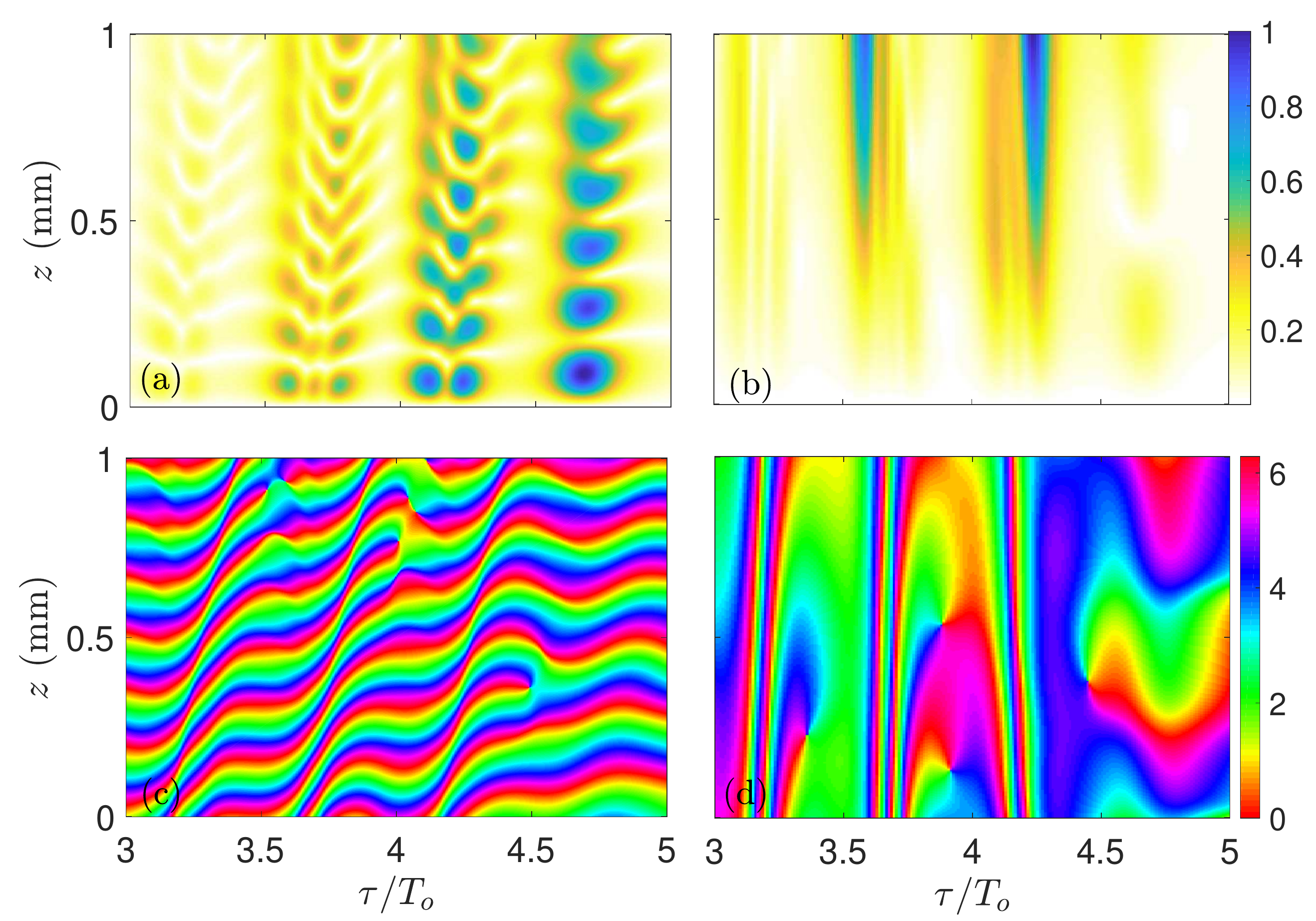}
\caption{Spatiotemporal buildup of high-harmonic radiation in the quantum model in the low ionization probability case (a),(c) and the intermediate ionization probability case (b),(d), with the parameters of Fig.~\ref{fig:specGs}. Frequencies in the range $[2U_p+|I_p|,\infty)$ were considered in the calculation of the filtered analytic field $\hat{E}_{ab}(z,\tau)$ and the filtered analytic dipole velocity $\hat{\overline{v}}_{ab}(z,\tau)$, where the post-processing consisted of multiplying $E$ by a $\sin^2$ window. (a),(b) Amplitude of the high-harmonic part of the field, normalized to the maximum amplitude recorded in each simulation. (c),(d) Phase $\phi_{ab}$ of the high-harmonic emission, in radians, computed from the phase of high-harmonic part of the dipole velocity.}\label{fig:phase_emission}
\end{figure*}
Now, we consider the buildup of high-harmonic radiation during the laser pulse propagation through ground-state atoms.
In Fig.~\ref{fig:phase_emission}a-b, we have plotted the amplitude of the field $E$ in the quantum model for frequencies $\omega > 2U_p+|I_p|$ in the  $(z,\tau)$ plane, for the low and intermediate ionization probability regimes, respectively.
Note that $U_p$ is larger in the intermediate ionization probability regime, because of the higher initial peak intensity of the pulse.
We observe very different behavior in the two cases.
In the low ionization probability regime, the maximum amplitude of the high-harmonic radiation oscillates considerably during propagation, a phenomenon known as Maker fringes \cite{Heyl11}, limiting the coherent buildup of the high harmonics.
On the other hand, in the intermediate ionization probability regime, we observe two bursts of radiation, around $\tau = 3.6T_o$ and $\tau = 4.2T_o$, which build up continuously throughout propagation.
The improved coherent buildup in this regime compared to the low ionization probability regime is also reflected by the higher intensity plateau in the spectrum at $z=1\mm$ in Fig.~\ref{fig:specGs}b compared to Fig.~\ref{fig:specGs}a.

This behavior indicates differing phase-matching conditions in each regime, and we confirm this by computing the phase of the high-harmonic emission.
We have plotted the phase of the high-harmonic part of the dipole velocity, $\phi_{ab}(z,\tau)$, for the frequency range $\omega > 2U_p+I_p$ in Figs.~\ref{fig:phase_emission}c-d.
In order for the radiation to build up coherently over a given propagation distance, the phase of the radiation contained in $\overline{v}$ must not vary much over that distance.
In the low ionization probability regime, we see that the phase of emission varies significantly throughout propagation, possibly at a constant rate which depends on $\tau$.
This explains the oscillations in the amplitude of the radiated field at these frequencies observed in Fig.~\ref{fig:phase_emission}a.
On the other hand, in the intermediate ionization probability regime, for certain fixed $\tau$, we see bands of phase of which are almost constant in $z$.
In particular, this is consistent with the values of $\tau$ for which the bursts of high-harmonic radiation are seen to build up.

\begin{figure*}
\includegraphics[width=0.85\textwidth]{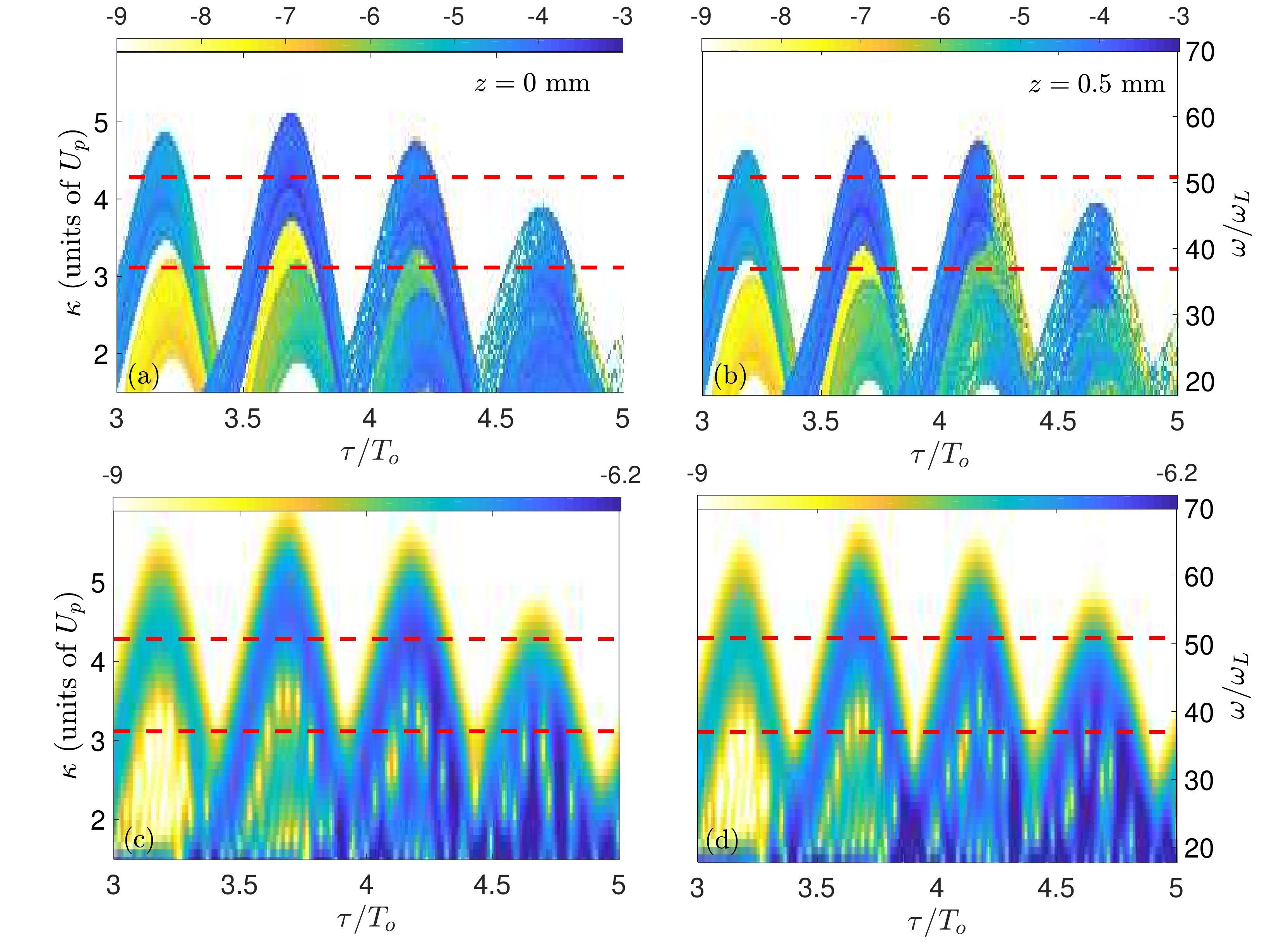}
\caption{Electron dynamics at $z=0$ (a),(c) and $z=0.5\mm$ (b),(d) during laser-pulse propagation through a ground-state gas in the intermediate ionization probability regime, i.e.\ $I=9\times10^{13}\Wcm$, $\omega_L=0.0378 \au$, and $\rho=10^{18}\cc$. (a),(b) Recollision flux $R(\kappa,\tau;z)$ from the classical model. (c),(d) Spectrogram of the dipole acceleration $d_a(\tau)$ from the quantum model. The spectrograms were computed using a $\cos^4$ window of duration $0.15T_o$. The dotted lines indicate $2U_p+|I_p|$ and $3.17U_p+|I_p|$. The left axes, indicating the recollision kinetic energy $\kappa$, are related to the right axes, indicating the radiated frequency $\omega$, by $\kappa = \omega$.}\label{fig:recollisionsGs}
\end{figure*}
The classical model does not provide any obvious explanation of the phase properties of the quantum high-harmonic emission.
In Fig.~\ref{fig:recollisionsGs}, we compare the recollision flux from the classical model to the spectrogram of high-harmonic emission from the quantum model in the intermediate-ionization regime.
As in the scattering-propagation experiment, we see a strong correspondence between the classical and quantum calculations, even after propagation to $z=0.5\mm$.
Comparing Figs.~\ref{fig:recollisionsGs}c and \ref{fig:recollisionsGs}d, we see that between $z=0$ and $z=0.5\mm$, the intensity of emission at given times and frequencies has not changed significantly.
In contrast, Fig.~\ref{fig:phase_emission}d shows that at certain times, the phase of the emission has changed significantly.
It is likely that semiclassical arguments \cite{Lewe94,Gaar08,Sand99} can be used to bridge the gap between the classical model and the quantum high-harmonic phase.
In particular, quantities like the action and recollision time play key roles in determining the phase of high-harmonic emission, and they can be extracted from the classical calculations.
Tracking the evolution of these quantities throughout propagation may be a promising avenue for identifying mechanisms of phase-matching with the classical model.

\section{Propagation of a nearly-LP pulse}\label{sec:EP}
Lastly, we use the 2D model to study the propagation of an EP pulse through a ground-state atomic gas.
Specifically, we examine the stability of the polarization of LP pulses.
In the 2D model [Eq.~\eqref{eq:EOMfield}], a pulse which is initially LP along the $x$ direction remains LP along this direction when propagating through a gas of atoms initially in their ground state, or any initial state satisfying the symmetry $(y,v_y) \rightarrow (-y,-v_y)$.
However, if one perturbs an incident LP pulse with a small ellipticity, does that perturbation grow during propagation?
This is a natural question, whose answer determines the robustness of the physical picture provided by the 1D atom-field models that we have focused on in this paper.
Indeed, if the perturbation were to grow, then LP propagation would be unstable and one would always need to consider models of at least two dimensions, even in the LP case.

\begin{widetext}
To address this question, we consider an incident pulse of the form
\begin{equation*}
\bE_0(\tau) = \begin{cases}
\frac{E_0}{\sqrt{1 + \xi_0^2}} \sin^2(\frac{\pi \tau}{T_m})\left[\cos(\omega_L \tau) \hat{\bf x} + \xi_0 \sin(\omega_L \tau)\hat{\bf y} \right] & {\rm for}\,\,\ 0 < \tau < T_m, \\
0 & {\rm for}\,\, T_m < \tau < \tau_f.
\end{cases}
\end{equation*}
\end{widetext}
Here, $\xi_0$ is the initial ellipticity of the pulse.
Since $\xi_0 = 0 $ would be LP, nearly LP pulses correspond to the case $|\xi_0| \ll 1$.
We have normalized the field amplitude by $\sqrt{1 + \xi_0^2}$ to maintain the relationship between $E_0$ and the field intensity.
The atoms are now modeled in 2D via Eq.~\eqref{eq:EOMq} for the quantum model or Eq.~\eqref{eq:EOMc} for the classical model.
We take a softening parameter $a = 0.8 \au$
For the quantum model, we numerically compute the ground state using imaginary-time propagation, leading to a ground-state energy of $I_p=-0.4991 \au$ 
This is nearly the same ground-state energy as for our 1D calculations.
For the classical initial state, we use the 2D analog of Eq.~\eqref{eq:classicalGs}, with $g_\sigma$ (Eq.~\eqref{eq:gSigma}) as the initial energy distribution, and the boundaries of the energy range taken as $h_{\min} = -1/a$ and $h_{\max} = -0.25\au$
The parameters of $g_\sigma$ for the 2D model are obtained by optimizing the ionization probabilities of the 2D classical model compared to the 2D quantum model for an LP external pulse of varying intensities, as in the 1D case.
This yields $k = 100.5\au$ and $h_m=-0.3813\au$
We remark that these parameters are close to those obtained in the 1D case.

\begin{figure*}
\includegraphics[width=0.85\textwidth]{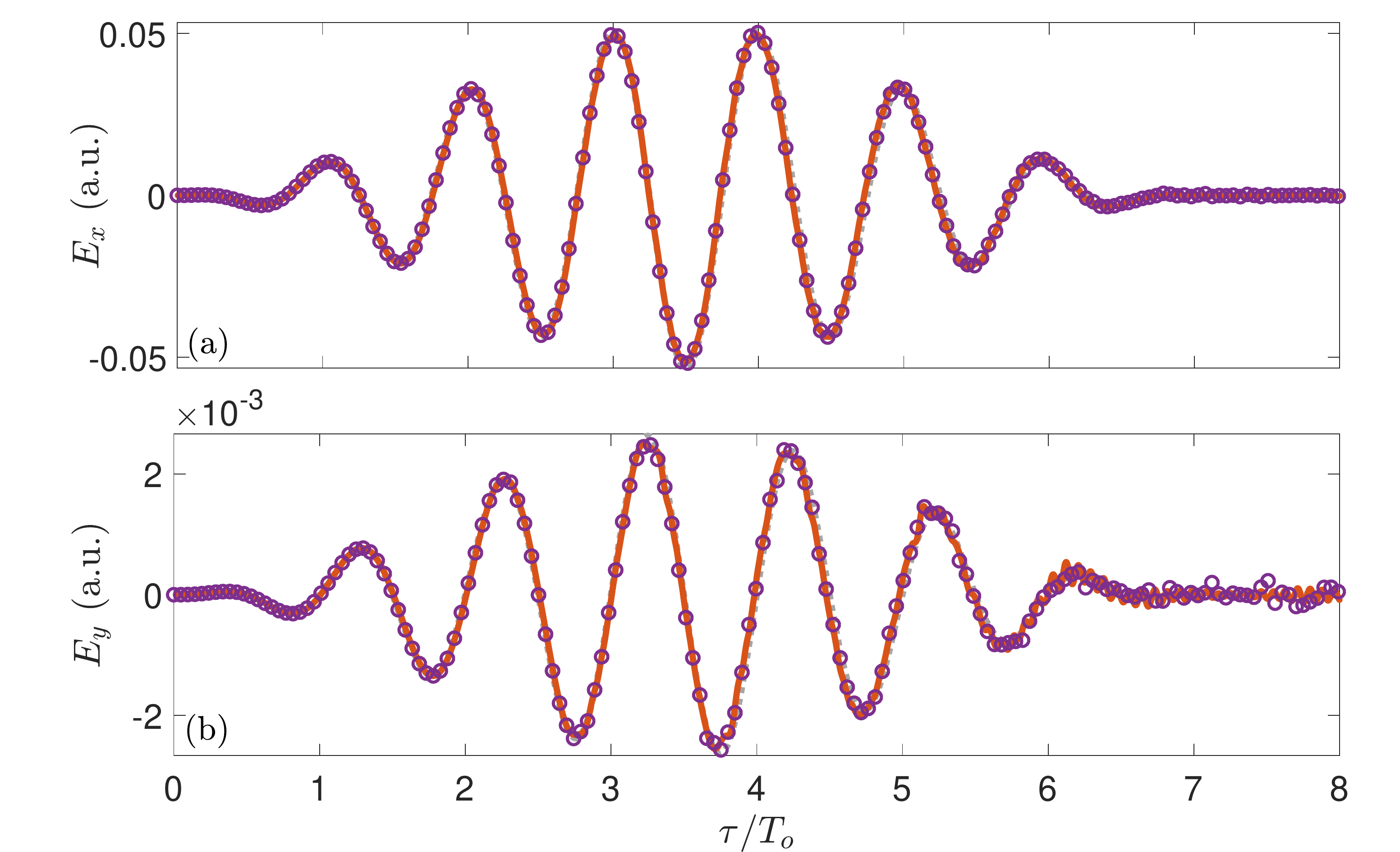}
\caption{Elliptically-polarized electric field $\bE(z,\tau)$ after propagation to $z=1\mm$ in a ground-state atomic gas with density $\rho = 10^{18} \cc$, for an incident pulse with  peak intensity $I=10^{14} \Wcm$, carrier frequency $\omega_L=0.0378$, and ellipticity $\xi_0 = 0.05$. The solid orange curves correspond to the quantum model and the purple circles correspond to the classical model. (a) The field along the major axis, $E_x(z,\tau)$. (b) The field along the minor axis, $E_y(z,\tau)$. The initial field, plotted as the grey dotted lines, is mostly covered by the field after propagation. The spatial step used is $\Delta z = 16.7 \lambda_L = 20\,\,\mu{\rm m}$.}\label{fig:2Dfields}
\end{figure*}

Figure \ref{fig:2Dfields} shows the result of the propagation of the nearly-LP pulse to $z=1\mm$ in a ground-state gas in the intermediate ionization probability regime.
The peak intensity of the pulse is selected as $I=10^{14} \Wcm$ and the initial ellipticity is $\xi_0 = 0.05$.
Under these conditions, the initial ionization probability in the quantum model is $P_{\rm ion}(T_m) = 5.8\%$ and in the classical model is $P_{\rm ion}(T_m) = 3.4\%$.
These values are comparable to the ionization probabilities in our 1D calculations in the intermediate ionization regime, though somewhat smaller.
We have nevertheless used the same density of $\rho=10^{18} \cc$ here, which means the free electron density is lower in this case than the 1D case.
After $1\mm$ of propagation, the change to the major-axis component of the nearly-LP field, shown in Fig.~\ref{fig:2Dfields}a, looks qualitatively similar to the change to the LP field in the 1D model, shown in Fig.~\ref{fig:fundSigI9E13}a.
That is, there is very little reshaping overall, with some blueshifting visible at the falling edge of the pulse.
There is also an excellent agreement between the quantum and classical models, as in the 1D case.
Turning our attention to the field component along the minor axis, plotted in Fig.~\ref{fig:2Dfields}b, we observe a similar behavior as for $E_x$.
The field $E_y$ computed from the classical model contains stronger high frequency oscillations at the end of the pulse than its quantum counterpart.
We have determined that this is due in part to the numerical error associated in solving Eq.~\eqref{eq:EOMc}---i.e. it can be improved by using a finer discretization of the distribution function.
Even at this level of accuracy, the agreement between the classical and quantum calculations is quite good, but the classical calculation has the significant added  virtue of being about $1.7$ times faster than the quantum calculation.
This matter is discussed further in the appendix.

\begin{figure*}
\includegraphics[width=0.85\textwidth]{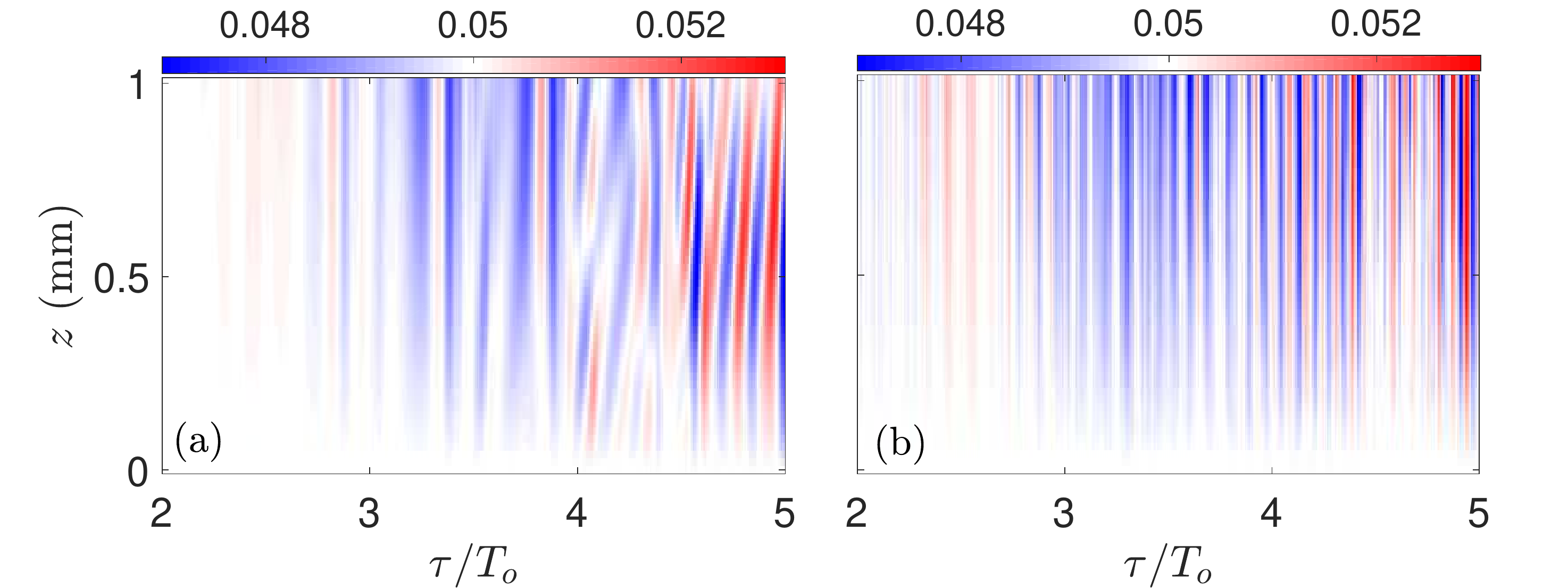}
\caption{Spatiotemporal ellipticity $\xi(z,\tau)$ of the electric field $\bE(z,\tau)$ for the quantum model (a) and the classical model (b), with the parameters of Fig.~\ref{fig:2Dfields}. The ellipticity is indicated by the color scale, which is centered on the initial ellipticity $\xi_0 = 0.05$.}\label{fig:ellipticity}
\end{figure*}
Because both fields change so little over the course of the propagation, we find that the spatiotemporal ellipticity of the field in the middle of the pulse remains close to its initial value.
This is illustrated in Fig.~\ref{fig:ellipticity}, where $\xi(z,\tau)$ is plotted for both the quantum and classical models.
The fine features between the two models differ, with the quantum model (Fig.~\ref{fig:ellipticity}a) predicting nonmonotonic oscillations of $\xi(z,\tau)$ throughout propagation.
Still, both models predict very small fluctuations, on the order of $0.06 \xi_0$, around the initial ellipticity throughout propagation.
Both models also predict that the offset angle of the polarization ellipse, $\theta(z,\tau)$, remains close to its initial value of zero, satisfying  $|\theta(z,\tau)| < 6\times 10^{-3}$ for $2T_o < \tau < 5T_o$ (not shown).
We have obtained similar results in the low ionization probability regime, with a pulse of peak-intensity $I=6\times10^{13} \Wcm$, gas density $\rho=2 \times 10^{19}\cc$, and the same initial ellipticity as the present case (not shown).
There, we note that the classical calculation yields even more prominent high-frequency oscillations in the field than those seen in Fig.~\ref{fig:2Dfields}, indicating a need for a more accurate solution of Eq.~\eqref{eq:EOMc}.
Also, in this regime, the classical model significantly overestimates the group velocity of the laser pulse compared to the quantum model, in contrast to the corresponding 1D model for a similar set of parameters, i.e.\ those of Fig.~\ref{fig:fundSigI5E13}.
It is possible that the 2D initial condition distribution of the classical model may be further optimized to improve the agreement with the quantum model on this phenomenon.
Nevertheless, our calculations provide evidence for the stability of the laser polarization near LP through experimentally-relevant propagation distances.
This justifies the use of purely 1D models for the investigation of the propagation of LP pulses.

\section{Conclusion}\label{sec:concl}
In summary, we have presented an in-depth study of the behavior of first-principles reduced models for the propagation of intense laser pulses in atomic gases with densities on the order of $10^{17}$--$10^{19} \cc$ over a distance of $1\mm$, which are the parameter ranges of many experiments.
We mainly focused on the simplest possible model: a linearly-polarized, one-dimensional laser field propagating through a gas of one-dimensional model atoms, which can be treated either quantum-mechanically or classically.
In a previous work, we showed that the quantum and classical models exhibit quantitative agreement when the electron is initially ionized \cite{Berm18}.
Here, we identified a proxy for their quantitative agreement when the electron is initially in the ground state: the intensity-dependent ionization probability.
We proposed a ground-state initial condition for the classical model which provides optimal agreement with the quantum model for this single-atom observable.
This in turn led to very similar behavior between the two models on the macroscopic level as well.
In particular, the quantum and classical models exhibit good quantitative agreement on their predictions of pulse energy loss, blueshift, subluminal group velocity, and low-order harmonic generation in the low-to-intermediate ionization probability regimes.
Therefore, we have demonstrated that the classical model is a viable tool for the simultaneous first-principles simulation of coupled laser pulse-electron dynamics and visualization of the electron dynamics in phase space.

We used this tool to investigate harmonic generation and the stability of nearly-LP pulses.
High-harmonic generation was examined, with an emphasis on the scattering-propagation experiment.
The extension of the high-harmonic cutoff by propagation through the gas, reported in Ref.~\cite{Berm18} (and also reported in another context in Ref.~\cite{Lori08}), was explained in detail using the phase space perspective afforded by the classical model.
We also reported on calculations of harmonic generation from ground-state atoms.
The classical and quantum models were in quantitative agreement for the intensity and phase of low-order harmonics in the low-ionization probability regime, and the classical model was exploited to identify bounded electron motion as the mechanism for the radiation.
When we considered an increased ionization probability (by increasing the peak intensity of the incident laser pulse), the times of disagreement between the classical and quantum models on the radiation pointed to interference effects as the dominant radiation mechanism at those times, as opposed to bounded electron motion or tunneling ionization \cite{Sere14}.
Lastly, we used the 2D models to show that the initial polarization of a nearly-LP pulse is stable with respect to propagation effects up to intermediate ionization probabilities.
Hence, the results from the 1D models for LP fields emphasized in this paper are expected to be robust with respect to increases of the model dimension.

Our work paves the way for trajectory-based control strategies which explicitly include, and potentially exploit, propagation effects.
For example, attosecond steering of the electron trajectories after ionization has proven to be a viable method for controlling HHG \cite{Chip09,Briz13,Haes14} and THz generation \cite{Mart15}.
However, up until now, the robustness of these methods to propagation effects has needed to be verified {\it a posteriori}, rather than having been built-in from the outset.
On the other hand, recent experiments demonstrate that propagation effects may be harnessed to effect unprecedented enhancements of HHG  \cite{John18,Card18}.
By linking the classical motion of the electrons with the reshaping of the field during propagation, our reduced classical models provide a tool for the development of new schemes to control the properties of the generated radiation.

\section*{Acknowledgements}
We acknowledge Fran\c{c}ois Mauger for many stimulating discussions.
S.A.B. acknowledges Francesco Fedele and Denys Dutykh for extensive advice on numerical schemes for the Liouville equation.
The project leading to this research has received funding from the European Union's Horizon 2020 research and innovation program under the Marie Sk{\l}odowska-Curie grant agreement No 734557.
S.A.B. and T.U. acknowledge funding from the NSF (Grant No. PHY1602823).
S.A.B. acknowledges funding from the Georgia Tech College of Sciences for extended visits to Marseille.
 
\appendix
\section{Numerical implementation}\label{A:Numerics}
\subsection{Schr{\"o}dinger equation}\label{A:Schr}
The TDSE \eqref{eq:EOMq} was solved using a second-order operator splitting scheme in both 1D and 2D cases \cite{Band13_2}.
Derivatives of the wave function with respect to the electron position (i.e.\ for the application of the momentum and kinetic energy operators) were performed in the Fourier domain.
Absorbing boundary conditions were employed \cite{Lori09}, consisting of sending the wave function smoothly to zero within $32\au$ of each domain boundary using a $\cos^{1/8}$ function.
For the 1D case, the computational domain selected was $x \in [-1800,1800] \au$, discretized with a spatial step size of $\Delta x = 5/16 \au$
A fixed time-step of $\Delta \tau = 0.1\au$ was used.
We verified that for this set of integration parameters, the high-harmonic spectrum of a single atom in an external, monochromatic field (the one used as the initial field in Sec.~\ref{sec:scattering}) was converged.
That is, we compared the dipole velocity spectrum with these parameters to a spectrum calculated with either (i) a larger domain, (ii) a smaller $\Delta x$, or (iii) a smaller $\Delta \tau$, and each of these was indistinguishable from the spectrum with the above set of parameters.
For the 2D case, the computational domain selected was $(x,y) \in [-120,120] \times [-80,80] \au$, discretized with a spatial step size of $\Delta x = \Delta y = 7/16 \au$
A fixed time-step of $\Delta \tau = 0.1\au$ was used.
These parameters do not provide a fully converged spectrum for a single atom in an external field, but they were used nevertheless because a more accurate calculation would have taken too long, and here we focus on the near-fundamental frequencies in the 2D case.

Indeed, going from 1D to 2D greatly increases the computation time for the TDSE.
For example, integrating the 1D TDSE for a pulse of duration $\tau_f = 8T_o$ with a time step of $\Delta \tau = 0.1\au$, a domain size of $x \in [-120,120] \au$, and $\Delta x = 7/16 \au$ takes $0.07\,\,\min$, while doing the same for the 2D TDSE with the parameters given in the preceding paragraph takes $5.23\,\,\min$, i.e.\ over $700$ times longer.
For the calculation, we used a standard implementation of the fast-Fourier transform on an $8$-core desktop computer.
Because the TDSE calculation needs to be performed repeatedly to advance the laser field, this increase in computation time greatly increases the time of a propagation calculation.

\subsection{Liouville equation}\label{sec:Liouville}
To solve Liouville Eq.~\eqref{eq:EOMc}, we employ the particle-in-cell (PIC) scheme described in \cite{Evst13}.
We describe the scheme for the 1D case for simplicity, though the same scheme was used in 2D.
In a nutshell, the distribution function $f(x,v,z,\tau)$ at a fixed $z$ is discretized at $\tau=0$ on a uniform grid in phase space, and subsequently each grid point follows the characteristics, or particle trajectories, of Eq.~\eqref{eq:EOMc}.
More precisely, we represent $f$ by $N$ particles with trajectories $(x_j(z,\tau),v_j(z,\tau))$, such that
\begin{equation}\label{eq:f_discrete}
f(x,v,z,\tau) = \sum_{j = 1}^N w_j \delta(x - x_j(z,\tau)) \delta(v - v_j(z,\tau)).
\end{equation}
The particle trajectories obey the equations of motion of a classical electron in the combined Coulomb and laser fields, i.e.\
\begin{align*}
\partial_\tau x_j & = v_j, \\
\partial_\tau v_j & = -\partial_x V(x_j) - E(z,\tau).
\end{align*}
These may be derived from the single-particle time-dependent Hamiltonian
\begin{equation}\label{eq:Ham_electron}
H(x,v,z,\tau) = \frac{v^2}{2} + V(x) + E(z,\tau)x,
\end{equation}
where $x$ and $v$ are canonically conjugate and $z$ acts as a label.
Taking advantage of the fact that Hamiltonian \eqref{eq:Ham_electron} is separable into kinetic and potential energy terms, we compute the trajectories $(x_j(z,\tau),v_j(z,\tau))$ using a third-order explicit symplectic scheme \cite{Ruth83}.
We used a fixed time-step of $\Delta \tau = 0.1\au$, as in the quantum case.
This value of $\Delta \tau$ allows a point-wise comparison of the electric fields computed in the classical and quantum models, and we verified that the single-atom dipole velocity spectrum for an external monochromatic field (again, the initial field of Sec.~\ref{sec:scattering}) with this $\Delta \tau$ is converged.

Because $E(z,\tau)$ is computed numerically at every $z$ by solving Eq.~\eqref{eq:EOMfield}, it is only known at discrete values of $\tau$.
In other words, an explicit expression for $E(z,\tau)$ that may be evaluated at any $\tau$ is unavailable.
On the other hand, the scheme used for obtaining $(x_j(z,\tau),v_j(z,\tau))$ requires the evaluation of $E(z,\tau)$ at times in between adjacent time steps.
To compute these values, it is ideal to discretize $E(z,\tau)$ with the same time steps as those used in the trajectory calculation, with spacing $\Delta \tau$.
Then, intermediate values are obtained by quadratic interpolation of $E(z,\tau)$ using the values of $E(z,\tau)$ at the nearest available time steps.
That is, when advancing the trajectories from $\tau_m$ to $\tau_m + \Delta \tau$, the requisite intermediate-time values of $E$ are obtained by quadratic interpolation of $E(z,\tau_m-\Delta \tau),E(z,\tau_m),$ and $E(z,\tau_m+\Delta \tau)$.
Quadratic interpolation provides the intermediate values of $E$ to second-order accuracy in $\Delta \tau$, which is sufficiently accurate to retain the third-order accuracy of the time integration scheme.

\begin{figure*}
\includegraphics[width=0.85\textwidth]{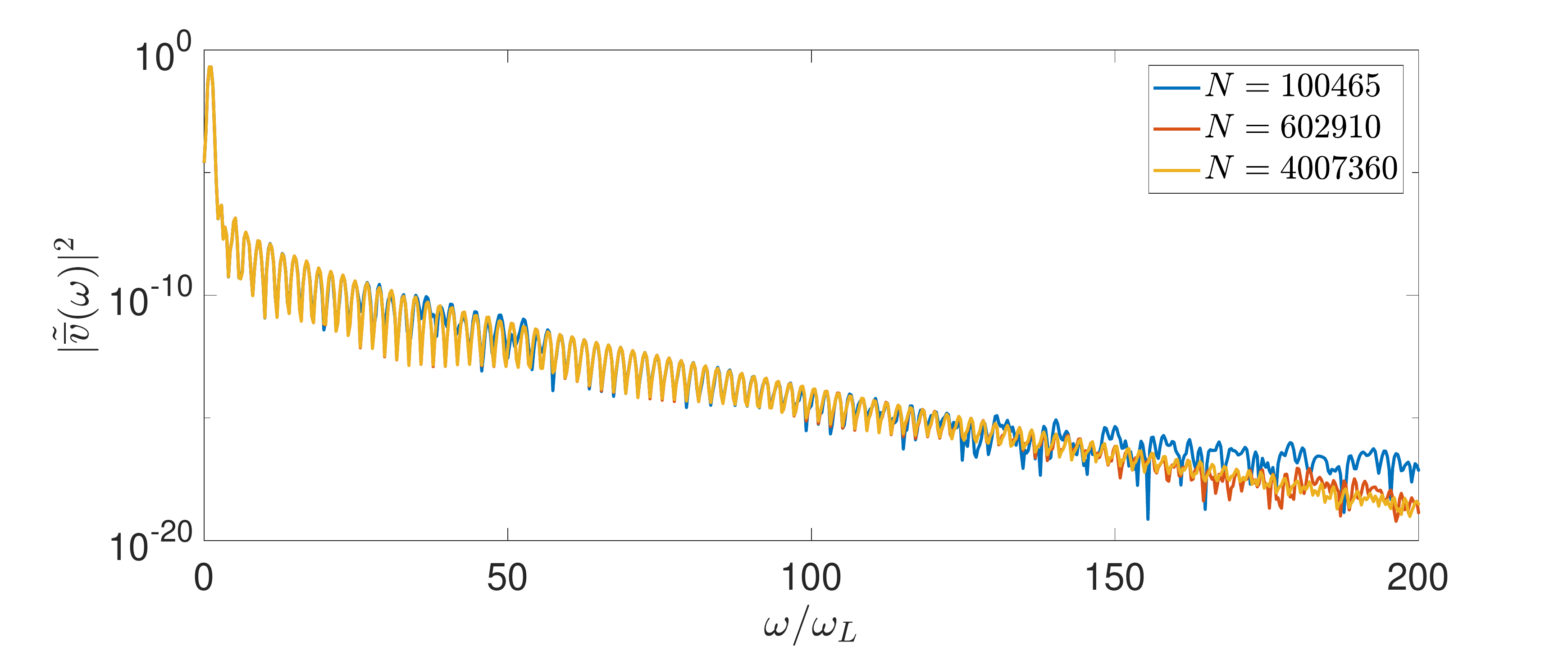}
\caption{Convergence of the classical single-atom dipole velocity spectrum with increasing $N$ for the incident field $E_0(\tau)$ and electron initial conditions of Fig.~\ref{fig:fundScat}.  A $\sin^4$ window  was applied to $\overline{v}(\tau)$  prior to the calculation of the power spectrum. Each curve corresponds to a different value of $N$ used in the discretization \eqref{eq:f_discrete} of the distribution function.}\label{fig:N_convergence}
\end{figure*}

In Eq.~\eqref{eq:f_discrete}, trajectory $j$'s contribution to $f$ is weighted by $w_j$, which is determined by the trajectory's initial condition $(x_j(z,0),v_j(z,0))$ and the initial distribution function $f_0(x,v)$.
$N$ of these initial conditions are selected from a uniform, equally-spaced grid of points on the $(x,v)$ phase space.
The boundaries of this grid are selected such that $f_0$ is sufficiently large (i.e.\ non-negligble) for points within the boundaries, and the initial conditions $(x_j(z,0),v_j(z,0))$ kept are those within the boundaries of the grid and with $f_0(x_j(z,0),v_j(z,0)) > 0$.
Hence, $N$ depends on how many points are contained in the area of the grid where $f_0(x,v)>0$.
In practice, we estimate the necessary resolution of our grid such that the number of grid cells in the nonzero-$f_0$ area is approximately equal to a target number of trajectories $N_{\rm goal}$.
We typically choose a round number for $N_{\rm goal}$, and this yields an actual number $N$ of trajectories which is close to $N_{\rm goal}$.
For example, the values of $N$ reported in the legend of Fig.~\ref{fig:N_convergence}  correspond to $N_{\rm goal} = 10^{5}$, $N_{\rm goal} = 6\times10^{5}$, and $N_{\rm goal} = 4\times10^{6}$,  respectively.
Finally, the weights are given by
\begin{align*}
& w_j = \frac{f_0(x_j(z,0),v_j(z,0))}{\sum_{j'} f_0(x_{j'}(z,0),v_{j'}(z,0))}.
\end{align*}
Selecting the initial conditions on a uniform grid, rather than performing a Monte-Carlo simulation, leads the spectrum of the dipole velocity $\overline{v}(z,\tau)$ to converge more quickly with increasing $N$ \cite{Uzdi10}.

The choice of $N$ determines how many harmonics are accurately resolved by the classical calculation.
In Fig.~\ref{fig:N_convergence}, we show classical dipole velocity spectra obtained from the solution of Eq.~\eqref{eq:EOMc} with different values of $N$.
The calculation was performed with the initial electric field $E_0(\tau)$ and the electron initial conditions $f_0(x,v)$ of the scattering experiment (see Sec.~\ref{sec:scattering}).
Increasing $N$ is equivalent to increasing the fineness of the grid from which the particle initial conditions are sampled from, because we make the boundaries of the grid independent of $N$.
Naturally, this should improve the representation of the distribution function.
As a consequence, we see in Fig.~\ref{fig:N_convergence} that more and more harmonics are converged as $N$ increases.
Comparing the spectrum for $N=100465$ to the one with $N=602910$, we see that they are in agreement up until about $25\omega_L$, after which there are some deviations from $25$--$60\omega_L$ and significant deviations for $\omega > 120\omega_L$.
This indicates that the spectrum with $N=100465$ is converged up until about $25\omega_L$.
Similarly, the $N=602910$ spectrum is seen to be converged up to about $135\omega_L$ by comparison with the $N=4007360$ spectrum.
Evidently, a very large number of particles is required to accurately calculate the high harmonics in the classical model, which however are of very low intensity.
All the ground-state atom calculations presented in main text employ $N_{\rm goal} = 10^5$, while the calculations for the scattering-propagation experiment employ $N_{\rm goal} = 4\times10^6$.
Hence, the high harmonics of the presented classical spectra in the ground-state cases are not fully converged (though this does not seem to affect the lower harmonics and hence, the values of $E(z,\tau)$), while those for the scattering-propagation experiment should be nearly converged.

When the initial distribution $f_0$ is given by Eq.~\eqref{eq:classicalGs}, it is convenient to have an explicit expression for the normalization constant $N_h$.
According to Eq.~\eqref{eq:energyNorm}, $N_h$ is the normalization constant for a uniform initial distribution with fixed energy $h$, given by $\delta(h-H(x,v))$.
Using the fact that the Heaviside step function is the antiderivative of the Dirac delta function, we have
\begin{equation*}
N_h = \left(\frac{d}{dh} \int \Theta(h - H(x,v)) {\rm d}x{\rm d}v \right)^{-1}
\end{equation*}
Now, the integrated quantity is simply the phase-space area $A$ of the orbit with energy $h$, so by the inverse function theorem, we have $N_h = dh/dA$.
Furthermore, since the area is essentially the action of the orbit, $dh/dA$ is the orbit's frequency, and thus we have
\begin{equation}\label{eq:normNu}
N_h = \frac{\nu(h)}{2\pi}, 
\end{equation}
where the factor of $2\pi$ is due to the fact that $\nu$ is the angular frequency.
Equation \eqref{eq:normNu} still requires an explicit expression for $\nu(h)$, and for all the calculations in this paper, we used the approximation given by Eq.~\eqref{eq:omegaA}.
Note that in the 2D case, the interpretation of $N_h$ is slightly different, but it is straightforward to evaluate the 2D version of the integral in Eq.~\eqref{eq:energyNorm} directly, leading to $N_h^{2{\rm D}} = [2\pi^2(h^{-2} - a^2)]^{-1}$.

Compared to the TDSE, it is less computationally costly to go from 1D to 2D for the Liouville equation solved with the PIC scheme.
For example, integrating the 1D Liouville equation for a pulse of duration $\tau_f = 8T_o$ with a time step of $\Delta \tau = 0.1\au$ and $N_{\rm goal} = 10^5$ using an optimized, parallelized code takes $0.81\,\,\min$, whereas doing the same for the 2D Liouville equation using a non-optimized, parallelized code takes $3.27\,\, \min$---only a factor of $4$ longer.
For this calculation, we used the same $8$-core desktop computer as for the timed TDSE calculations, and we parallelized the integration of different trajectories using OpenMP.
Hence, the classical model has a much more favorable scaling of computational cost as the system dimension is increased, as expected.

\subsection{Unidirectional pulse equation}\label{sec:numericsUni}
The numerical methods of the previous sections are used to evaluate $\overline{v}$, which is the source term of the unidirectional pulse-propagation equation \eqref{eq:EOMfield} in the moving frame.
Hence, given $E(z,\tau)$, we can calculate $\overline{v}(z,\tau)$ and advance the electric field in space by $\Delta z$ to obtain $E(z+\Delta z,\tau)$.
To do this, we discretize $E(z,\tau)$ by Fourier transform, i.e.
\begin{equation}
\tilde{E}_k(z) = \frac{1}{\tau_f}\int_0^{\tau_f} E(z,\tau) \exp\left[-i\left(\frac{2\pi k}{\tau_f}\right)\tau\right]{\rm d} \tau.
\end{equation}
for integer values of $k$.
We take $|k| \leq k_{\max}$, where $k_{\max}$ depends on the number of time steps $n = \lfloor \tau_f/\Delta \tau \rfloor$ used in the time-discretization of Eqs.~\eqref{eq:EOMq} or \eqref{eq:EOMc}.
It is given by
\begin{equation*}
k_{\max} = \begin{cases}
n/2 & \text{for } n \text{ even}, \\
(n-1)/2 & \text{for } n \text{ odd}.
\end{cases}
\end{equation*}
Applying the Fourier transform to both sides of Eq.~\eqref{eq:EOMfield}, we obtain
\begin{equation}\label{eq:EOMfieldFourier}
\partial_z \tilde{E}_k = \frac{2\pi\rho}{c}\tilde{\overline{v}}_k(z)
\end{equation}
for each $k$.
Hence, we have converted the PDE \eqref{eq:EOMfield} into a finite set of coupled ODEs \eqref{eq:EOMfieldFourier}.
We solve this system of ODEs using the two-step implicit Adams-Moulton method in Predict-Evaluate-Correct-Evaluate (PECE) mode \cite{Hair93}, which is of third-order accuracy in the spatial step $\Delta z$.
Justification for the use of this method is provided in the next subsection.
In order to evaluate $\overline{v}(z,\tau)$, one needs the electric field at the discrete time positions $\tau_m = m\Delta \tau$, which is obtained directly from the discrete inverse-Fourier transform of $\tilde{E}_k(z)$.
\begin{figure*}
\includegraphics[width=0.85\textwidth]{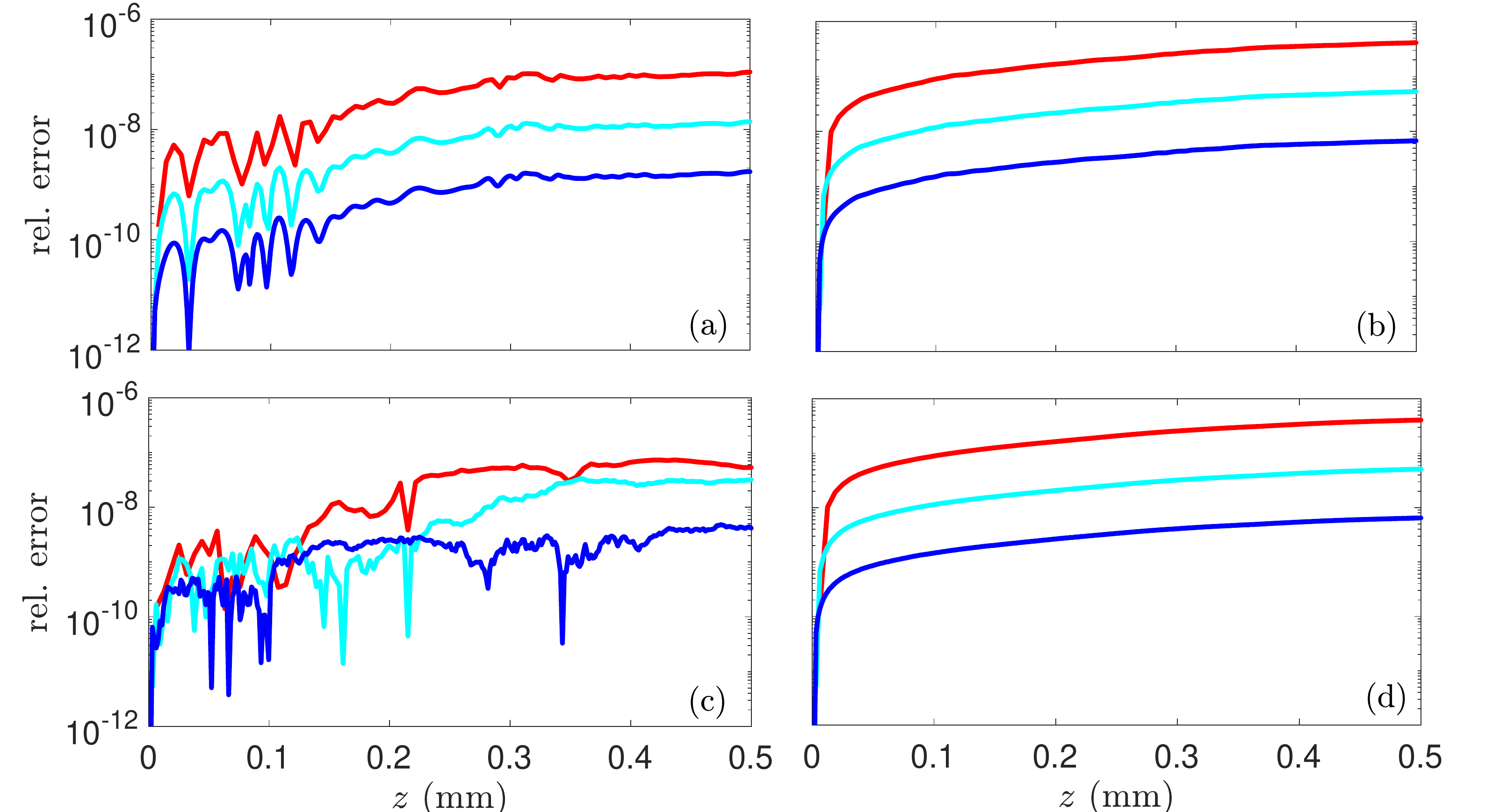}
\caption{Comparison of the relative energy error for the propagation simulations with a third-order Runge-Kutta scheme (RK3) (a),(c) and the two-step Adams-Moulton scheme (AM2) (b),(d). The scattering-propagation experiment setup was used for the field and particle initial conditions, with the parameters of Fig.~\ref{fig:fundScat}. Each curve  was computed with a different $\Delta z$. Red curves: $\Delta z = (21/4)\lambda_L = 6.33\,\,\mu{\rm m}$. Cyan curves: $\Delta z = (21/8) \lambda_L = 3.16\,\,\mu{\rm m}$. Blue curves: $\Delta z = (21/16)\lambda_L = 1.58 \,\,\mu{\rm m}$. (a),(b) Quantum model. (c),(d) Classical model, where $N = 602910$ particles were used for the solution of the Liouville equation.}\label{fig:energy_convergence}
\end{figure*}

As an aside, we remark that it is important to use local interpolation, such as the quadratic interpolation described in Sec.~\ref{sec:Liouville}, to obtain the values of $E(z,\tau)$ in between time steps, rather than evaluating the discrete inverse-Fourier transform at intermediate times.
That is, one may construct an approximation to $E(z,\tau)$ from the finite set Fourier components of $E$, as
\begin{equation}\label{eq:efourierapprox}
E(z,\tau) \approx \sum_{|k| \leq k_{\max}} \tilde{E}_k(z) \exp\left[ i \left(\frac{2\pi k}{\tau_f}\right)\tau \right].
\end{equation}
In principle, one could then use this approximation to obtain $E(z,\tau)$ at arbitrary times, including the intermediate times required for the integration of Eq.~\eqref{eq:EOMc}.
However, in general, $E(z,\tau)$ is not periodic on its domain $\tau \in [0,\tau_f]$, and as a result, Eq.~\eqref{eq:efourierapprox} exhibits Gibbs oscillations.
This leads to incorrect approximations of $E(z,\tau)$ at values in between the time steps, especially near $\tau = 0$ and $\tau = \tau_f$.
This issue may be successfully avoided by using local interpolation of the field.

\subsubsection{Verification}
Here, we present evidence for the accuracy of our numerical computations.
In the moving frame, there are no conserved quantities, but we can build one using Eq.~\eqref{eq:energy}.
We augment our system of equations \eqref{eq:EOMfieldFourier} for the field modes with an equation for a variable $\HH(z)$ representing the mean field energy density, satisfying
\begin{equation}\label{eq:EOMhvar}
\partial_z \HH = \frac{\rho}{c\tau_f}\int_0^{\tau_f}\overline{v}(z,\tau)E(z,\tau){\rm d}\tau.
\end{equation}
Hence, $h(z) - U_{\rm EM}(z)$ should be conserved during propagation, where $U_{\rm EM}(z)$ is computed using Eq.~\eqref{eq:Uem}.
Specifically, the integrand of Eq.~\eqref{eq:Uem} is discretized at times $\tau_m = m\Delta \tau$, with the values of the field $E(z,\tau_m)$ obtained by discrete inverse-Fourier transform of the modes $\{\tilde{E}_k\}$, and it is summed using the trapezoidal rule.
The right-hand side of Eq.~\eqref{eq:EOMhvar} is computed similarly.
We take $\HH(0) = U_{\rm EM}(0)$ and monitor the accuracy of our simulations through the error $|\HH(z) - U_{\rm EM}(z)|/\HH(0)$.

The error for the quantum and classical models is plotted in Fig.~\ref{fig:energy_convergence} for simulations in the scattering-propagation experiment setup (see Sec.~\ref{sec:scattering}) for different values of $\Delta z$.
The values of $\Delta z$ are reported in the caption of Fig.~\ref{fig:energy_convergence}.
We also compare the behavior of a third-order Runge-Kutta method (RK3) \cite{Hair93} with the two-step Adams-Moulton method (AM2) employed throughout the main text.
We use the RK3 method here because it is of the same order as the AM2 method.
These plots show that $\HH(z) - U_{\rm EM}(z)$ is a good indicator for the numerical accuracy of our computations, in the sense that the error in the conservation of this quantity has the expected scaling with $(\Delta z)^3$: when $\Delta z$ is divided by $2$, the error goes down by a factor of $8$.
This scaling is observed in all cases, except the classical RK3 case (Fig.~\ref{fig:energy_convergence}c).
In general, the behavior of the energy error for the AM2 method---a gradual growth at a constant rate---is more typical than the behavior of the error for the RK3 method, which exhibits some erratic oscillations.
The cause of this behavior for the RK3 method is unclear, and may be a sign of stiffness of the model equations.
Also, while both methods have a comparable error for each $\Delta z$, the AM2 method only requires two evaluations of $\overline{v}(z,\tau)$ for each space step (in PECE mode), compared to three for the RK3 method.
Because evaluating $\overline{v}(z,\tau)$ is the computationally-expensive step of solving the model equations, this is a significant advantage of AM2.
For these reasons, we selected the AM2 method for the integration of the model equations.

\begin{figure*}
\includegraphics[width=0.75\textwidth]{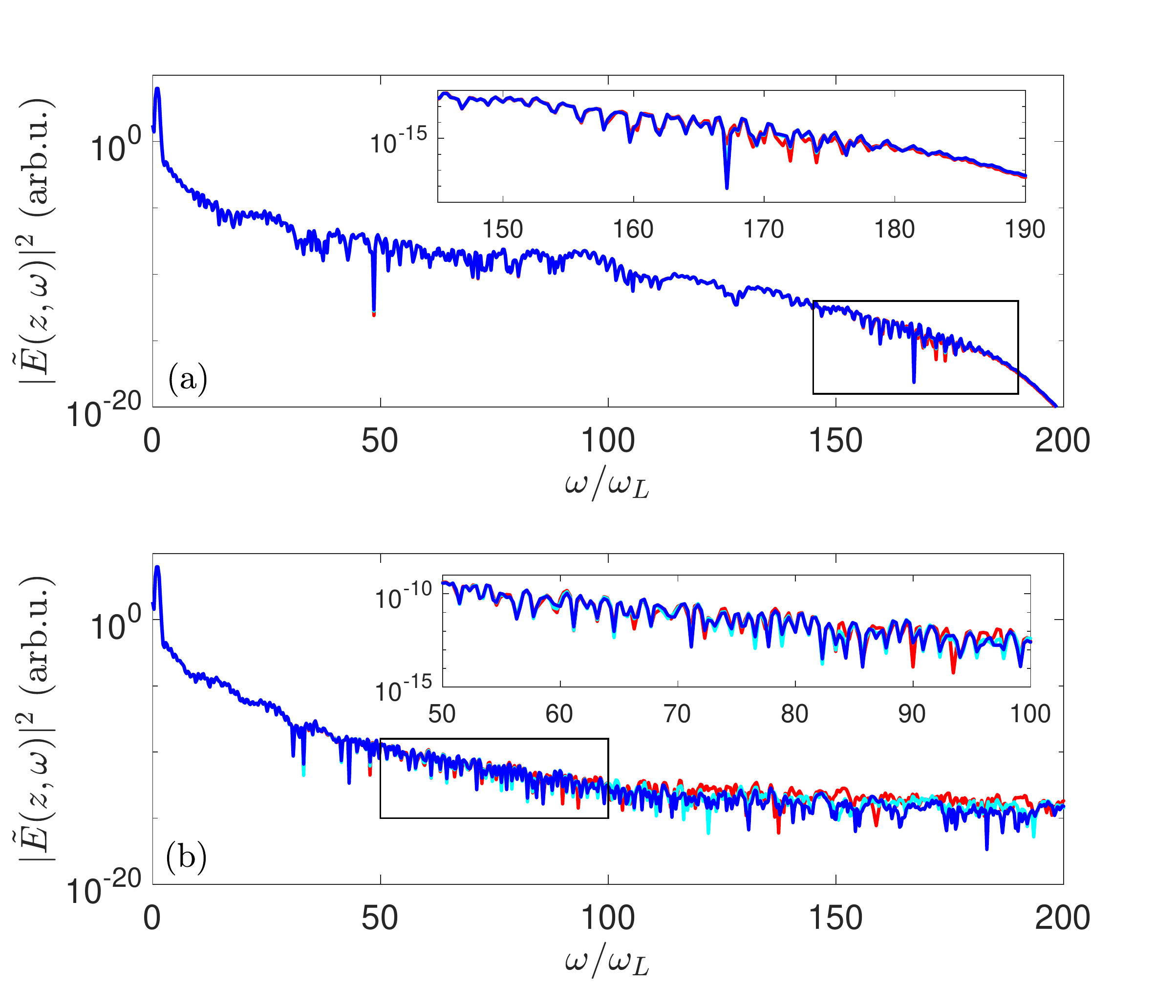}
\caption{Comparison of the electric field spectra after propagation to $z=0.5\mm$ (using the two-step Adams-Moulton method) in quantum (a) and classical (b) cases. The scattering-propagation experiment setup was used for the field and particle initial conditions, with the parameters of Fig.~\ref{fig:fundScat}. A $\sin^4$ window was applied to $E(z,\tau)$ prior to computation of the spectrum. The insets show magnifications of the rectangles, where the discrepancies between calculations with different $\Delta z$ begin to be observable. Red curves: $\Delta z = (21/4)\lambda_L = 6.33\,\,\mu{\rm m}$. Cyan curves: $\Delta z = (21/8) \lambda_L = 3.16\,\,\mu{\rm m}$. Blue curves: $\Delta z = (21/16)\lambda_L = 1.58
\,\,\mu{\rm m}$. For the classical calculation, $N = 602910$ particles were used for the solution of the Liouville equation.}\label{fig:specConvergence}
\end{figure*}

\begin{figure*}
\includegraphics[width=0.85\textwidth]{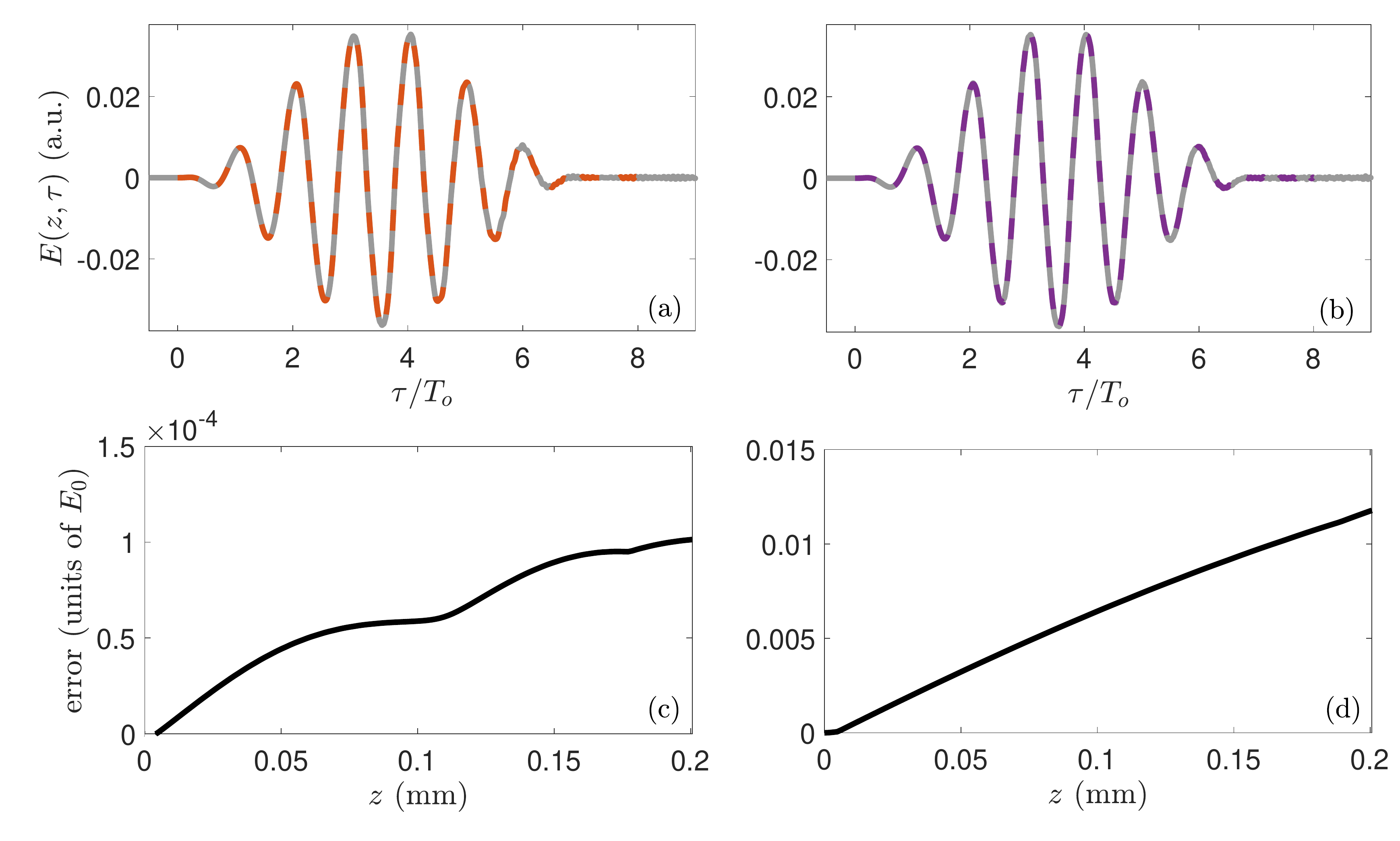}
\caption{Effect of the time domain size on unidirectional pulse propagation with ground-state atoms for the quantum (a),(c) and classical (b),(d) models. The parameters are those of Fig.~\ref{fig:fundSigI5E13}, corresponding to the low ionization probability regime. (a),(b) Electric field $E(z,\tau)$ at $z=0.2\mm$ computed for two different time domains. The solid grey curves are for the domain $\tau \in [-0.5T_o,9T_o]$, and the dashed colored curves are the domain $\tau \in [0,8T_o]$. (c),(d) Error between the fields in the two calculations as a function of $z$. For the classical calculation, $N = 101210$ particles were used for the solution of the Liouville equation.}\label{fig:longbox}
\end{figure*}

Next, we check the convergence of our calculations with respect to $\Delta z$ for the spectrum of $E$ after propagation to a given $z$.
Figure \ref{fig:specConvergence} shows the spectra of $E(z,\tau)$ at $z=0.5\mm$ in the quantum and classical models, with each of the $\Delta z$ of Fig.~\ref{fig:energy_convergence} used for propagation.
In both calculations, we see that the spectra are in good agreement for each $\Delta z$, indicating that this range of $\Delta z$ is small enough to obtain converged results.
Furthermore, we see in both cases that the spectra are indistinguishable up to a particular high frequency, after which the differences between the spectra with different $\Delta z$ are visible.
For the quantum case (Fig.~\ref{fig:specConvergence}a), this frequency is $\omega \approx 170\omega_L$, near the high harmonic cutoff, where the spectrum with $\Delta z = (21/4)\lambda_L$ (red curve) is seen to depart from the spectra computed with smaller $\Delta z$ (cyan and blue curves).
The calculations with $\Delta z = (21/8)\lambda_L$ and $\Delta z = (21/16)\lambda_L$ are indistinguishable from each other, indicating that the quantum calculation has converged with $\Delta z = (21/8) \lambda_L$.

Meanwhile, for the classical case, it is seen in Fig.~\ref{fig:specConvergence}b that $\omega \approx 80 \omega_L$ is the frequency where the $\Delta z = (21/4)\lambda_L$ spectrum begins to depart from the other two spectra.
Further, the $\Delta z = (21/8) \lambda_L$ spectrum is only in excellent agreement with the $\Delta z = (21/16) \lambda_L$ spectrum up until $\omega \approx 100\omega_L$, after which some small deviations are visible.
Thus, the classical calculation is not completely converged with respect to $\Delta z$ for these frequencies.
Note that here, $N = 602910$ particles were used in the classical calculation, indicating that the Liouville equation is converged up to $\omega \approx 135 \omega_L$, as shown in Fig.~\ref{fig:N_convergence}.
Thus, while we cannot draw any conclusions about frequencies greater than $135 \omega_L$, we can say that the frequencies from $100$--$135\omega_L$ experience propagation dynamics on scales smaller than than those resolved by $\Delta z = (21/8) \lambda_L$.
Nevertheless, the behavior of these high harmonics of very low intensity does not seem to influence the lower frequency components of the field, because the latter are the same for each $\Delta z$.
Therefore, we trust the results of the simulations for the lower, more intense harmonics in both the quantum and classical cases, even if we have not fully converged the highest harmonics.
In any case, it is the highest-intensity parts of the spectrum which are determinant for the electron dynamics, as discussed in the main text.

Lastly, we consider the effect of changing the time domain size for simulations in which a pulse propagates through atoms initiated in the ground state.
If $\tau = 0$ is the time at which the incident pulse starts, meaning $E(z,\tau)=0$ for all $\tau < 0$, then extending the time domain over which the simulation takes places to values of $\tau < 0$ should not influence the results.
For times $\tau$ that the field is zero, so too should be the dipole velocity $\overline{v}(z,\tau)$, so that $\partial_z E=0$.
Also, if the atoms are in a stationary state before the onset of the laser pulse, then we should have $\partial_\tau \psi = -i \epsilon \psi $ and $\partial_\tau f =0$ in the quantum and classical cases, respectively, where $\epsilon$ is the energy of the quantum state.
Thus, in theory, integrating the electron fields for times $\tau<0$ should not influence their subsequent evolution.
On the other hand, for a given final time $\tau_f$, changing the domain size to include times $\tau > \tau_f$ should not change the results for $\tau \leq \tau_f$.
This is because, for an arbitrary time $\tau'$, $\overline{v}(z,\tau')$ depends on $E(z,\tau)$ for all $\tau < \tau'$, but no $\tau \geq \tau'$.

Figure \ref{fig:longbox} shows the degree to which our numerical schemes respect these properties of the equations by comparing two different domain sizes: $\tau \in [-0.5T_o,9T_o]$ on the one hand and $\tau \in [0,8T_o]$ on the other.
Here, $\rho = 2\times10^{19} \cc$, $\Delta z = 1.6\,\,\mu{\rm m}$, and the initial conditions for the field and particles are those of Fig.~\ref{fig:fundSigI5E13}.
Figures ~\ref{fig:longbox}a and \ref{fig:longbox}b show that, overall, the fields from the two calculations are in agreement after propagation to $z=0.2\mm$, for the times during which the domains overlap.
Additionally, the field computed on the longer domain has remained zero for $\tau < 0$ throughout propagation.
We define the error as $\max_\tau |E_{\rm long}(z,\tau) - E_{\rm short}(z,\tau)|$, where ``long" and ``short" refer to the fields calculated with the longer and shorter domains, respectively, and the maximum is taken over the time domain at which the two calculations overlap.
This error is plotted in Figs.~\ref{fig:longbox}c and \ref{fig:longbox}d.
Note that we have subtracted off the error at $z=0$, which is nonzero because the time steps in the two calculations were not exactly aligned.
By the end of the simulations, the quantum model error relative to $E_0$ is about $10^{-4}$, while for the classical model the relative error is about $10^{-2}$.
In both cases, the errors are due to the deviations of numerical representations of the electron initial conditions $\psi_0(x)$ and $f_0(x,v)$ from true stationary states.
This leads to differing electron fields at $\tau = 0$ in the long and short domain calculations.
These errors can be reduced by reducing the time step $\Delta \tau$, and by reducing the spatial step $\Delta x$ in the quantum calculation or increasing the number of particles $N$ in the classical calculation.

\subsubsection{Selection of $\Delta z$}
How one selects $\Delta z$ in practice depends on what phenomena one is interested in.
For example, if one is interested in Maker fringes of high harmonics, as seen in Fig.~\ref{fig:phase_emission}a, then $\Delta z$ should be small compared to the wavelength of the harmonic intensity oscillations.
On the other hand, the near-fundamental frequencies and lower-order harmonics tend to evolve over much longer length scales, as seen in Fig.~\ref{fig:low_harmonics}, and the higher harmonics do not seem to influence their evolution.
Therefore, a larger $\Delta z$ may be acceptable for investigating these phenomena (as well as a reduced number of particles for the classical calculation).
In the main text, we err on the side of caution for the 1D simulations by selecting very small values of $\Delta z$, comparable to those used in Figs.~\ref{fig:energy_convergence} and \ref{fig:specConvergence}, where we are sure the high-harmonic parts of the quantum and classical (with enough particles) spectra are converged.
For the 2D simulations, where we only focus on the ellipticity of the full field, we select a much larger $\Delta z$ in order to speed up the calculations.
In any case, the hypothesis underlying the reduced models used in this paper is that the field evolves slowly compared to $\lambda_L$.
If the phenomena of interest begin to occur over smaller length scales, then a more general model---namely, one that includes backward-propagating waves---should be used.

\end{document}